\documentclass[3p, 11pt, authoryear]{elsarticle}

\makeatletter
\def\ps@pprintTitle{%
    \let\@oddhead\@empty
    \let\@evenhead\@empty
    \let\@evenfoot\@oddfoot
    }
\makeatother

\makeatletter
\long\def\@makefntext#1{%
  \parindent=0pt%
  \noindent
  \textsuperscript{\@thefnmark}\,#1%
}
\makeatother

\usepackage[table]{xcolor} 
\usepackage{tabularx}
\usepackage{colortbl}
\usepackage{amsmath}
\usepackage{amssymb}
\usepackage{amsthm}
\usepackage{mathrsfs}
\usepackage[mathscr]{eucal}
\usepackage{bm}
\usepackage{graphicx}
\usepackage[final]{hyperref}
\usepackage[ruled,vlined]{algorithm2e}
\usepackage{enumitem}
\usepackage{array}
\usepackage{booktabs}
\usepackage{multirow}
\usepackage{makecell}
\usepackage[title]{appendix}
\usepackage{empheq}
\usepackage[capitalise, nameinlink, noabbrev]{cleveref}
\usepackage{cases}
\usepackage{textcomp}
\usepackage{gensymb}
\usepackage{subcaption}
\hypersetup{
	colorlinks=true,       
	linkcolor=darkred,        
	citecolor=blue,        
	filecolor=magenta,     
	urlcolor=blue         
}
\usepackage{float}
\usepackage{fancyhdr}
\usepackage[colorinlistoftodos,prependcaption,textsize=normalsize]{todonotes}
\usepackage{regexpatch}

\usepackage{tcolorbox} 

\usepackage{csvsimple}

\usepackage{longtable} 

\makeatletter
\xpatchcmd{\@todo}{\setkeys{todonotes}{#1}}{\setkeys{todonotes}{inline,#1}}{}{}
\newcommand{\publicationnotice}{%
{\normalfont\small
\vspace{-20pt}
\begin{center}
    Published in {\it Journal of the Mechanics and Physics of Solids}\\
    DOI: \url{https://doi.org/10.1016/j.jmps.2026.106920}
\end{center}
\vspace{10pt}}}
\xpatchcmd{\MaketitleBox}
  {\footnotesize\itshape\elsaddress\par\vskip36pt}
  {\footnotesize\itshape\elsaddress\par\vskip36pt\publicationnotice}
  {}
  {\PackageError{arxiv-update}{Could not insert publication notice into title block}{}}
\makeatother

\fancypagestyle{arxivheader}{%
  \fancyhf{}
  \fancyhead[L]{\footnotesize\bfseries\itshape Galloway and Jha}
  \fancyhead[R]{\footnotesize\bfseries\itshape Joint Material--Structural Optimization of hMSMs}
  \fancyfoot[C]{\thepage}
}

\usepackage{etoolbox}
\usepackage{courier}
\usepackage{listings}
\definecolor{codegreen}{rgb}{0,0.6,0}
\definecolor{codegray}{rgb}{0.5,0.5,0.5}
\definecolor{codepurple}{rgb}{0.58,0,0.82}
\definecolor{backcolour}{rgb}{0.91, 0.97, 0.99}

\lstdefinestyle{mystyle}{
    backgroundcolor=\color{backcolour},   
    commentstyle=\color{codegreen},
    keywordstyle=\color{magenta},
    numberstyle=\tiny\color{codegray},
    stringstyle=\color{codepurple},
    basicstyle=\ttfamily\footnotesize,
    breakatwhitespace=false,         
    breaklines=true,                 
    captionpos=b,                    
    keepspaces=true,                 
    numbers=left,                    
    numbersep=5pt,                  
    showspaces=false,                
    showstringspaces=false,
    showtabs=false,                  
    tabsize=2
}

\lstdefinestyle{mystyle2}{
    backgroundcolor=\color{backcolour},
    basicstyle=\ttfamily\footnotesize,
    breakatwhitespace=false,         
    breaklines=true,                 
    captionpos=b,                    
    keepspaces=true,                 
    numbers=left,                    
    numbersep=5pt,                  
    showspaces=false,                
    showstringspaces=false,
    showtabs=false,                  
    tabsize=2
}

\usepackage{orcidlink}

\usepackage{tikz}
\usetikzlibrary{arrows.meta,positioning,shapes.geometric}

\newcommand{\norm}[1]{\left\lVert #1 \right\rVert}

\newcommand{\tr}{\mathrm{tr}\,}

\newtheorem{remark}{Remark}

\newcommand{\blue}[1]{#1}

\newcommand{\dd}{\,\textup{d}}

\newcommand{\bzero}{\bm{0}}

\newcommand{\ba}{\bm{a}}
\newcommand{\bb}{\bm{b}}

\newcommand{\bn}{\bm{n}}

\newcommand{\br}{\bm{r}}

\newcommand{\bt}{\bm{t}}
\newcommand{\bu}{\bm{u}}

\newcommand{\bx}{\bm{x}}

\newcommand{\bB}{\bm{B}}
\newcommand{\bC}{\bm{C}}

\newcommand{\bE}{\bm{E}}
\newcommand{\bF}{\bm{F}}

\newcommand{\bI}{\bm{I}}

\newcommand{\bM}{\bm{M}}

\newcommand{\bP}{\bm{P}}

\newcommand{\bR}{\bm{R}}

\newcommand{\bU}{\bm{U}}

\newcommand{\bX}{\bm{X}}

\newcommand{\btau}{\bm{\tau}}

\renewcommand{\arraystretch}{1.75}

\crefformat{appendix}{#2#1#3}
\Crefname{lstlisting}{Listing}{Listings}

\graphicspath{{figures/}}

\begin{document}

\begin{frontmatter}

\title{Model-Informed Joint Material-Structural Optimization of Hard-Magnetic Soft Materials}

\author[aff1]{Ian Galloway\,\orcidlink{0009-0002-6428-8624}}
\ead{Ian.Galloway@mines.sdsmt.edu}

\author[aff1]{Prashant K. Jha\,\orcidlink{0000-0003-2158-364X}\corref{cor1}}
\ead{prashant.jha@sdsmt.edu; pjha.sci@gmail.com}

\address[aff1]{Department of Mechanical Engineering, South Dakota School of Mines and Technology, Rapid City, SD 57701, USA}

\cortext[cor1]{Corresponding author}

\begin{abstract}
This work develops a model-informed framework for predictive analysis and optimal design of hard-magnetic soft materials (hMSMs). These materials undergo contact-free, field-driven deformation, making them attractive for soft robotics, adaptive structures, and bio-inspired systems. Accurate prediction requires effective structure--property relations, while optimal design requires simultaneous control of structural density, magnetic particle distribution, and remanent magnetization direction. To address these issues, this work makes two main contributions. First, classical rigid-inclusion relations, a Hill self-consistent relation, and constrained-kinematics models are placed into a unified effective shear-modulus framework for particle-filled elastomers. \blue{Six shear-modulus relations that depend on particle volume fraction, together with one default control relation, are combined with three strain-energy density functions to obtain 21 constitutive models.} The results show that the strain-energy density form has a relatively small effect for the actuation problems considered, whereas the effective shear-modulus relation can significantly affect deformation when magnetic material overlaps with regions undergoing substantial deformation. Experimental stress--strain data are then used to select a representative shear-modulus relation, with the Mooney relation giving the lowest errors among the candidates for the dataset and metrics considered. Second, the selected constitutive model is incorporated into a joint material--structural optimization formulation with separate design fields for structural density, magnetic particle volume fraction, and remanent magnetization direction; the \blue{particle volume fraction} field affects both effective stiffness and magnetic loading, while the applied magnetic field is prescribed. Rotational, translational, restorative, and three-dimensional morphing examples show that the framework handles different active design fields, objectives, and single- or multi-load-case formulations, producing non-intuitive hMSM designs with prescribed deformation responses. The framework is implemented in the open-source \texttt{CEADpx/MatTO} repository.
\end{abstract}

\begin{keyword}
Hard-magnetic soft materials, topology optimization, constitutive model selection, structure--property relations, magneto-mechanical coupling, joint material--structural design
\MSC[2020] 74P10 \sep 74P15 \sep 74Q15
\end{keyword}

\end{frontmatter}

\pagestyle{arxivheader}

\section{Introduction}

Hard-magnetic soft materials (hMSMs) are magnetically active composites consisting of hard-magnetic particles embedded within a compliant elastomeric matrix. Unlike soft-magnetic materials, hMSMs retain a remanent magnetization after magnetization and can therefore undergo large, programmable deformations when subjected to external magnetic fields \cite{zhao2019mechanics,kim2022magnetic,lu2024mechanics,narayanan2024hard}. Their contact-free actuation, high compliance, and potential compatibility with soft biological environments make them attractive for applications in soft robotics, shape-morphing structures, adaptive systems, biomedical devices, bio-inspired mechanisms, and soft adhesives \cite{alapan2020reprogrammable,bastola2021shape,chen2022evoking,abbasi2023snap,wu2020multifunctional,lucarini2022recent,zhao2022smart,li2025magnetic}.

The design of hMSM structures raises two closely connected issues. The first is predictive modeling: how constitutive choices, including magnetic energy, strain-energy density, and effective structure--property relations, influence the predicted deformation response. This issue is particularly important here because the work brings together several effective-modulus relations for particle-filled elastomers, which can predict different stiffness values at the same particle volume fraction. The second is optimal design: how structural density, magnetic particle distribution, and remanent magnetization direction should be varied together to achieve prescribed actuation responses. The following subsections organize the introduction around these two issues and then summarize the scope and contributions of this work.

\subsection{Constitutive modeling of hMSMs: Effective shear-modulus and strain-energy density functions}

Predictive modeling of hMSMs requires a finite-strain magneto-mechanical formulation together with structure--property relations that describe how magnetic particle volume fraction modifies the effective elastic moduli. Energy-based continuum formulations have provided a useful foundation for modeling large magnetically induced deformation \cite{zhao2019mechanics,kadapa2022unified}, while magneto-viscoelastic and microstructure-informed formulations have been developed to capture additional features of magneto-active and hard-magnetic soft materials \cite{garcia2019magneto,saxena2013theory,saxena2014nonlinear,rambausek2022computational,stewart2023magneto,garcia2021microstructural}. However, the effective material response also depends strongly on particle volume fraction, particle arrangement, and particle--matrix interactions. Experimental and modeling studies show that filler content and microstructure can substantially modify stiffness, hysteresis, and magnetic actuation response \cite{kramarenko2015magnetic,sorokin2015hysteresis,garcia2021influence,moreno2022effects}. Thus, predictive hMSM modeling requires not only a magneto-mechanical energy formulation, but also appropriate structure--property relations for the effective elastic moduli.

\blue{Magnetic kinematics is a separate constitutive choice in hMSM models with a prescribed applied magnetic field \cite{zhao2019mechanics,mukherjee2021explicit,stewart2023magneto}. An $\bF$-based map can introduce stretch dependence of the magnitude of the remanent magnetic flux density, whereas an $\bR$-based map preserves magnitude and follows the macroscopic polar rotation \cite{danas2024stretchindependent}. \citet{katuseledoes} derive a shape-dependent law for inclusion rotation that recovers the polar map for a circular inclusion but predicts rotations that this map can miss for an inclined anisotropic inclusion under pure stretch. Separately, a Zeeman interaction alone does not generally constitute a complete nonlinear magnetoelastic energy \cite{dorfmann2024energy}. We adopt the ideal hMSM model used in prior mechanics and topology optimization studies \cite{zhao2019mechanics,zhao2022topology}. This approximation excludes self- and demagnetizing fields and therefore does not capture magnetic interactions between spatially separated regions. We examine how elastic constitutive relations that depend on particle volume fraction affect actuation and joint material--structural optimization. We retain the $\bF$ map for the constitutive comparisons and main design examples, and compare $\bF$- and $\bR$-based restorative-beam optimizations under otherwise identical settings in \cref{ss:beamSensitivity}.}

Several structure--property relations have been proposed for particle-filled elastomers and related composites. Classical rigid-inclusion theories describe the effect of particle volume fraction on effective stiffness through modified elastic moduli \cite{guth1945theory,mooney1951viscosity,kerner1956elastic}; a review of several classical results is provided in \cite{ahmed1990review}. Hill's self-consistent theory \cite{hill1965self} provides general relations between the effective elastic moduli, the elastic moduli of the constituent phases, and their respective volume fractions. This type of effective modulus--volume fraction relation has also been used to model \blue{adhesives made from magnetic elastomers} \cite{zhao2022smart}. A complementary viewpoint is provided by homogenization-based and constrained-kinematics models, which start from the observation that the deforming matrix surrounding rigid inclusions need not experience the same deformation invariant as the macroscopic composite \cite{lopez2013nonlinear2,alshammari2019addition}. In these models, particle reinforcement is represented by modifying the invariant supplied to the matrix strain-energy function, typically the first invariant of the right Cauchy--Green deformation tensor, together with a \blue{factor equal to the matrix volume fraction}. This contrast is useful for hMSMs because the magnetic particle volume fraction affects both elastic stiffness and magnetic actuation strength. In this work, these two viewpoints are brought into a common effective shear-modulus representation, enabling direct comparison and model selection.

In addition to the effective shear-modulus relation, the selected strain-energy density function can influence both predicted deformation and numerical robustness. Understanding whether different strain-energy forms and effective modulus models produce similar or distinct responses is therefore important for predictive accuracy and computational efficiency. For example, the isochoric--volumetric neo-Hookean form (\textbf{NH1}) in \cref{eq:NH1} and the St.~Venant--Kirchhoff energy (\textbf{SVK}) in \cref{eq:SVK} often require smaller load increments or fail to converge beyond moderate applied magnetic fields, whereas the reformulated neo-Hookean form (\textbf{NH2}) in \cref{eq:NH2} can permit larger load steps and larger maximum applied fields while producing similar deformation responses in the cases considered.

A central contribution is a unified effective-modulus framework in which classical rigid-inclusion models, Hill's self-consistent relation, and constrained-kinematics models are recast as seven candidate shear-modulus relations $G(\phi)$\blue{, where $\phi$ is the volume fraction of magnetic material}. Combined with three strain-energy density functions, this gives 21 candidate constitutive models and provides a basis for assessing when model choice changes the predicted hMSM response. The sensitivity study is therefore not limited to a single benchmark: uniformly magnetized and \blue{tip-magnetized} beams separate the effects of magnetic loading and particle placement, while wheel, gripper, and three-dimensional actuation examples test whether the same trends persist across different deformation modes. For the tested actuation regimes, the two neo-Hookean forms, \textbf{NH1} and \textbf{NH2}, and the St.~Venant--Kirchhoff model, \textbf{SVK}, give nearly identical deformation measures, whereas the choice of $G(\phi)$ relation becomes important when magnetic material overlaps with regions undergoing substantial deformation. This comparison does not establish weak sensitivity to strain-energy choices for more strongly nonlinear matrices or other loading regimes. The subsequent model-selection step uses experimental stress--strain data from \cite{garcia2021influence} to compare the candidate $G(\phi)$ relations against both extracted shear-modulus values and full true stress--strain responses. For the dataset and metrics considered, the Mooney relation gives the lowest errors and is therefore used with \textbf{NH2} in the joint material--structural optimization examples.

\subsection{Topology optimization and joint material--structural design of hMSMs}

Topology optimization and related inverse-design approaches have been used to distribute material and magnetization in soft magnetic actuators for prescribed deformation, locomotion, and target shapes \cite{lum2016shape,wang2021evolutionary,wu2020evolutionary,lloyd2020data,sundaram2019topology,tian2020designing}. For hMSMs, \citet{zhao2022topology} developed a finite-strain, adjoint-based formulation that optimizes structural topology and weights associated with prescribed remanent-magnetization vectors and can also optimize uniform applied magnetic fields under structural-volume and stress constraints. More recently, \citet{perezgarcia2026topology} combined an experimentally calibrated model of remanence-induced anisotropy with coupled magnetostatic--mechanical finite-element analysis, including the surrounding vacuum, and optimized material distribution, particle content, remanent magnetization, and actuation inputs. Their experiments demonstrate that remanence-induced anisotropy can affect both actuation and optimized designs. These studies establish that simultaneous optimization of structural, material, and magnetic variables is not unique to the present work.

The present formulation combines constitutive-model selection with joint material--structural optimization. \blue{The shear-modulus relations compared in the first part} of this work are placed in a common $G(\phi)$ form and evaluated using published stress--strain data. The selected relation is then embedded in a finite-strain, adjoint-based formulation with separate design fields for structural density $\rho$, particle volume fraction $\phi$, and remanent direction $\theta_{B^r}$. The \blue{particle volume fraction} field changes both effective stiffness through $G(\phi)$ and magnetic loading and is subject to its own \blue{constraint on the domain-averaged particle volume fraction} and local upper bound, while \blue{the energy interpolation ties the magnetic energy to the presence of structural material}. The distinguishing feature emphasized here is therefore the integration of the \blue{selected $G(\phi)$ relation} with \blue{an independently constrained particle volume fraction field}, rather than simultaneous optimization of topology and magnetization as a general capability. The formulation is demonstrated using different active subsets of the design fields in rotational and translational problems, a restorative beam subjected to two opposing load cases, and a three-dimensional morphing surface. The feasibility of realizing such spatially varying designs is supported by magnetically assisted multimaterial printing, which enables local control of shape, composition, and particle orientation \cite{kokkinis2015multimaterial}, and voxel-encoding printing, which enables spatial programming of magnetization density and direction in hMSMs \cite{wu2020evolutionary}.

Although these examples use an isotropic $G(\phi)$ relation and a prescribed applied field, the joint optimization formulation is not tied to these particular constitutive and field choices. Remanence-induced anisotropy and microstructural particle effects can be incorporated through an enriched free energy, with additional internal variables where needed, while self- and demagnetizing fields can be included by augmenting the state problem with a magnetic potential and the surrounding domain, as in \citet{perezgarcia2026topology}. Such extensions would modify the equilibrium equations and adjoint sensitivities but would retain the design fields, constraints, and overall optimization structure introduced here. The present study adopts the ideal-hMSM setting used in prior experimentally validated mechanics and optimization studies \cite{zhao2019mechanics,zhao2022topology} to focus on the coupling among particle concentration, effective stiffness, and magnetic loading in the joint design problem. \blue{A generalization of this joint optimization approach to other stimulus-responsive soft materials, including magneto-active and liquid crystal elastomers, is presented in \citet{galloway2026framework}.}

\subsection{Scope and contributions}

This work has two connected parts. In the first part, effective-modulus relations and constrained-kinematics models for particle-reinforced elastomers are placed into a common shear-modulus form $G(\phi)$, producing six nontrivial structure--property relations together with one default control relation. These seven relations are combined with three strain-energy density functions to obtain 21 constitutive models for sensitivity analysis of hMSM deformation. Experimental stress--strain data from \cite{garcia2021influence} are then used to select a representative shear-modulus relation for the design studies. The second part develops a joint material--structural optimization framework in which magnetic particle volume fraction, remanent magnetization direction, and structural density can be optimized selectively or simultaneously.

\vspace{6pt}
\noindent\textbf{Main contributions:}
\begin{enumerate}
    \item A unified effective shear-modulus representation that brings classical rigid-inclusion relations, a Hill self-consistent relation specialized to rigid inclusions and \blue{a nearly incompressible matrix}, and constrained-kinematics models into a common $G(\phi)$ framework.

    \item A systematic sensitivity analysis of 21 constitutive models obtained by combining seven effective shear-modulus relations with three strain-energy density functions.

    \item Selection of the $G(\phi)$ relation using two data-based metrics: error relative to extracted shear-modulus values and \blue{error between simulated and experimental true stress--strain responses} \cite{garcia2021influence}.

    \item A joint material--structural optimization framework for hMSMs that couples structural density, magnetic particle volume fraction, and continuous remanent magnetization direction through a nonlinear finite-strain magneto-mechanical energy formulation. \blue{The particle volume fraction field affects both effective stiffness and magnetic loading and has independent bounds on its domain average and local values.}

    \item Demonstrations of rotational and translational hMSM designs, a restorative beam under two opposing load cases, and a three-dimensional load-bearing morphing surface, using different active design fields and objective functions.

    \item An open-source implementation in \lstinline{CEADpx/MatTO} to support reproducibility, extension, and broader use in hMSM and field-responsive soft-material design studies.
\end{enumerate}

The effective shear-modulus framework considered here is not limited to hMSMs. \blue{Similar $G(\phi)$ relations} are used for conventional filled elastomers and magnetorheological elastomers, where filler content and particle arrangement can strongly affect the effective stiffness \cite{guth1945theory,jolly1996mrmaterials,ivaneyko2011lattice}. Different research communities often use different reinforcement or homogenization models for this same problem. Therefore, comparing these relations in a common form and selecting a suitable model using experimental data can also be useful for other filled-elastomer systems.

\subsection{Organization of the article}

\cref{s:setup} presents the continuum formulation and constitutive models. \cref{s:modelDependence,s:modelSelection} compare the candidate models and select an effective shear-modulus relation using published data. \cref{s:topOptimFormulation} introduces the joint material--structural optimization formulation, and \cref{s:applications} presents the optimization examples and sensitivity studies. \cref{s:conclusion} summarizes the findings and future directions; the appendices provide supporting derivations and additional numerical results.

\subsection{Software availability}\label{ss:softwareAvailability}

The computational framework is implemented in the open-source \lstinline{CEADpx/MatTO} software repository\footnote{\lstinline{CEADpx/MatTO}: \blue{\url{https://github.com/CEADpx/matto/tree/v0.3.0}}} \cite{galloway2026matto}. \blue{The repository contains the finite-element code, scripts for comparing constitutive models, and the optimization examples used in this work.}

\section{Continuum formulation of hard-magnetic soft materials (hMSMs)}\label{s:setup}
Deformation of hard-magnetic soft materials depends on finite-strain elasticity, magnetic actuation, and the effective material properties induced by magnetic particle inclusions. As a starting point, an energy-based continuum formulation for hMSMs following \cite{zhao2019mechanics} is considered. To isolate the elastic constitutive effects, the form of the magnetic energy is kept fixed, while two elastic modeling choices are varied: the strain-energy density function and the effective shear-modulus relation $G(\phi)$, where $\phi$ is the magnetic particle volume fraction.

The strain-energy density controls the nonlinear elastic response for a given set of moduli. To understand the extent to which the strain-energy density form affects the predicted response, three energy models are analyzed. The first two are variants of neo-Hookean energy, and the third is the St.~Venant--Kirchhoff model. In contrast, the effective shear modulus $G(\phi)$ accounts for particle-induced stiffening of the elastomeric matrix. Several effective-modulus models have been employed in the literature, and it is important to compare them objectively so that an appropriate model can be identified and selected. With this in mind, a unified effective-modulus form is developed through $G(\phi)$. This includes classical rigid-inclusion relations due to Guth, Mooney, and Kerner \cite{guth1945theory,mooney1951viscosity,kerner1956elastic}, Hill's self-consistent theory \cite{hill1965self} \blue{specialized to rigid inclusions in a nearly incompressible matrix}, and constrained-kinematics models for rigid inclusions in elastomers \cite{lopez2013nonlinear2,alshammari2019addition,garcia2021microstructural}. A review of several classical particulate-reinforcement models is given in \cite{ahmed1990review}.

A key point of this section is that these different approaches can be written in a common effective-modulus form through $G(\phi)$. In particular, the constrained-kinematics models, which are originally written by modifying the deformation invariant, such as the trace of the right Cauchy--Green deformation tensor, are recast here as effective shear-modulus relations. This gives seven $G(\phi)$ models that can be combined with the three strain-energy density functions, leading to 21 constitutive models. \cref{s:modelDependence} studies the sensitivity of forward magnetic actuation to these choices, and \cref{s:modelSelection} uses experimental stress--strain data to select a representative $G(\phi)$ relation for the joint material--structural optimization studies.

\subsection{Notation, assumptions, and kinematics}\label{ss:notation}

\cref{tab:notation} summarizes the main notation used throughout the article. Unless stated otherwise, all fields are defined with respect to the reference configuration $\Omega_0$. Bold lowercase symbols denote vectors, and bold uppercase symbols denote second-order tensors, with a few exceptions, such as using $\bX$ to denote the reference position vector. The constitutive formulation uses the following assumptions:
\begin{itemize}
    \item[\textbf{A1.}] \textbf{Quasistatic conservative response.}
    Deformations are treated within finite-strain continuum mechanics. Inertial, viscous, and other dissipative or time-dependent effects are neglected.

    \item[\textbf{A2.}] \blue{\textbf{Magnetic interaction model.}
    The magnetic contribution to the free energy is a Zeeman-type interaction between the remanent magnetic flux density and the externally applied magnetic field. The energy can use either $\bF$ or $\bR$, where $\bR$ is the rotation tensor in the polar decomposition $\bF=\bR\bU$ and $\bU$ is the right stretch tensor. The constitutive comparisons and main design examples use the $\bF$-based energy. To demonstrate that the optimization framework can use either map, we also optimize the restorative beam using the $\bR$-based energy and compare the two designs in \cref{ss:beamSensitivity}. Dipole--dipole interaction energies are not included.}

    \item[\textbf{A3.}] \textbf{Prescribed magnetic fields.}
    The applied magnetic flux density $\bB^a$ is prescribed as spatially constant and is not obtained by solving Maxwell's equations. The remanent magnetic flux density has prescribed magnitude $B^r$, while its direction $\theta_{B^r}(\bX)$ may be prescribed or treated as a design field. Consequently, self- and demagnetizing fields generated by the material are not included.

    \item[\textbf{A4.}] \textbf{Homogenized isotropic elastic response.}
    Magnetic particles are not resolved individually. Their effect is represented through the local volume fraction $\phi(\bX)$, which modifies the effective shear modulus $G(\phi)$. Near-incompressibility is imposed by taking the bulk modulus much larger than the shear modulus. This description assumes that local concentration adequately represents particle-induced stiffness change.
\end{itemize}

Let $\Omega_0 \subset \mathbb{R}^d$ and $\Omega \subset \mathbb{R}^d$ denote the reference and current configurations of the body, respectively, where $d=2,3$ is the spatial dimension. The deformation is described by the motion $\boldsymbol{\chi}:\Omega_0 \rightarrow \Omega$, which maps a material point $\bX\in\Omega_0$ to its spatial position $\bx=\boldsymbol{\chi}(\bX)$ in the current configuration. \blue{The displacement field, the deformation gradient, the right Cauchy--Green deformation tensor, and Green--Lagrange strain tensor are}
\begin{equation*}
    \bu(\bX)=\bx-\bX,
    \quad
    \bF=\nabla_{\bX}\boldsymbol{\chi}
    =\bI+\nabla_{\bX}\bu\blue{,} \quad 
    \bC=\bF^T\bF,
    \quad
    \bE=\frac{1}{2}(\bC-\bI).
\end{equation*}
The Green--Lagrange strain tensor is decomposed into volumetric and deviatoric parts as
\begin{equation}\label{eq:ESplit}
    \bE_v=\frac{\tr\bE}{3}\bI,
    \qquad
    \bE_d=\bE-\bE_v,
    \qquad
    \bE=\bE_v+\bE_d .
\end{equation}
It follows that
\begin{equation}\label{eq:trE}
    \tr(\bE^2)
    =
    \tr(\bE_d^2)
    +
    \frac{1}{3}(\tr\bE)^2 .
\end{equation}
The key invariants used throughout the continuum formulation are $J=\det\bF$ and $I_1=\tr(\bC)$.

\begin{table}[H]
    \centering
    \vspace{-5pt}
    \caption{Notation used in the continuum mechanics and joint material--structural optimization framework.}
    \label{tab:notation}
    \vspace{-5pt}
    \small
    \setlength{\tabcolsep}{6pt}
    \renewcommand{\arraystretch}{1.35}
    \begin{tabular}{p{0.18\linewidth} p{0.76\linewidth}}
        \hline
        \textbf{Symbol} & \textbf{Description} \\
        \hline
        $\Omega_0$, $\Omega$ & Reference and current configurations of the body \\
        $\bX$, $\bx$ & Material and spatial points in the reference and current configurations \\
        $\boldsymbol{\chi}$, $\bu$ & Motion map, $\boldsymbol{\chi}:\Omega_0\rightarrow\Omega$, and displacement field, $\bu(\bX)=\bx-\bX$ \\
        $\bF$ & Deformation gradient, $\bF=\nabla_{\bX}\boldsymbol{\chi}=\bI+\nabla_{\bX}\bu$ \\
        \leavevmode\blue{$\bM(\bF)$, $\bR$} & \leavevmode\blue{Magnetic mapping tensor and rotation tensor in $\bF=\bR\bU$, respectively} \\
        $\bC=\bF^T\bF$ & Right Cauchy--Green deformation tensor \\
        $\bE=(\bC-\bI)/2$ & Green--Lagrange strain tensor \\
        $\bE_v$, $\bE_d$ & Volumetric and deviatoric parts of $\bE$ \\
        $J$, $I_1$ & Jacobian determinant, $J=\det\bF$, and first invariant of $\bC$, $I_1=\tr(\bC)$ \\
        $W$, $W_{\mathrm{elas}}$, $W_{\mathrm{magn}}$ & Total, elastic, and magnetic free-energy densities \\
        $\widehat{W}$, $\widehat{W}_{\mathrm{elas}}$, $\widehat{W}_{\mathrm{magn}}$ & Design-dependent total, elastic, and magnetic free-energy densities \\
        $\mathcal{R}$, $\widehat{\mathcal{R}}^{(\ell)}$ & Weak residuals for the forward and design-dependent equilibrium problems \\
        $\bP$, $\widehat{\bP}$ & First Piola--Kirchhoff stress tensors for the forward and design-dependent formulations \\
        $\bb$, $\bar{\bt}$, $\bar{\bu}$ & Body force, prescribed traction, and prescribed displacement \\
        $\Gamma_u$, $\Gamma_t$ & Displacement and traction parts of the boundary \\
        $\bB^a$, $\bB^r$ & Externally applied and remanent magnetic flux densities \\
        $B^a$, $B^r$ & Magnitudes of the externally applied and remanent magnetic flux densities \\
        $\phi(\bX)$, $\theta_{B^r}(\bX)$ & Magnetic particle volume fraction and remanent magnetization direction \\
        $\rho(\bX)$ & Structural density design variable \\
        $\bar{\rho}$ & Filtered and projected physical density field \\
        $\phi_{\mathrm{phys}}$, $\theta_{B^r,\mathrm{phys}}$ & \leavevmode\blue{Physical} magnetic particle volume fraction and remanent magnetization direction \\
        $G_0$, $K_0$ & Shear and bulk moduli of the host matrix \\
        $G(\phi)$, $K(\phi)$ & Effective shear and bulk moduli of the particle-filled composite \\
        $\widehat{G}$, $\widehat{K}$ & Design-dependent effective shear and bulk moduli \\
        $k_E$, $\phi_{\max}$ & Einstein coefficient used in the effective-modulus relations and maximum particle volume fraction \\
        $\mu_0$ & Magnetic permeability of vacuum \\
        $Q_m$, $D_m$ & Displacement measure for model $m$ and its relative percentage difference from the default model \\
        $\mathrm{rE}$, $\mathrm{rRMSE}$ & Relative error in extracted $G(\phi)$ data and relative root-mean-square error in stress--strain response \\
        $Q_h$, $\mathcal{H}$, $\mathcal{P}_\rho$ & Raw design space, Helmholtz-filter operator, and density projection operator \\
        $r_\rho$, $r_\phi$, $r_\theta$ & Filter radii for $\rho$, $\phi$, and $\theta_{B^r}$ \\
        $v_{\max}$, $\Phi_{\max}$ & \leavevmode\blue{Upper bounds on domain-averaged structural and particle volume fractions} \\
        $f$, $f_{\mathrm{rot}}$, $f_{\mathrm{tr}}$, $f_{\mathrm{comp}}$ & Generic, rotational, translational, and compliance objective functions \\
        \hline
    \end{tabular}
\end{table}

\subsection{Magneto-mechanical energy and equilibrium}\label{ss:energyEquilibrium}

The material response is derived from a total free-energy density consisting of elastic and magnetic contributions,
\begin{equation*}
    W(\bX,\bF,\phi,\bB^r,\bB^a)
    =
    W_{\mathrm{elas}}(\bX,\bF,\phi)
    +
    W_{\mathrm{magn}}(\bX,\bF,\phi,\bB^r,\bB^a).
\end{equation*}
\blue{The elastic energy is introduced in \cref{ss:strainEnergyModels}. The magnetic contribution is written as
\begin{equation}\label{eq:magneticEnergy}
    W_{\mathrm{magn}}
    =-\frac{\phi(\bX)}{\mu_0}
    \big[\bM(\bF)\bB^r(\bX)\big]\cdot\bB^a.
\end{equation}
Here, $\bM(\bF)$ is the tensor that maps the reference remanent magnetic flux density in the magnetic interaction term. This work primarily uses $\bM(\bF)=\bF$, which gives the model of \citet{zhao2019mechanics}. In the restorative-beam example in \cref{ss:beamSensitivity}, we also take $\bM(\bF)=\bR(\bF)$, where $\bR$ is the rotation tensor in the right polar decomposition $\bF=\bR\bU$, with $\bU=(\bF^T\bF)^{1/2}$. This choice rotates the remanent vector without changing its magnitude. The map is fixed for each optimization. The two-dimensional expression for $\bR(\bF)$ is given in \labelcref{s:appMagneticMaps}.

In \cref{eq:magneticEnergy}, $\mu_0=4\pi\times10^{-7}~\mathrm{N/A^2}$ is the magnetic permeability of vacuum, $\phi:\Omega_0\to[0,\phi_{\max}]$ is the magnetic particle volume fraction, and $\bB^a$ is the externally applied magnetic flux density. Also, $\bB^r$ is the intrinsic remanent magnetic flux density of the particles, so that $\phi\bB^r$ is the effective remanent magnetic flux density of the composite in this homogenized model. The optimization examples use $\phi_{\max}=0.30$.}

\begin{figure}[h]
    \centering
    \includegraphics[width=0.4\linewidth]{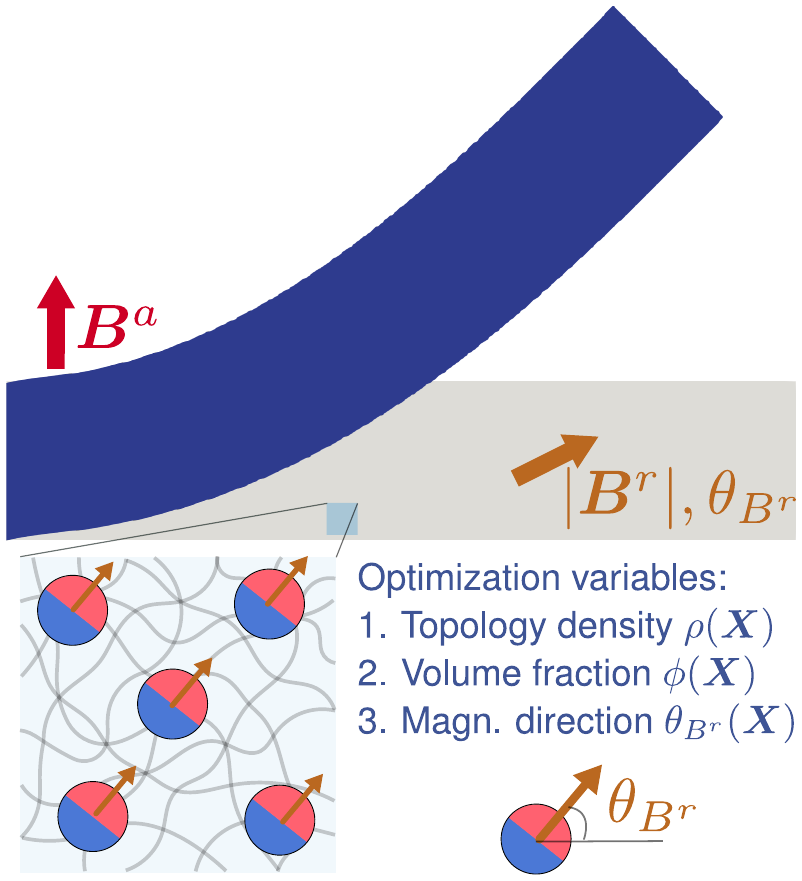}
    \caption{Undeformed and deformed hard-magnetic soft material beam with a representative volume element containing polymer fibers and magnetic particles. In the joint material--structural optimization formulation, the design fields are the structural density $\rho$, magnetic particle volume fraction $\phi$, and remanent magnetization direction $\theta_{B^r}$.}
    \label{fig:DesignVariableSetup}
\end{figure}

\blue{The applied magnetic flux density $\bB^a$ is prescribed. The remanent magnetic flux density has prescribed magnitude $B^r$ and orientation $\theta_{B^r}(\bX)$, which may be specified or treated as a design field:
\begin{equation*}
    \bB^r=B^r[\cos\theta_{B^r},\sin\theta_{B^r}]^T\quad\text{(2D)},
    \qquad
    \bB^r=B^r[\cos\theta_{B^r},\sin\theta_{B^r},0]^T\quad\text{(3D)}.
\end{equation*}
The three-dimensional example in \cref{ss:design3d} allows fully three-dimensional deformation while constraining the remanent direction to the $x$--$y$ plane.}
A schematic of the hMSM structure and the design variables used later in the joint optimization formulation is shown in \cref{fig:DesignVariableSetup}.

Given $(\phi,\bB^r,\bB^a,\bb,\bar{\bt},\bar{\bu})$, the strong form is to find $\bu$ such that
\begin{equation*}
\begin{alignedat}{3}
    \nabla_{\bX}\cdot\bP+\bb \;&= \mathbf{0},
        &&\qquad \bX \in \Omega_0, \\
    \bu \;&= \bar{\bu},
        &&\qquad \bX \in \Gamma_u \subset \partial\Omega_0, \\
    \bP\bn \;&= \bar{\bt},
        &&\qquad \bX \in \Gamma_t := \partial\Omega_0 \setminus \Gamma_u .
\end{alignedat}
\end{equation*}
Here, $\bb$ is the body force per unit reference volume, $\bar{\bu}$ is the prescribed displacement on $\Gamma_u$, $\bar{\bt}$ is the prescribed traction on $\Gamma_t$, and $\bn$ is the outward unit normal on the reference boundary. The first Piola--Kirchhoff stress tensor is obtained from the total free energy as
\begin{equation}
    \bP = \frac{\partial W}{\partial \bF}.
\end{equation}

To state the weak form, the trial and test spaces are defined as
\begin{equation*}
    V_u
    =
    \left\{
    \bu\in [H^1(\Omega_0)]^d:
    \bu=\bar{\bu}\ \text{on } \Gamma_u
    \right\},
    \qquad
    V_v
    =
    \left\{
    \delta\bu\in [H^1(\Omega_0)]^d:
    \delta\bu=\mathbf{0}\ \text{on } \Gamma_u
    \right\}\,.
\end{equation*}
Here, the prescribed displacement $\bar{\bu}$ enters through the trial space $V_u$, while the displacement boundary $\Gamma_u$ enters both the trial and test spaces. The Sobolev space $H^1(\Omega_0)$ is defined as
\begin{equation*}
    H^1(\Omega_0)
    =
    \left\{
    v \in L^2(\Omega_0) :
    \nabla v \in [L^2(\Omega_0)]^d
    \right\},
\end{equation*}
and $[H^1(\Omega_0)]^d$ denotes the corresponding vector-valued space.

\blue{The residual functional for $\bu\in V_u$, $\delta\bu\in V_v$, and prescribed data $(\phi,\bB^r,\bB^a,\bb,\bar{\bt},\bar{\bu})$ is}
\begin{equation}\label{eq:weakResidual}
\begin{aligned}
    \mathcal{R}
    &\left(
        \bu,\delta\bu;
        \phi,\bB^r,\bB^a,
        \bb,\bar{\bt},\bar{\bu}
    \right)
    \\
    &:=
    \int_{\Omega_0}
    \bP:\nabla_X\delta\bu\,dV
    -
    \int_{\Omega_0}
    \bb\cdot\delta\bu\,dV
    -
    \int_{\Gamma_t}
    \bar{\bt}\cdot\delta\bu\,dS ,
\end{aligned}
\end{equation}
where $\bP=\partial W/\partial \bF$ and therefore depends on the prescribed fields through the free-energy density $W$.

The weak form is to find $\bu\in V_u$ such that
\begin{equation*}
    \mathcal{R}
    \left(
        \bu,\delta\bu;
        \phi,\bB^r,\bB^a,
        \bb,\bar{\bt},\bar{\bu}
    \right)
    =
    0,
    \qquad
    \forall \delta\bu\in V_v\,.
\end{equation*}
This variational form is used in the nonlinear finite-element implementation for the forward simulations and topology optimization studies. To compute the first Piola--Kirchhoff stress, automatic differentiation in the FEniCSx finite-element framework \cite{baratta2023dolfinx} is used.

\subsection{Elastic strain-energy density functions}\label{ss:strainEnergyModels}

This subsection defines the elastic strain-energy density functions used in the model-dependence study. Three forms are analyzed numerically: two compressible neo-Hookean forms and one St.~Venant--Kirchhoff form. The effective shear modulus $G(\phi)$ is left unspecified here and is introduced through the structure--property relations in \cref{ss:Gmodels}.

To model near-incompressibility, the bulk modulus of the host material is taken to be much larger than the shear modulus,
\begin{equation}\label{eq:bulkModulus}
    K_0 = 500G_0,
\end{equation}
where $G_0$ and $K_0$ denote the shear and bulk moduli of the host matrix. This gives
\begin{equation*}
    \nu_0
    =
    \frac{3K_0-2G_0}{6K_0+2G_0}
    =
    0.499 .
\end{equation*}
Throughout this work, the effective bulk modulus is taken as
\begin{equation}\label{eq:Kphi}
    K(\phi)=K_0 .
\end{equation}
The Lam\'e parameter is related to the shear and bulk moduli as
\begin{equation*}
    \lambda(\phi)
    =
    K(\phi)-\frac{2G(\phi)}{3}.
\end{equation*}

The first elastic model is the isochoric--volumetric compressible neo-Hookean energy
\begin{flalign}
\qquad \textbf{(Strain Energy: NH1)} \qquad
&&
W_{\mathrm{elas}}
=
\frac{G(\phi)}{2}
\left(J^{-2/3}I_1-3\right)
+
\frac{K(\phi)}{2}(J-1)^2 ,
&&
\label{eq:NH1}
\end{flalign}
where $J=\det\bF$ and $I_1=\tr \bC$.

The second elastic model is another compressible neo-Hookean energy \cite{korobeynikov2026two},
\begin{flalign}
\qquad \textbf{(Strain Energy: NH2)} \qquad
&&
W_{\mathrm{elas}}
=
\frac{G(\phi)}{2}
\left(I_1-3-2\ln J\right)
+
\frac{K(\phi)}{2}(J-1)^2 .
&&
\label{eq:NH2}
\end{flalign}
Compared with \textbf{NH1}, this form avoids the isochoric factor $J^{-2/3}$ and was found to be more robust in the numerical examples considered.

The third elastic model is the St.~Venant--Kirchhoff energy, which is quadratic in the Green--Lagrange strain:
\begin{equation*}
    W_{\mathrm{elas}}
    =
    G(\phi)\tr(\bE^2)
    +
    \frac{\lambda(\phi)}{2}
    \left(\tr\bE\right)^2 .
\end{equation*}
Using the split in \cref{eq:ESplit} and the identity in \cref{eq:trE}, this energy can be written as
\begin{equation*}
    W_{\mathrm{elas}}
    =
    G(\phi)\tr(\bE_d^2)
    +
    \left[
        \frac{\lambda(\phi)}{2}
        +
        \frac{G(\phi)}{3}
    \right]
    \left(\tr\bE\right)^2 .
\end{equation*}
Since the term in brackets is $K(\phi)/2$, the equivalent form for the St.~Venant--Kirchhoff energy is
\begin{flalign}
\qquad \textbf{(Strain Energy: SVK)} \qquad
&&
W_{\mathrm{elas}}
=
G(\phi)\tr(\bE_d^2)
+
\frac{K(\phi)}{2}
\left(\tr\bE\right)^2 .
&&
\label{eq:SVK}
\end{flalign}

Together, \textbf{NH1}, \textbf{NH2}, and \textbf{SVK} define the three strain-energy density functions used in the model-dependence study. The effective shear-modulus relations $G(\phi)$ are introduced next.

\subsection{Effective shear-modulus relations}\label{ss:Gmodels}

The magnetic particle volume fraction $\phi$ strongly influences the elastic stiffness of the composite. In this work, this effect is represented through an effective shear modulus $G(\phi)$, while the bulk modulus is kept fixed as in \cref{eq:Kphi}. Seven models for $G(\phi)$ are analyzed based on different theoretical approaches: one default or control model, four effective-modulus relations for rigid inclusions, and two relations obtained by recasting constrained-kinematics models into an effective-modulus form.

\paragraph{Rigid-inclusion approach}
The first group of models accounts for the stiffening effect of rigid particles by modifying the shear modulus directly. This group includes classical relations due to Guth, Mooney, and Kerner \cite{guth1945theory,mooney1951viscosity,kerner1956elastic}, together with Hill's self-consistent formula for the effective elastic moduli of two-phase composites \cite{hill1965self}. \blue{The Hill relation used here is derived in \labelcref{s:appHill} for rigid inclusions in a nearly incompressible matrix.}

The resulting effective shear-modulus models are
\begin{equation}\label{eq:GmodelsStrategy1}
    \begin{aligned}
        \textbf{(G Model: Default)} \hspace{80pt}
        &G(\phi) = G_0, \\
        \textbf{(G Model: Guth)} \hspace{80pt}
        &G(\phi) = G_0\left(1+k_E\phi+14.1\phi^2\right), \\
        \textbf{(G Model: Mooney)} \hspace{80pt}
        &G(\phi) = G_0\exp\left(\frac{k_E\phi}{1-1.35\phi}\right), \\
        \textbf{(G Model: Kerner)} \hspace{80pt}
        &G(\phi) = G_0\left(1+\frac{A\phi}{1-\phi}\right), \\
        \textbf{(G Model: Hill)} \hspace{80pt}
        &G(\phi) = \frac{G_0}{1-k_E\phi}.
    \end{aligned}
\end{equation}
Here, $k_E=2.5$ is the Einstein dilute-limit coefficient for rigid spherical inclusions in an incompressible matrix. The \textbf{Default} model is included as a control case and ignores particle-induced stiffening. For the Kerner relation, the coefficient $A$ is given by \cite[Equation 8]{ahmed1990review} for $\nu = 0.5$:
\begin{equation*}
    A
    =
    \frac{15(1-\nu)}{8-10\nu} = k_E = 2.5\,.
\end{equation*}

\begin{remark}
    As a volume fraction in a particle-filled elastomer, $\phi$ lies below $1$. The constitutive models impose a stricter mathematical restriction: the \textbf{Mooney} relation is singular at $\phi=1/1.35$, and the \textbf{Hill} relation is singular at $\phi=1/2.5=0.4$. Thus, comparisons involving all seven $G(\phi)$ relations require $\phi<0.4$, pointwise if $\phi$ varies spatially. The optimization examples use the smaller local upper bound $\phi_{\max}=0.30$.
\end{remark}

\paragraph{Constrained-kinematics approach}
The second strategy accounts for rigid inclusions by modifying the invariant $I_1$ to represent constrained matrix deformation. In contrast to the rigid-inclusion models above, this approach is originally expressed at the level of the strain-energy density rather than directly through an effective shear modulus. The first model considered in this group is the homogenized incompressible $I_1$-based strain-energy model of Lopez-Pamies \cite{lopez2013nonlinear2},
\begin{equation}\label{eq:LPHomogenized}
    \Psi_{\mathrm{LP}}(I_1,\phi)
    =
    (1-\phi)
    \Psi_0\left(
        \frac{I_1-3}{(1-\phi)^{7/2}}+3
    \right),
\end{equation}
where $\Psi_0(I_1)$ is an $I_1$-based strain-energy function for the host material. The factor $(1-\phi)$ accounts for the volume fraction of deforming matrix material, while the modified invariant passed to $\Psi_0(\cdot)$ accounts for the constrained kinematics induced by rigid inclusions.

To connect this constrained-kinematics form to an effective shear-modulus relation, consider the neo-Hookean-type deviatoric energy used in \textbf{NH2}, ignoring the volumetric penalty term,
\begin{equation*}
    \Psi_0(I_1)=\frac{G_0}{2}(I_1-3).
\end{equation*}
Substituting this choice into \cref{eq:LPHomogenized} gives
\begin{equation*}
    \Psi_{\mathrm{LP}}(I_1,\phi)
    =
    (1-\phi)\frac{G_0}{2}
    \left(
        \frac{I_1-3}{(1-\phi)^{7/2}}
    \right)
    =
    \frac{1}{2}\frac{G_0}{(1-\phi)^{5/2}}(I_1-3).
\end{equation*}
This naturally leads to an effective shear modulus, referred to here as \textbf{LP},
\begin{flalign*}
\qquad \textbf{(G Model: LP)} &&
G(\phi)
=
\frac{G_0}{(1-\phi)^{5/2}},
&& 
\end{flalign*}
so that the corresponding effective strain energy simplifies to
\begin{equation*}
    \Psi_{\mathrm{LP}}(I_1,\phi)
    =
    \frac{G(\phi)}{2}(I_1-3)\,.
\end{equation*}
Thus, the constrained-kinematics model has the same form as $\Psi_0=G_0(I_1-3)/2$, but with $G_0$ replaced by the effective shear modulus $G(\phi)$.

This observation allows the constrained-kinematics model to be interpreted as a structure--property relation for $G(\phi)$, placing it in the same framework as the models in \cref{eq:GmodelsStrategy1}. This transition from a modified-invariant formulation to an effective-modulus formulation is important because it allows the \textbf{LP} relation to be combined with different strain-energy density functions, including \textbf{NH1} in \cref{eq:NH1} and \textbf{SVK} in \cref{eq:SVK}. 

More recently, the Lopez-Pamies approach was extended in \cite{garcia2021microstructural} to incorporate the structure--property relation proposed in \cite{alshammari2019addition}. They consider the following form of the energy:
\begin{equation*}
    \Psi_{\mathrm{LPA}}(I_1,\phi)
    =
    (1-\phi)
    \Psi_0\left(\chi(\phi)(I_1-3)+3\right),
\end{equation*}
where $\chi(\phi)$ captures the influence of $\phi$ on the modified invariant. It is given by \cite{alshammari2019addition}
\begin{equation*}
    \chi(\phi)
    =
    1+0.67g\phi+1.62(g\phi)^2,
\end{equation*}
where $g$ is a coefficient fixed below. If we define a shear modulus function, referred to as \textbf{LPA}, as follows
\begin{flalign}
\qquad \textbf{(G Model: LPA)} &&
G(\phi)
=
G_0(1-\phi)\chi(\phi)\,,
&& \label{eq:LPA}
\end{flalign}
then, for $\Psi_0=G_0(I_1-3)/2$, the effective strain energy becomes
\begin{equation*}
    \Psi_{\mathrm{LPA}}(I_1,\phi)
    =
    \frac{G(\phi)}{2}(I_1-3).
\end{equation*}
Again, the effective strain energy has the same form as $\Psi_0$, but with the shear modulus replaced by $G(\phi)$. The same argument used above to extend the \textbf{LP} relation to \textbf{NH1} and \textbf{SVK} can therefore be applied to the \textbf{LPA} relation. Thus, \textbf{LPA} is also treated as a structure--property relation for $G(\phi)$ and is combined with all three strain-energy density functions.

\begin{remark}
The recasting of the \textbf{LP} and \textbf{LPA} constrained-kinematics models as effective shear-modulus relations is a key step in this work because it converts modified-invariant models into additional $G(\phi)$ relations that can be compared directly with rigid-inclusion models. For the specific choice $\Psi_0(I_1)=G_0(I_1-3)/2$ used above, this recasting is algebraically exact. The effective shear-modulus relations that emerge from this recasting can then be used as structure--property relations in other strain-energy forms, including \textbf{NH1} and \textbf{SVK}.
\end{remark}

It remains to fix the constant $g$ in $\chi(\phi)$. In \cite{alshammari2019addition}, $g$ appears as a shape or cluster factor in the reinforcement function
\begin{equation*}
    \chi(\phi)
    =
    1+0.67g\phi+1.62(g\phi)^2 .
\end{equation*}
In the present work, the \textbf{LPA} form is fixed by setting the constant $g$ rather than fitting it using the data. The value of $g$ is chosen so that the linear term in the amplification function $\chi(\phi)$ matches the classical Einstein/Guth coefficient $k_E=2.5$ used in the rigid-inclusion relations. This gives
\begin{equation*}
    0.67g=k_E,
    \qquad
    g=\frac{k_E}{0.67}\approx 3.73 .
\end{equation*}

\begin{remark}
    With $g = 3.73$, the amplification function $\chi(\phi)$ has the same first-order reinforcement coefficient as the classical rigid-inclusion model of Guth \cite{guth1945theory}. The final \textbf{LPA} effective modulus $G(\phi)=G_0(1-\phi)\chi(\phi)$ also includes the \blue{matrix volume fraction $(1-\phi)$} inherited from the constrained-kinematics formulation. Therefore, \textbf{LPA} with $g = 3.73$ is not intended to match the dilute-limit slope of the Guth relation; however, it provides a conservative structure--property relation that retains the classical first-order amplification inside $\chi(\phi)$. The parameter $g$ may also be determined by fitting to experimental data as done in \cite{alshammari2019addition}. In this work, we keep $g = 3.73$ fixed.
\end{remark}

\subsection{Constitutive relations summary}\label{ss:constitutiveSummary}

In summary, this work considers three elastic strain-energy density functions and seven effective shear-modulus relations $G(\phi)$. Combining each strain-energy density with each $G(\phi)$ relation gives a total of $21$ candidate constitutive models. \cref{tab:modelSummary} in \cref{s:modelSelection} summarizes the shear-modulus models and reports the model-selection metrics.

\section{Numerical investigations of model dependence in hMSM deformation}
\label{s:modelDependence}

This section investigates how the constitutive choices introduced in \cref{s:setup} affect magnetically actuated deformation of hMSMs. The magnetic energy formulation and governing equations are kept fixed, while the elastic strain-energy density function and effective shear-modulus relation $G(\phi)$ are varied. The goal is to determine when these constitutive choices produce meaningful differences in the predicted response. The numerical studies begin with cantilever beams, where loading magnitude and magnetic-material placement can be varied in a controlled way, and then consider wheel and gripper geometries with more complex deformation modes. Model dependence is quantified by comparing deformation fields and averaged displacement metrics across the candidate constitutive models.

\subsection{Cantilever beam benchmarks}

The cantilever beam benchmarks are used to separate two possible sources of constitutive model sensitivity. The first study varies the applied magnetic field magnitude for a uniformly magnetized beam, while the second localizes the magnetic material near the beam tip. Together, these cases distinguish sensitivity caused by actuation strength from sensitivity caused by spatial overlap between magnetic material and the primary bending region.

\subsubsection{Problem setup}
\label{ss:beamSetup}

\begin{figure}[h]
    \centering
    \includegraphics[width=0.95\linewidth]{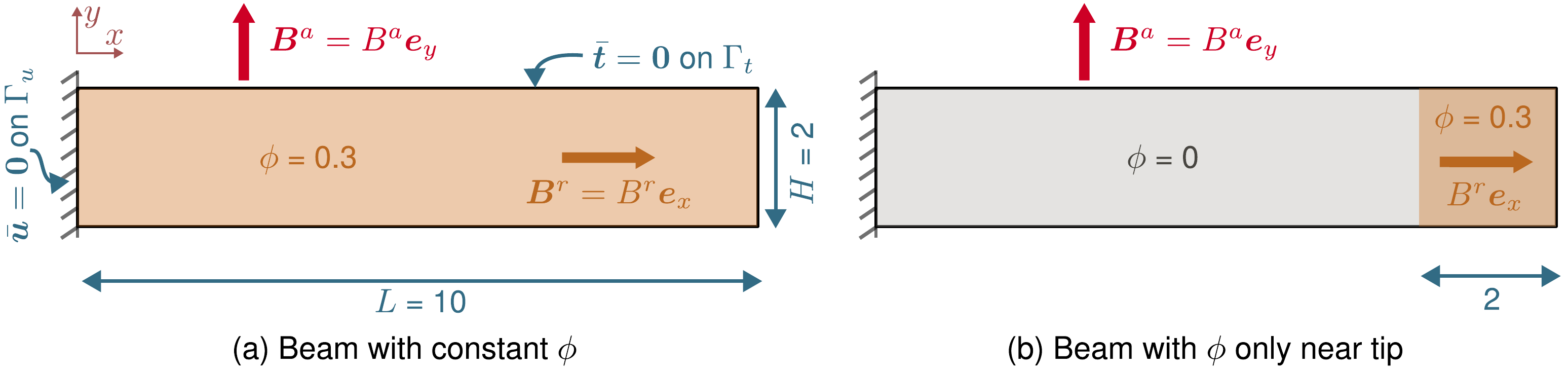}
    \caption{Cantilever beam benchmarks used to study model dependence. Both beams have the same dimensions and are clamped on the left boundary with the remaining boundary traction free. A spatially uniform applied magnetic field is imposed upward, while the remanent magnetic flux density is prescribed along the beam axis with $B^r=100~\mathrm{mT}$. Matrix shear modulus is $G_0 = 100~\mathrm{kPa}$, while the bulk modulus is given by \cref{eq:bulkModulus}. In (a), magnetic material is uniformly distributed with $\phi=0.3$. In (b), magnetic material is localized near the beam tip, with $\phi=0.3$ for $X>8~\mathrm{mm}$ and $\phi=0$ otherwise. Lengths are in millimeters.}
    \label{fig:beamSetup}
\end{figure}

The geometry, boundary conditions, magnetic field direction, remanent magnetization direction, and magnetic particle distribution for the two beam benchmarks are provided in \cref{fig:beamSetup}. For each loading condition, the strain-energy density functions in \cref{ss:strainEnergyModels} and the effective shear-modulus relations in \cref{ss:Gmodels} are evaluated.

For the uniformly magnetized beam in \cref{fig:beamSetup}(a), three applied magnetic field magnitudes are considered: $25$, $100$, and $400~\mathrm{mT}$, representing small, medium, and large loading, respectively. For the tip-magnetized beam in \cref{fig:beamSetup}(b), the applied magnetic field magnitude is fixed at $400~\mathrm{mT}$.

\begin{figure}[h]
    \centering
    \includegraphics[width=0.8\linewidth]{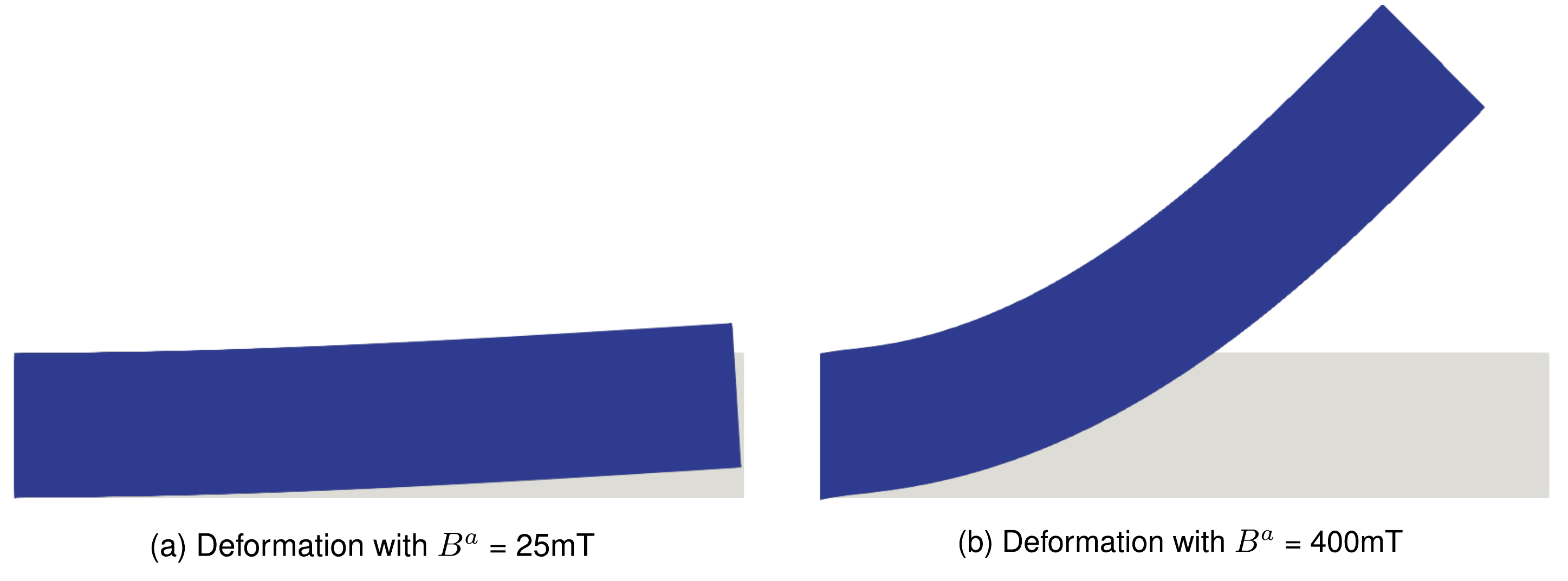}
    \caption{Representative deformation response of the uniformly magnetized cantilever beam under small magnetic actuation (a) and large magnetic actuation (b). The results correspond to the Mooney model with \textbf{NH2} strain energy and the uniformly magnetized beam in \cref{fig:beamSetup}(a).}
    \label{fig:beamDeformation}
\end{figure}

Representative deformed configurations under small and large magnetic actuation for the uniformly magnetized beam are shown in \cref{fig:beamDeformation}.

A mesh convergence study for the beam benchmark is provided in \labelcref{s:appConvergence}. Based on that study, a characteristic mesh size of $h_c = 0.08~\mathrm{mm}$ is used for all beam simulations in this subsection.

\subsubsection{Results}

To compare the deformation responses, a measurement portion of the reference boundary, denoted by $\Gamma_{\mathrm{meas}}\subset\partial\Omega_0$, is prescribed. In the finite-element implementation, this boundary is represented by the corresponding set of measurement points or nodes, denoted by the index set $\mathcal{I}_{\mathrm{meas}}$. For a given constitutive model $m$, let $\bu_m$ denote the equilibrium displacement field. The magnitude of the average displacement vector over the measurement boundary is taken as the quantity of interest:
\begin{equation}\label{eq:QoI}
    Q_m
    =
    \norm{
        \frac{1}{|\Gamma_{\mathrm{meas}}|}
        \int_{\Gamma_{\mathrm{meas}}}
        \bu_m(\bX)\,\dd S
    }
    \approx
    \norm{
        \frac{1}{N_{\mathrm{meas}}}
        \sum_{i\in\mathcal{I}_{\mathrm{meas}}}
        \bu_m(\bX_i)
    }\,,
\end{equation}
where $\norm{\ba} = \sqrt{\ba \cdot \ba}$ and $N_{\mathrm{meas}}=|\mathcal{I}_{\mathrm{meas}}|$ is the number of measurement points. The corresponding quantity for the \textbf{Default} model is denoted by $Q_D$.

The absolute relative percentage difference between model $m$ and the default model is then defined as
\begin{equation}\label{eq:PercentDiff}
    D_m
    =
    \left|
    \frac{Q_m-Q_D}{Q_D}
    \right|
    \times 100 .
\end{equation}
The quantity $D_m$ measures the relative effect of the selected structure--property model on the predicted deformation, using the default model $G(\phi)=G_0$ as the control case.

For the cantilever beam benchmark, the measurement boundary is the right edge of the beam,
\begin{equation*}
    \Gamma_{\mathrm{meas}}
    =
    \{\blue{[L,Y]^T}\in\partial\Omega_0:\,0\leq Y\leq H\},
\end{equation*}
where $H$ is the beam height; see \cref{fig:beamSetup}. The corresponding discrete set $\mathcal{I}_{\mathrm{meas}}$ consists of the measurement points located on this right edge.

\begin{figure}[h]
    \centering
    \includegraphics[width=0.7\linewidth]{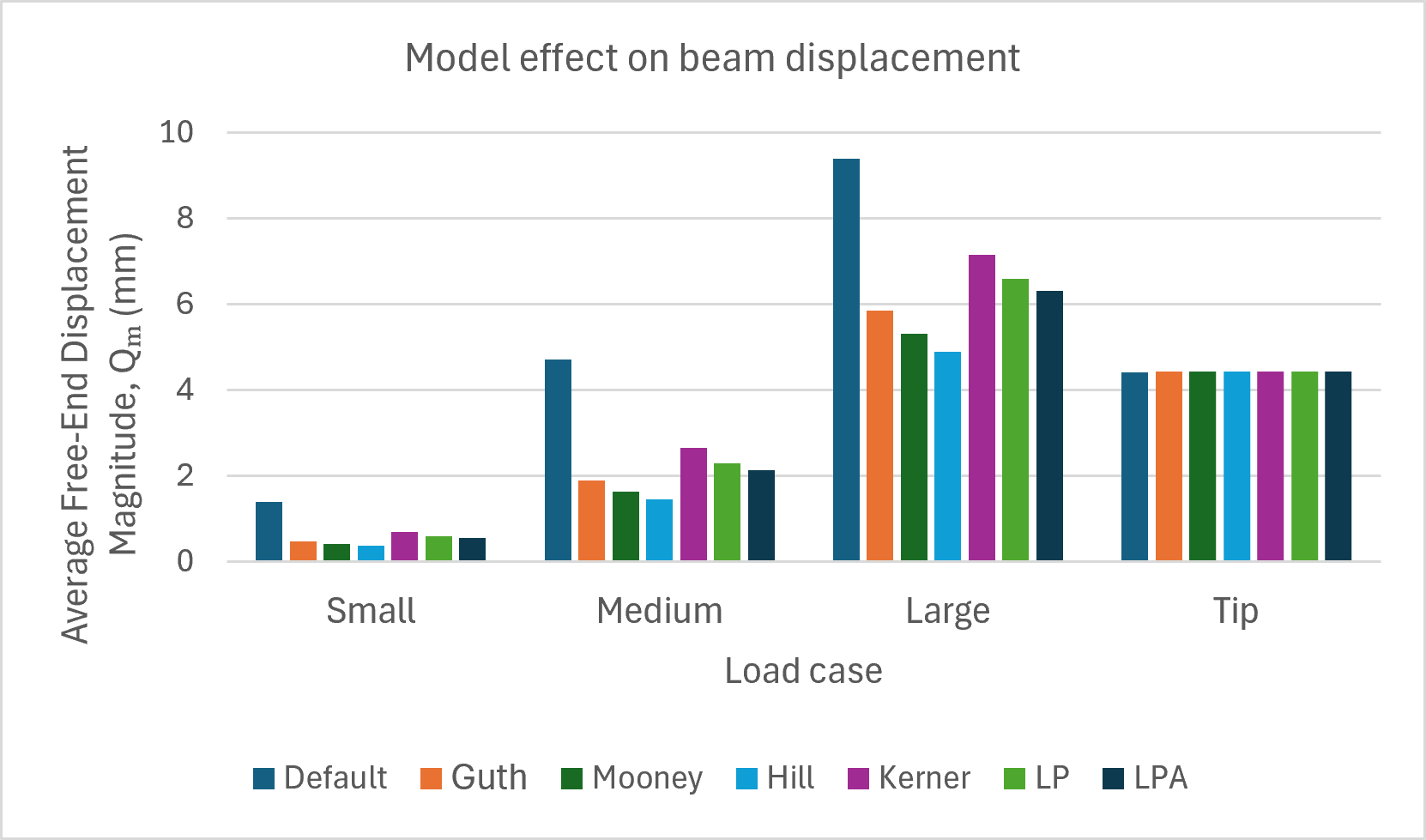}
    \caption{Average free-end displacement magnitude $Q_m$ from \cref{eq:QoI} predicted by the different structure--property models for the cantilever beam benchmarks. The first three results, from left to right, correspond to the uniformly magnetized beam in \cref{fig:beamSetup}(a) subjected to increasing applied magnetic field magnitudes $B^a$. The rightmost result corresponds to the tip-magnetized beam in \cref{fig:beamSetup}(b) subjected to $B^a=400~\mathrm{mT}$. Both beams are $10\times2~\mathrm{mm}$, clamped on the left, and use $h_c=0.08~\mathrm{mm}$, $G_0=100~\mathrm{kPa}$, $K_0=500G_0$, and $B^r=100~\mathrm{mT}$ along the beam axis. The uniform case has $\phi=0.30$ throughout, while the tip case has $\phi=0.30$ for $X>8~\mathrm{mm}$ and zero elsewhere. Results use \textbf{NH2} and all seven $G(\phi)$ relations; the upward applied fields are $25$, $100$, and $400~\mathrm{mT}$ for the uniform case.}
    \label{fig:beamResults}
\end{figure}

Comparisons among the strain-energy density functions for a fixed $G(\phi)$ relation show little variation: \textbf{NH1} and \textbf{NH2} give nearly identical responses, while \textbf{SVK} differs by less than $1\%$. Therefore, only the results from the \textbf{NH2} energy model are presented in the following discussion to isolate the influence of the effective shear-modulus relation $G(\phi)$. The average free-end displacement magnitude $Q_m$ predicted by the different effective shear-modulus relations is presented in \cref{fig:beamResults}.

For the uniformly magnetized beam in \cref{fig:beamSetup}(a), all loading cases in \cref{fig:beamResults} exhibit the same qualitative ordering. The default model predicts the largest displacement, followed by the Kerner, LP, LPA, Guth, Mooney, and Hill models in decreasing order of displacement magnitude. This ordering is consistent with the effective stiffness predicted by the different $G(\phi)$ relations: larger values of $G(\phi)$ produce smaller deformation under the same magnetic loading.

The rightmost result in \cref{fig:beamResults} corresponds to the tip-magnetized beam in \cref{fig:beamSetup}(b). In contrast to the uniformly magnetized cases, the predicted displacements are nearly identical across the structure--property models. This result suggests that the overlap between magnetic material and highly deforming bending regions plays an important role in determining the sensitivity to the selected $G(\phi)$ relation.

The metric $D_m$ is summarized in \cref{tab:beamRelativeDifference}. A clear reduction in structure--property model sensitivity is observed as the loading magnitude increases, with the largest deviations occurring for the small loading condition and progressively smaller differences observed for the medium and large loading cases. The Hill and Mooney models consistently produce the largest deviations from the default model, while the Kerner model remains the closest among the modified structure--property relations. The LP and LPA models exhibit intermediate behavior, producing larger displacements than the Guth, Mooney, and Hill models but smaller displacements than the Kerner model. These results indicate that model sensitivity becomes increasingly important for small actuation responses and fine deformation control.

\begin{table}[H]
\centering
\caption{Relative percentage change $D_m$ in the beam displacement measure $Q_m$ with respect to the default model, computed using \cref{eq:PercentDiff}. The results are for \textbf{NH2} strain energy.}
\label{tab:beamRelativeDifference}
\small
\setlength{\tabcolsep}{4pt}
\begin{tabular}{llcccccc}
\toprule
\textbf{Beam setup} & $\blue{B^a}$ (mT) & \textbf{Guth} & \textbf{Mooney} & \textbf{Hill} & \textbf{Kerner} & \textbf{LP} & \textbf{LPA} \\
\midrule
Uniform, Fig.~\ref{fig:beamSetup}(a)
& $25$ (small)
& 65.1 & 70.0 & \cellcolor{blue!8}\textbf{73.4} & 49.6 & 57.2 & 60.4 \\

Uniform, Fig.~\ref{fig:beamSetup}(a)
& $100$ (medium)
& 59.9 & 65.3 & \cellcolor{blue!8}\textbf{69.1} & 43.4 & 51.3 & 54.7 \\

Uniform, Fig.~\ref{fig:beamSetup}(a)
& $400$ (large)
& 37.6 & 43.4 & \cellcolor{blue!8}\textbf{47.9} & 23.6 & 29.8 & 32.7 \\
Tip-localized, Fig.~\ref{fig:beamSetup}(b)
& \cellcolor{blue!8}$400$
& \cellcolor{blue!8}0.4 & \cellcolor{blue!8}0.4 & \cellcolor{blue!8}\textbf{0.5} & \cellcolor{blue!8}0.3 & \cellcolor{blue!8}0.3 & \cellcolor{blue!8}0.4 \\
\bottomrule
\end{tabular}
\end{table}

The tip-magnetized beam setup further shows that model sensitivity depends strongly on the spatial overlap between magnetic material and the primary bending region. As highlighted in the final row of \cref{tab:beamRelativeDifference}, all structure--property models predict nearly identical displacements for the tip-magnetized case, with relative differences no greater than $0.5\%$. Thus, when magnetic material is removed from the primary bending region, the influence of the selected $G(\phi)$ relation becomes negligible for this benchmark.

\subsection{Wheel and gripper benchmarks}\label{s:WheelGripperBench}

The wheel and gripper benchmarks extend the cantilever study to geometries with different deformation modes and different magnetic-material placement.

\subsubsection{Problem setup}

\begin{figure}[h]
\centering
\includegraphics[width=\linewidth]{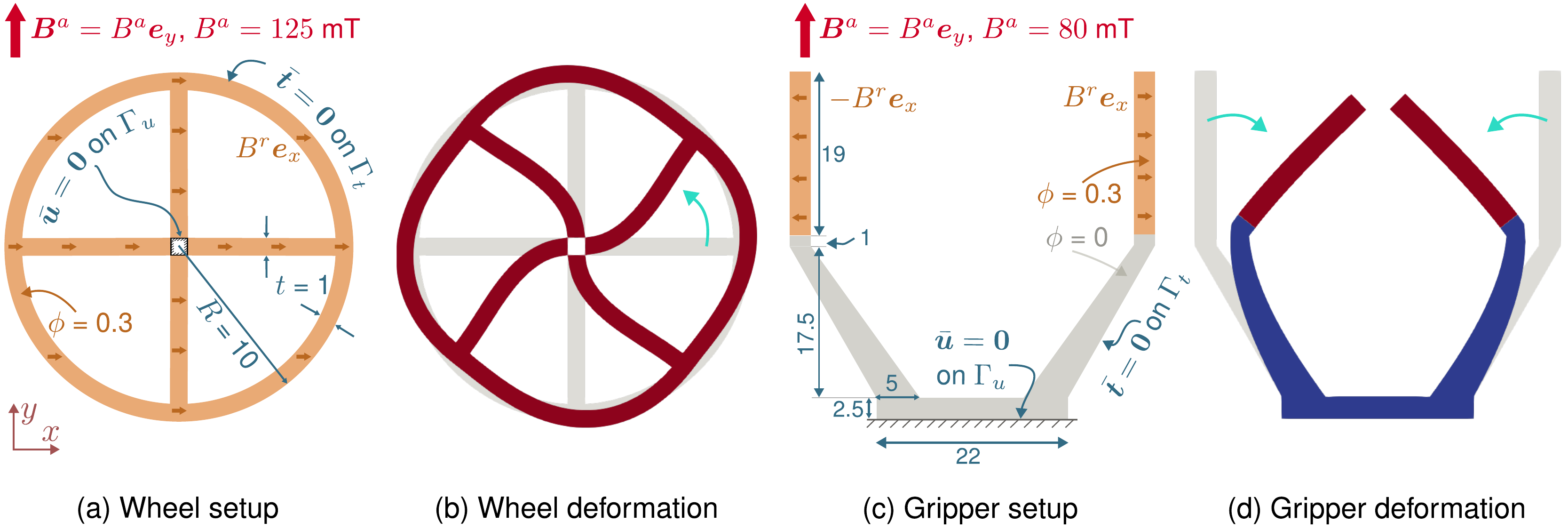}
\caption{Setup and deformed configurations for the wheel and gripper benchmarks. Panel (a) presents the wheel setup, with the hub boundary fixed and all remaining boundaries traction free. The remanent magnetic flux density magnitude is $B^r = 100~\mathrm{mT}$. Panel (b) gives the resulting rotational deformation. Panel (c) presents the gripper setup, where magnetic material is localized in the gripping arms. The bottom boundary is fixed, while all remaining boundaries are traction free. Opposite remanent magnetic flux density directions are prescribed in the two gripping arms with $B^r=40~\mathrm{mT}$. Panel (d) gives the resulting closing deformation. For both benchmarks, $G_0$ is fixed to $G_0 = 100~\mathrm{kPa}$ and $K_0$ is given by \cref{eq:bulkModulus}. All lengths are in millimeters.}
\label{fig:wheelGripperSetup}
\end{figure}

The wheel and gripper setups are provided in \cref{fig:wheelGripperSetup}. For both benchmarks, the strain-energy density functions in \cref{ss:strainEnergyModels} and the effective shear-modulus relations in \cref{ss:Gmodels} are evaluated. The wheel is subjected to a spatially uniform upward applied magnetic field of $B^a=125~\mathrm{mT}$, and the remanent magnetic flux density magnitude is fixed at $B^r=100~\mathrm{mT}$. For the gripper, the applied field magnitude is $B^a=80~\mathrm{mT}$, and the remanent magnetic flux density magnitude is $B^r=40~\mathrm{mT}$. Mesh convergence studies for the wheel and gripper benchmarks are provided in \labelcref{s:appConvergence}. Based on those studies, characteristic mesh sizes of $h_c=0.06~\mathrm{mm}$ and $h_c=0.08~\mathrm{mm}$ are used for the wheel and gripper benchmarks, respectively.

\subsubsection{Results}

The same displacement measure $Q_m$ and relative percentage change $D_m$ defined in \cref{eq:QoI,eq:PercentDiff} are used for the wheel and gripper benchmarks. Only the measurement set changes between the two cases. For the wheel, the measurement points are selected near the right outer rim,
\begin{equation*}
    \Gamma_{\mathrm{meas}}^{\mathrm{wheel}}
    =
    \left\{
    \bX = \blue{[X, Y]^T}\in \partial \Omega_0:
    X^2+Y^2 = R^2,\;
    X>0,\;
    |Y|\leq 4t
    \right\}\,,
\end{equation*}
where $R$ is the outer radius and $t$ is the rim thickness. For the gripper, the measurement points are selected on the top edge of the right gripping arm,
\begin{equation*}
    \Gamma_{\mathrm{meas}}^{\mathrm{gripper}}
    =
    \left\{
    \bX = \blue{[X, Y]^T}\in\partial\Omega_0:
    Y=Y_{\mathrm{tip}},\;
    X_{\min}\leq X\leq X_{\max}
    \right\}\,,
\end{equation*}
where $X_{\min} = 18.5~\mathrm{mm}$, $X_{\max} = 21~\mathrm{mm}$, and $Y_{\mathrm{tip}} = 39.82~\mathrm{mm}$. The corresponding discrete sets $\mathcal{I}_{\mathrm{meas}}$ consist of the selected measurement points in these regions.

\begin{figure}[h]
\centering
\begin{subfigure}[b]{0.48\textwidth}
    \centering
    \includegraphics[width=\textwidth]{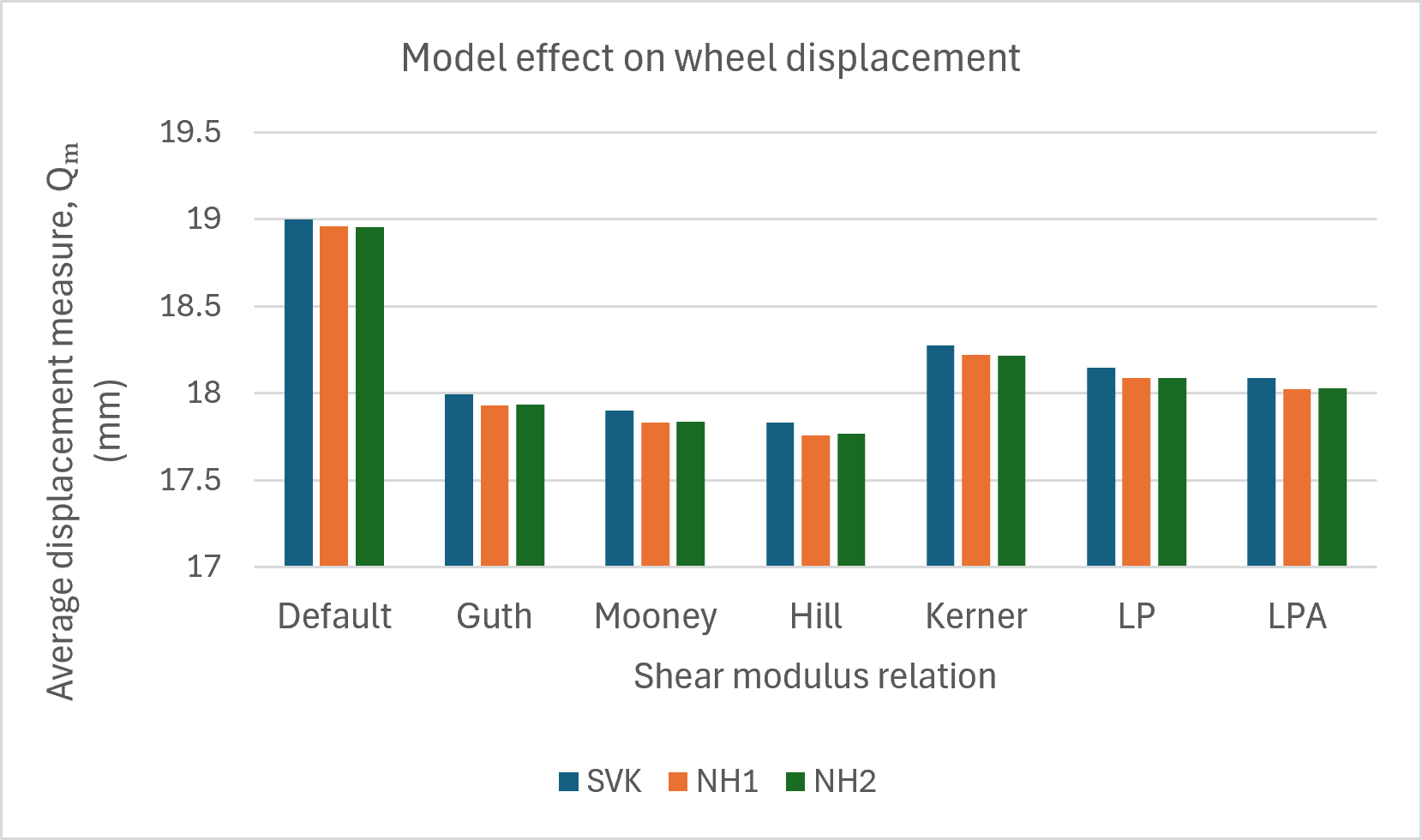}
    \caption{Wheel}
    \label{fig:WheelGraph}
\end{subfigure}
\hfill
\begin{subfigure}[b]{0.48\textwidth}
    \centering
    \includegraphics[width=\textwidth]{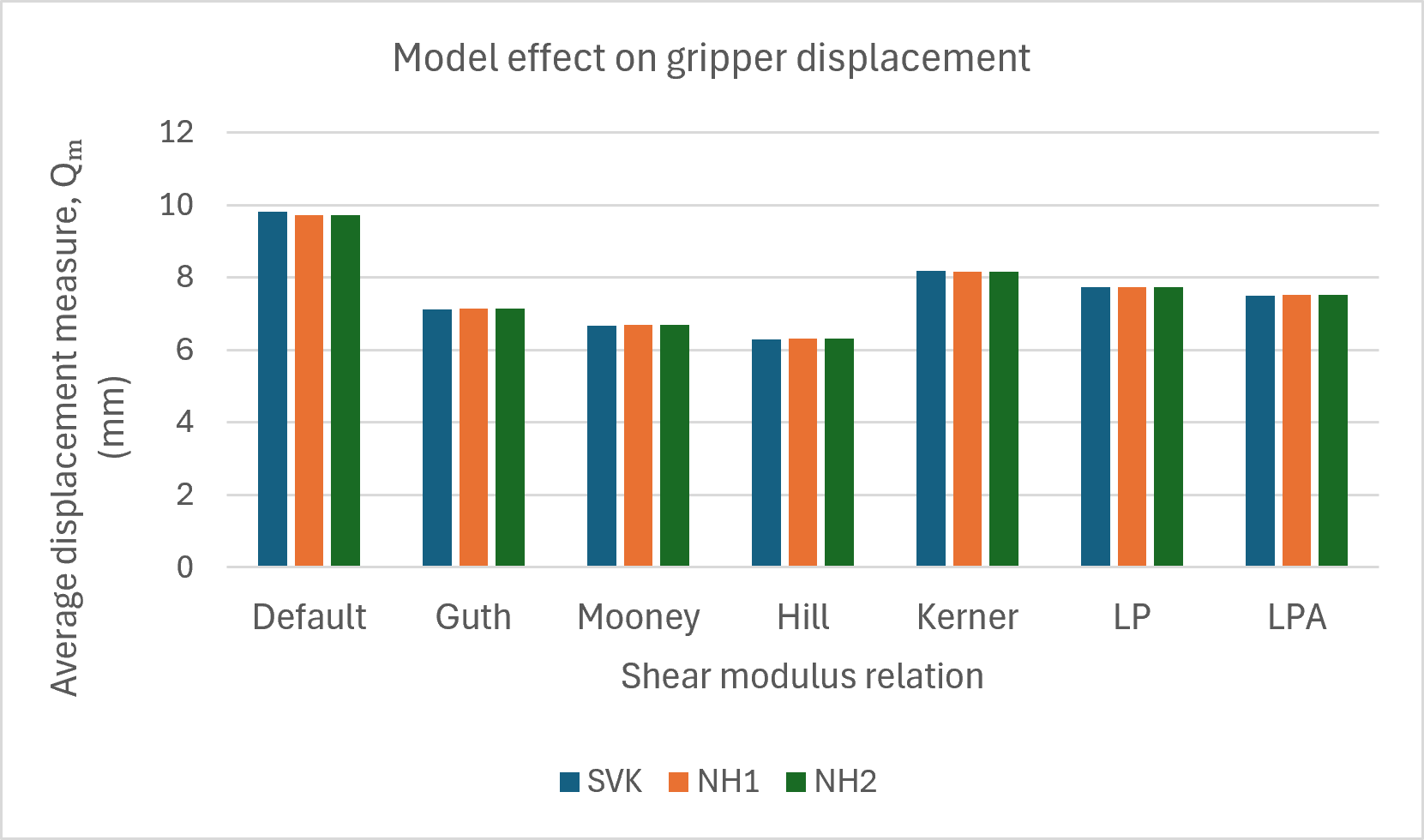}
    \caption{Gripper}
    \label{fig:GripperGraph}
\end{subfigure}
\caption{Average displacement magnitude predicted by different strain-energy density functions and effective shear-modulus relations for the wheel and gripper benchmarks.}
\label{fig:WheelGripperGraph}
\end{figure}

The displacement measures $Q_m$ predicted by the different strain-energy density functions and effective shear-modulus relations are presented in \cref{fig:WheelGripperGraph}. Similar to the cantilever beam benchmarks, the strain-energy density function has a relatively small effect: \textbf{NH1} and \textbf{NH2} give nearly identical responses, while \textbf{SVK} differs only slightly. Therefore, the quantitative comparison in \cref{tab:wheelGripperRelativeDifference} uses \textbf{NH2} to isolate the influence of the effective shear-modulus relation $G(\phi)$.

The results in \cref{fig:WheelGripperGraph} confirm the qualitative ordering observed in the cantilever beam benchmarks. The default model predicts the largest deformation measure, while the modified structure--property models predict smaller values because they increase the effective stiffness through $G(\phi)$. The spread among the structure--property models is visibly larger for the wheel benchmark than for the gripper benchmark.

The metric $D_m$ is summarized in \cref{tab:wheelGripperRelativeDifference}. The wheel benchmark is substantially more sensitive to the selected structure--property relation than the gripper benchmark. For the wheel, the relative percentage change ranges from $15.9\%$ for the Kerner model to $35.1\%$ for the Hill model. For the gripper, the corresponding values remain much smaller, ranging from $3.9\%$ to $6.3\%$.

\begin{table}[H]
\centering
\caption{Relative percentage change $D_m$ from \cref{eq:PercentDiff} in the displacement measure $Q_m$ with respect to the \textbf{Default} model for the wheel and gripper benchmarks.}
\label{tab:wheelGripperRelativeDifference}
\small
\begin{tabular}{lcccccc}
\toprule
\textbf{Geometry} & \textbf{Guth} & \textbf{Mooney} & \textbf{Hill} & \textbf{Kerner} & \textbf{LP} & \textbf{LPA} \\
\midrule
Wheel   & 26.6 & 31.2 & \cellcolor{blue!8}\textbf{35.1} & 15.9 & 20.5 & 22.7 \\
\rowcolor{blue!8}
Gripper & 5.4  & 5.9  & \textbf{6.3}  & 3.9  & 4.6  & 4.9  \\
\bottomrule
\end{tabular}
\end{table}

The larger sensitivity of the wheel is consistent with the magnetic-material distribution in \cref{fig:wheelGripperSetup}(a). Magnetic material is distributed throughout the wheel, including regions that undergo substantial deformation during actuation. Therefore, changes in $G(\phi)$ directly affect those deforming portions of the wheel and produce noticeable differences in $Q_m$.

In contrast, the gripper setup in \cref{fig:wheelGripperSetup}(c) localizes magnetic material in the gripping arms, while much of the deformation occurs in the compliant hinge regions near the base. As shown in \cref{tab:wheelGripperRelativeDifference}, this separation reduces the relative effect of the selected $G(\phi)$ relation. This behavior is consistent with the tip-magnetized cantilever beam in \cref{fig:beamSetup}(b), where the absence of magnetic material in the primary bending region makes the model dependence negligible.

\subsection{Three-dimensional beam actuation tests}\label{ss:ZhaoActuationTests}

The preceding benchmarks examine model dependence in two-dimensional geometries. As a final numerical test, a three-dimensional beam actuation problem is considered to check whether the same trends persist in a fully three-dimensional deformation setting.

\subsubsection{Problem setup}

\begin{figure}[h]
    \centering
    \includegraphics[width=\linewidth]{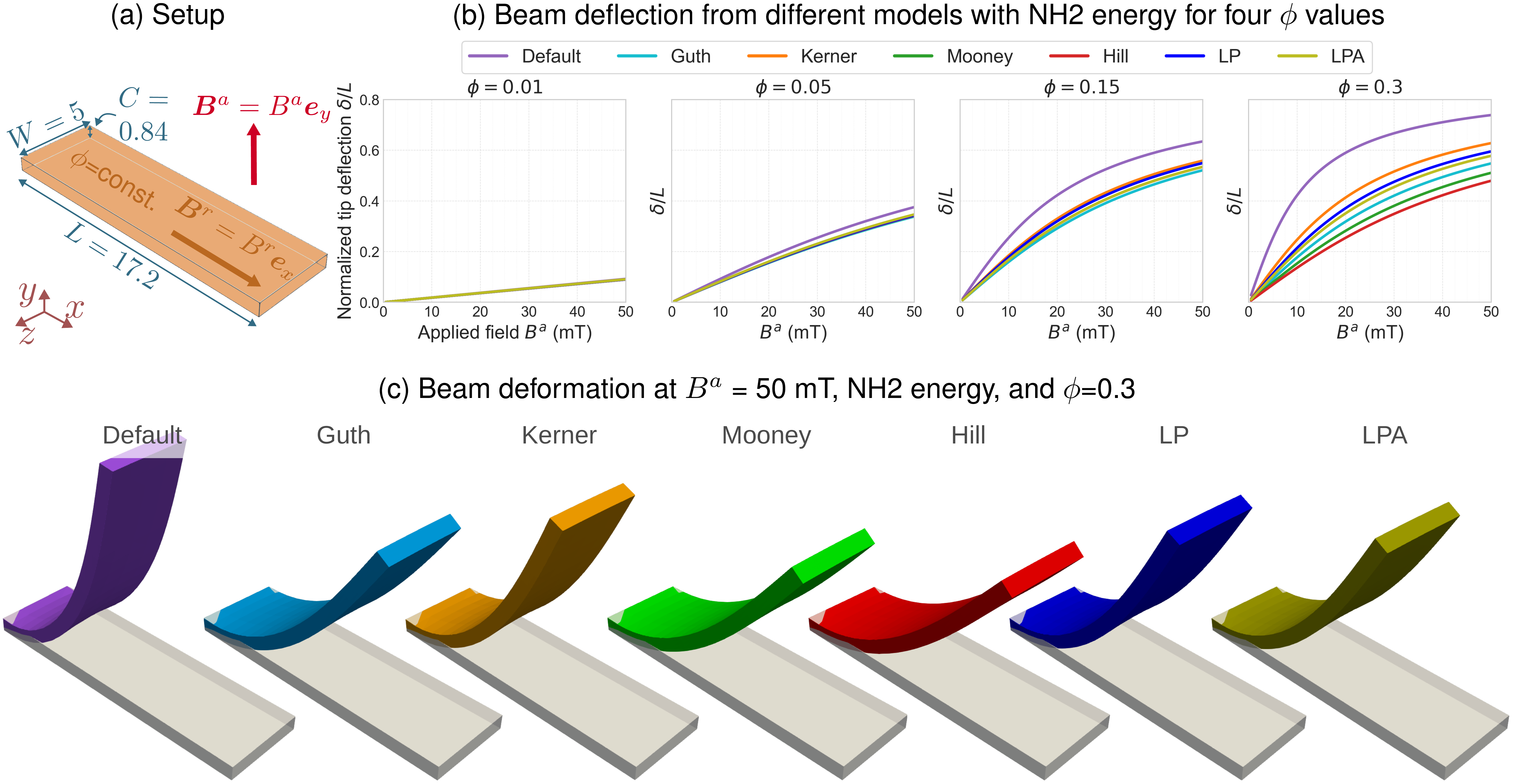}
    \caption{Three-dimensional actuation cuboid benchmark. Panel (a) presents the problem setup. Here, $G_0 = 303\,\mathrm{kPa}$, $K_0$ is given by \cref{eq:bulkModulus}, $B^r = 143\,\mathrm{mT}$, and $B^a = 50\,\mathrm{mT}$ is applied over 100 load increments. The domain is discretized with a structured mesh of $20 \times 4 \times 4$ elements. The left face (at $x= 0$) is fixed and the remaining boundary is traction free. Lengths are in millimeters. Panel (b) includes the normalized tip deflection $\delta_m/L$ versus $B^a$ plots for \textbf{NH2} with the seven $G(\phi)$ models. Panel (c) visualizes the deformed configurations at $B^a=50~\mathrm{mT}$ and $\phi=0.30$ for \textbf{NH2}.}
    \label{fig:actuationCuboidSetup}
\end{figure}

The setup, based on the benchmark example in \cite{zhao2019mechanics}, is provided in \cref{fig:actuationCuboidSetup}(a). The magnetic material is uniformly distributed with volume fraction $\phi \in \{0.01,\,0.05,\,0.15,\,0.30\}$. The response is measured using the normalized vertical tip deflection
\begin{equation*}
    Q_m = \frac{\delta_m}{L}
    =
    \frac{u_{y,m}(\bX_{\mathrm{tip}})}{L},
\end{equation*}
where $u_{y,m}(\bX_{\mathrm{tip}})$ is the vertical displacement of the monitored free-end point $\bX_{\mathrm{tip}} = \blue{[L, W, C]^T}$ (see \cref{fig:actuationCuboidSetup}(a)) predicted by model $m$.

\subsubsection{Results}

\begin{figure}[h]
    \centering
    \includegraphics[width=\linewidth]{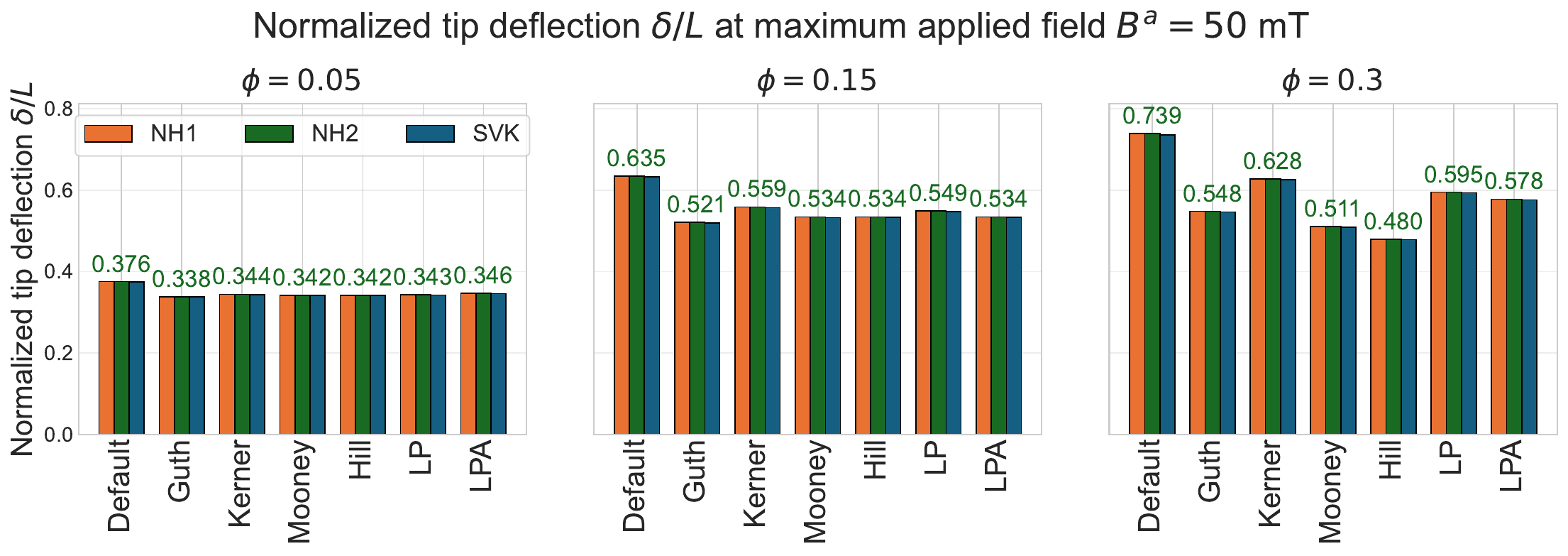}
    \caption{Normalized tip deflection $\delta_m/L$ at $B^a=50~\mathrm{mT}$ for $\phi=0.05$, $0.15$, and $0.30$. For each $G(\phi)$ model, the grouped bars compare \textbf{NH1}, \textbf{NH2}, and \textbf{SVK}. Numerical values above the green bars indicate the corresponding $\delta_m/L$ for \textbf{NH2}.}
    \label{fig:actuationEnergyBars}
\end{figure}

The effect of the strain-energy density function is presented in \cref{fig:actuationEnergyBars}. For each $\phi$ and each $G(\phi)$ model, the bars for \textbf{NH1}, \textbf{NH2}, and \textbf{SVK} are nearly identical. Thus, for this three-dimensional benchmark, the predicted actuation response is much more sensitive to the selected $G(\phi)$ relation than to the selected strain-energy density function, which is consistent with the two-dimensional beam, wheel, and gripper benchmarks. The ordering of the $G(\phi)$ models is dependent on $\phi$, as expected. 

For \textbf{NH2}, the normalized deflection as a function of the applied magnetic field is plotted in \cref{fig:actuationCuboidSetup}(b). At small particle volume fractions, the model predictions are close. As $\phi$ increases, the spread among the $G(\phi)$ models becomes more pronounced. \cref{fig:actuationCuboidSetup}(c) compares the corresponding deformed configurations at $B^a=50~\mathrm{mT}$ and $\phi=0.3$.

\subsection{Summary of key observations}

The numerical studies in this section lead to the following observations:
\begin{itemize}
    \item \textbf{The three elastic energy functions give similar responses in the tested cases.}
    Across the beam, wheel, gripper, and three-dimensional cuboid benchmarks, \textbf{NH1}, \textbf{NH2}, and \textbf{SVK} give nearly identical displacement or deformation measures for the same structure--property model under the loading conditions investigated. This result should not be generalized to more strongly nonlinear elastic models or other loading regimes.

    \item \textbf{The $G(\phi)$ model effect is significant.}
    The effective shear-modulus relation changes the predicted deformation, especially when magnetic material overlaps with regions undergoing substantial deformation.

    \item \textbf{Spatial placement matters.}
    The uniformly magnetized beam and the wheel demonstrate clear sensitivity to the selected $G(\phi)$ relation because magnetic material overlaps with regions undergoing substantial deformation.

    \item \textbf{Sensitivity is reduced when magnetic material is separated from the dominant deformation regions.}
    This behavior is observed in the \blue{tip-magnetized} beam and in the gripper, where the relative changes remain much smaller than in the uniformly magnetized beam and wheel.

    \item \textbf{The default model gives the largest deformation.}
    Since the default model uses $G(\phi)=G_0$, it ignores particle-induced stiffening and consistently predicts the largest displacement. Therefore, a structure--property relation informed by particle reinforcement is needed for the model-selection and optimization studies that follow.

    \item \textbf{NH2 is selected for later sections.}
    Since \textbf{NH2} gives responses similar to \textbf{NH1} and \textbf{SVK} while providing better numerical robustness in the examples considered, it is used as the fixed strain-energy density function in \cref{s:modelSelection} and in the topology optimization studies.
\end{itemize}

\section{Constitutive model selection}\label{s:modelSelection}

This section selects the effective shear-modulus relation $G(\phi)$ used in the joint material--structural optimization studies. The comparison uses the uniaxial true stress--strain data of \citet{garcia2021influence} for \blue{soft-magnetic composites containing carbonyl-iron particles at volume fractions of $0$, $0.05$, $0.15$, and $0.30$}. Because the tests are purely mechanical, they isolate the particle-induced stiffening represented by $G(\phi)$ without testing the magnetic actuation model or remanent-magnetization assumptions. The dataset therefore provides a basis for selecting a representative $G(\phi)$ relation for the subsequent design studies. However, it is a single dataset and does not vary particle size independently; it therefore cannot establish that the selected relation or model ranking applies generally to hard-magnetic composites or determine whether particle size changes the form of $G(\phi)$.

The full stress--strain comparison is also conditioned by the elastic energy used. Metric~2 embeds every candidate $G(\phi)$ relation in the same \textbf{NH2} energy so that differences can be attributed to the effective shear-modulus relation. However, \textbf{NH2} does not reproduce all of the high-strain stiffening visible in the measured curves. Consequently, Metric~2 ranks the candidate relations only within this constitutive setting; \blue{using a strain-energy function that accounts for finite chain extensibility}, such as an inverse-Langevin or Arruda--Boyce model, could change \blue{the stress--strain errors and the ranking of the $G(\phi)$ relations}.

The stress--strain curves reported in \cite[Fig.~1]{garcia2021influence} for the mixing ratio $5{:}1$ were digitized and processed as described in \labelcref{s:appModelSelection}. Because the small-strain region is sensitive to digitization and curve regularization, two digitization passes were used and averaged to obtain a reference dataset. \cref{fig:gonzalezDataExtract} shows the extracted stress--strain data and the corresponding small-strain shear-modulus estimates. Using the calibration window $\varepsilon_{xx}\in[0,\,0.15]$, the reference estimates for $G(\phi)$ in kPa are
\begin{equation*}
    G(0)=G_0=373.7,\quad
    G(0.05)=454.8,\quad
    G(0.15)=670.1,\quad
    G(0.30)=1321.8\,,
\end{equation*}
where $G_0$ is the shear modulus of the host matrix.

Two complementary comparisons are used. They are summarized below, with additional details provided in \labelcref{ss:metric1Details,ss:computeRMSE}.

\subsection{Selection metric 1: shear-modulus data}
\begin{figure}[h]
    \centering
    \includegraphics[width=0.8\linewidth]{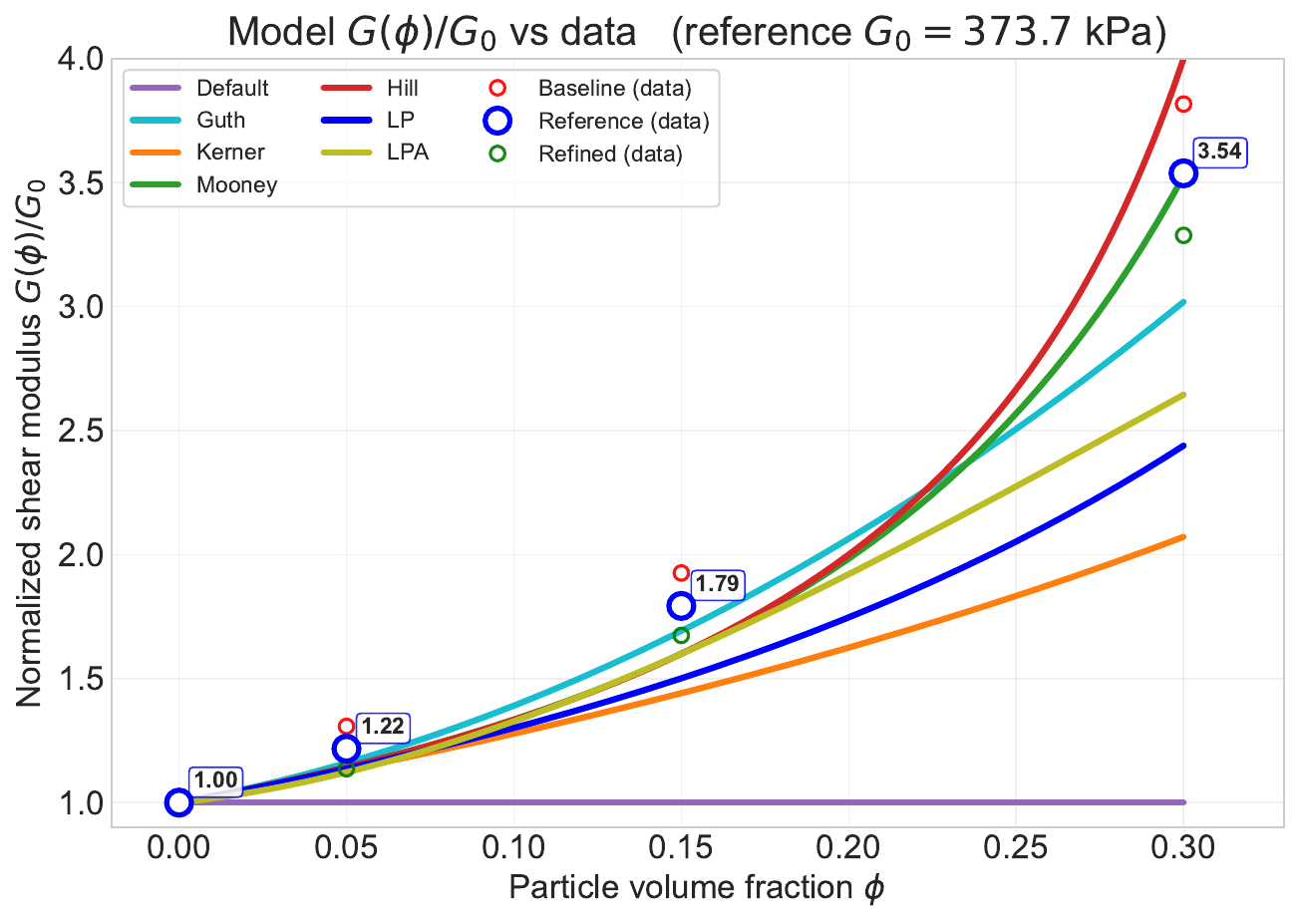}
    \caption{Comparison of the normalized shear-modulus relations $G(\phi)/G_0$ with the extracted modulus data. Solid curves denote the effective shear-modulus relations, while markers denote the baseline, refined, and reference estimates from the digitized experimental data.}
    \label{fig:metric1Gmodels}
\end{figure}

Each $G(\phi)$ model is compared directly with the extracted shear-modulus data. For each volume fraction, the relative error is defined as
\begin{equation}\label{eq:metric1rE}
    \mathrm{rE}(\phi, \mathrm{model})
    =
    \frac{
    \left|
    G_{\mathrm{model}}(\phi)
    -
    G^{\mathrm{data}}_{\varepsilon_{\max}}(\phi)
    \right|
    }
    {
    \left|
    G^{\mathrm{data}}_{\varepsilon_{\max}}(\phi)
    \right|
    } .
\end{equation}
The mean and standard deviation of $\mathrm{rE}$ over $\phi\in\{0,\,0.05,\,0.15,\,0.30\}$ are reported in \cref{tab:modelSummary}. \cref{fig:metric1Gmodels} shows the same comparison visually after normalizing by $G_0$. The \textbf{Default} model matches the data at $\phi=0$ by construction but cannot capture the stiffening at larger $\phi$. Among the non-default models, \textbf{Mooney} gives the lowest mean relative error, followed by \textbf{Guth} and \textbf{Hill}.

\subsection{Selection metric 2: stress--strain data}

The second metric compares simulated and extracted stress--strain curves directly. For each prescribed particle volume fraction $\phi$, effective shear-modulus relation $G(\phi)$, and matrix shear modulus $G_0$, a dog-bone specimen similar to that in \cite{garcia2021influence} is simulated using the \textbf{NH2} energy model and a spatially uniform particle volume fraction. The resulting true stress--strain curve is then compared with the extracted experimental curve. This comparison identifies the $G(\phi)$ relation that gives the smallest error across the tested values of $\phi$ and representative values of $G_0$.

The comparison uses $\phi\in\{0,\,0.05,\,0.15,\,0.30\}$ and $G_0\in\{373.7,\,400,\,420\}\,\mathrm{kPa}$. The relative root-mean-square error, dog-bone finite-element setup, averaging over $G_0$ and $\phi$, and representative stress--strain curves are provided in \labelcref{ss:uniaxialSetup,ss:computeRMSE}.

The Metric~2 results are summarized in \cref{tab:modelSummary}. The \textbf{Default} model gives the largest mean rRMSE, confirming that a $\phi$-dependent modulus relation is required. \blue{\textbf{Mooney} gives the smallest mean rRMSE ($0.248$), followed by \textbf{Hill} ($0.252$) and \textbf{Guth} ($0.255$).}

\begin{table}[H]
  \centering
  \caption{Summary of structure--property models for the effective shear modulus $G(\phi)$. 
  Each $G(\phi)$ relation is combined with the three strain-energy densities (\textbf{NH1}, \textbf{NH2}, and \textbf{SVK}), with the total energy including the magnetic contribution $W_{\mathrm{magn}}$ in \cref{eq:magneticEnergy}. The table reports model selection metrics from \cref{s:modelSelection}.
  Metric~1 reports the relative error $\mathrm{rE}$ between each $G(\phi)$ relation and the extracted shear-modulus data from \citet{garcia2021influence}.
  Metric~2 reports the relative root-mean-square error (rRMSE) between uniaxial simulations on dog-bone specimens and reference stress--strain curves. The model selection and the error calculations are discussed in \cref{s:modelSelection}.
  Mooney (blue shading) gives the lowest mean error in both metrics and is used in the optimization studies.
  Reference abbreviations:
  Guth: \citet{guth1945theory}; Kerner: \citet{kerner1956elastic}; Mooney: \citet{mooney1951viscosity}; A\&J: \citet{ahmed1990review}; Hill: \cite{hill1965self}; LP: \citet{lopez2013nonlinear2}; Gonzalez: \citet{garcia2021microstructural}; and Alsham: \citet{alshammari2019addition}; Gonzalez data: \citet{garcia2021influence}.
  }
  \label{tab:modelSummary}
  \vspace{-4pt}
  {
  \footnotesize
  \setlength{\tabcolsep}{6pt}
  \renewcommand{\arraystretch}{1.8}
  \begin{tabular}{@{}%
    l
    >{\raggedright\arraybackslash}p{0.1\textwidth}
    >{\raggedright\arraybackslash}p{0.25\textwidth}
    >{\raggedright\arraybackslash}p{0.08\textwidth}
    c c @{\hspace{1em}} c c
  @{}}
  \toprule
  \textbf{G model} &
  \textbf{Approach} &
  \textbf{$G(\phi)$} &
  \textbf{Refs.} &
  \multicolumn{4}{c}{\textbf{Model error using experimental data}} \\
  \cmidrule(lr){5-8}
  & & & &
  \multicolumn{2}{c}{\textbf{Metric~1}} &
  \multicolumn{2}{c}{\textbf{Metric~2}} \\
  & & & &
  \multicolumn{2}{c}{\scriptsize $\mathrm{rE}$ of $G(\phi)$} &
  \multicolumn{2}{c}{\scriptsize rRMSE of $\sigma_{xx}$ (uniaxial)} \\
  \cmidrule(lr){5-6}\cmidrule(lr){7-8}
  & & & &
  \;\;\;\;\textbf{Mean} & \textbf{Std} &
  \;\;\;\;\;\;\textbf{Mean} & \textbf{Std} \\
  \midrule
  Default &
  --- &
  $G_0$ &
  --- &
  0.335 & 0.313 & 0.455 & 0.189 \\
  \midrule
  Guth &
  \makecell[l]{Effective\\modulus} &
  \makecell[l]{$G_0 \left( 1 + k_E \phi + 14.1 \phi^2 \right)$\\
   $k_E = 2.5$} &
  \makecell[l]{Guth\\A\&J} &
  0.062 & 0.061 & 0.255 & 0.021 \\
  \midrule
  Kerner &
  \makecell[l]{Effective\\modulus} &
  \makecell[l]{$G_0 \left( 1 + A \phi / (1-\phi) \right)$\\ $A = k_E = 2.5$} &
  \makecell[l]{Kerner\\A\&J} &
  0.170 & 0.182 & 0.327 & 0.090 \\
  \midrule
  Mooney &
  \makecell[l]{Effective\\modulus} &
  $G_0 \exp\!\left(k_E \phi / (1-1.35\phi)\right)$ &
  \makecell[l]{Mooney\\A\&J} &
  \cellcolor{blue!8}\textbf{0.043} & \cellcolor{blue!8}\textbf{0.051} &
  \cellcolor{blue!8}\textbf{0.248} & 0.019 \\
  \midrule
  Hill &
  \makecell[l]{Effective\\modulus} &
  \makecell[l]{$G_0 / (1 - k_E \phi)$} &
  \makecell[l]{Hill\\App.~\hyperref[s:appHill]{A}} &
  0.075 & 0.058 & 0.252 & \cellcolor{blue!8}\textbf{0.014} \\
  \midrule
  LP &
  \makecell[l]{Constrained\\kinematics} &
  $G_0 / (1-\phi)^{5/2}$ &
  LP &
  0.135 & 0.135 & 0.300 & 0.057 \\
  \midrule
  LPA &
  \makecell[l]{Constrained\\kinematics} &
  \makecell[l]{$G_0 (1-\phi)\,\chi(\phi)$\\
  $\chi(\phi) = 1 + 0.67 g \phi + 1.62 (g \phi)^2$ \\
   $g = 3.73$ } &
  \makecell[l]{Gonzalez\\Alsham} &
  0.110 & 0.105 & 0.281 & 0.040 \\
  \bottomrule
  \end{tabular}
  }
  \vspace{10pt}
\end{table}

\subsection{Summary of results and selected model}

For the mechanical dataset and comparison metrics used here, \textbf{Mooney} gives the lowest mean error in both the direct $G(\phi)$ comparison and the full stress--strain comparison; see \cref{tab:modelSummary}. \textbf{Hill} and \textbf{Guth} give similar stress--strain errors, while \textbf{Kerner}, \textbf{LP}, and \textbf{LPA} give larger errors under these metrics. The \textbf{Default} relation performs poorly because it does not account for particle-induced stiffening. We therefore use \textbf{Mooney} in the design examples, without assuming that it is the best relation for all hard-magnetic materials. The restorative-beam comparison in \cref{ss:beamSensitivity} tests how replacing it with \textbf{Guth} or \textbf{Hill} affects the attained compliance.

For the \textbf{LPA} relation in \cref{eq:LPA}, motivated by \citet{garcia2021microstructural}, the amplification parameter is fixed a priori at $g=3.73$ through $0.67g=k_E=2.5$, rather than calibrated using the Garcia-Gonzalez dataset. For Metric~2, \textbf{LPA} is embedded in the same \textbf{NH2} energy as the other candidate relations. Thus, for each prescribed $G_0$, all seven $G(\phi)$ relations are fixed before their errors are evaluated. The mean errors for \textbf{Mooney} and \textbf{LPA} are $0.043$ and $0.110$ in Metric~1 and $0.248$ and $0.281$ in Metric~2, respectively. By contrast, the agreement reported by \citet{garcia2021influence} was obtained using a more detailed eight-chain formulation whose chain parameters and amplification parameter were calibrated against the experimental data. The calibrated constitutive model of \citet{garcia2021influence} and the present comparison of fixed $G(\phi)$ relations therefore assess different modeling choices.

\section{Joint material--structural optimization formulation}\label{s:topOptimFormulation}

This section introduces the joint material--structural optimization formulation used in the design examples. It builds on the density-based topology optimization framework of FEniTop~\cite{jia2024fenitop}, using filtering, projection, adjoint sensitivity analysis, and MMA-based design updates. The formulation shares with \citet{zhao2022topology} a finite-strain equilibrium problem, density-dependent elastic and magnetic energies, adjoint sensitivities, and multiple operating conditions. Zhao and Zhang parameterize remanent magnetization through weights over prescribed candidate vectors, can also optimize a uniform applied field, and impose structural-volume and stress constraints. \citet{perezgarcia2026topology} combine topology and material optimization with experimentally calibrated remanence-induced anisotropy and a coupled magnetostatic treatment that includes the surrounding vacuum. In the present formulation, the applied field is prescribed, while structural density $\rho$, particle volume fraction $\phi$, and remanent direction $\theta_{B^r}$ are represented by separate design fields. In particular, $\phi$ is an independently constrained \blue{continuous particle volume fraction field} that enters both the selected effective shear-modulus relation in \cref{eq:effectiveGDesignGeneral} and the magnetic energy in \cref{eq:WhatMagn}; a continuous angle specifies the remanent direction.

The three design fields have different physical roles: $\rho$ controls the presence of structural material, \blue{$\phi$ specifies the local volume fraction of magnetic particles}, and $\theta_{B^r}$ controls the local remanent direction. They remain independent optimization variables, while their mechanical and magnetic effects are coupled through the design-dependent free-energy density. The density interpolation suppresses the elastic and magnetic contributions as $\rho$ approaches zero, and $\phi$ modifies both the effective shear modulus and the magnetic interaction energy. The resulting formulation can optimize any active subset of the three fields, or all three together, under single- or multi-load-case combinations of mechanical and magnetic loading. For a selected manufacturing process, the computed fields must be converted into a realizable design: $\rho$ into a geometry satisfying the available feature resolution, \blue{$\phi$ into the discrete or graded compositions supported by the selected manufacturing process}, and $\theta_{B^r}$ into \blue{magnetization directions that can be programmed}. The converted design should then be reanalyzed and, if necessary, reoptimized.

\subsection{Design variable parameterization}\label{ss:designVariableParameterization}

Three design fields are considered on the reference domain $\Omega_0$: the structural density $\rho$, the magnetic particle volume fraction $\phi$, and the remanent magnetization direction $\theta_{B^r}$. Any subset of these variables may be optimized, while the remaining fields are prescribed.

In the discrete implementation, the raw design variables are stored element-wise. Let $\{\Omega_e\}_{e=1}^{N_e}$ denote the finite-element partition of $\Omega_0$. One density value $\rho_e$, one magnetic particle volume fraction value $\phi_e$, and one remanent magnetization direction value $\theta_{B^r,e}$ are associated with each element $\Omega_e$. These element-wise values define piecewise-constant fields on $\Omega_0$. For notational simplicity, the same symbols $\rho$, $\phi$, and $\theta_{B^r}$ are used for these piecewise-constant design fields.

The common discrete space for the raw scalar design fields is
\begin{equation}\label{eq:rawDesignSpace}
    Q_h
    =
    \left\{
    q\in L^2(\Omega_0):
    q|_{\Omega_e}\in \mathbb{P}_0(\Omega_e),
    \quad e=1,\ldots,N_e
    \right\}\,.
\end{equation}
Here, $\mathbb{P}_k(\Omega_e)$ denotes the space of polynomial functions of degree at most $k$ on the element $\Omega_e$. In particular, $\mathbb{P}_0(\Omega_e)$ denotes the space of constant functions on $\Omega_e$.

For any raw scalar design field $q\in Q_h$ and filter radius $r_q>0$, the Helmholtz-filter operator is written as
\begin{equation}\label{eq:helmholtzFilter}
\mathcal{H}(\,\cdot\,;r_q):Q_h\to V_h^{\mathrm{HH}} .
\end{equation}
The filtered field
\begin{equation*}
    \tilde{q}=\mathcal{H}(q;r_q)
\end{equation*}
is the unique function in $V_h^{\mathrm{HH}}$ satisfying
\begin{equation}\label{eq:HHFilter}
    \int_{\Omega_0}
    \left(
        r_q^2 \nabla \tilde{q}\cdot\nabla v
        +
        \tilde{q}v
    \right)\,dV
    =
    \int_{\Omega_0}
    qv\,dV,
    \qquad \forall v\in V_h^{\mathrm{HH}} .
\end{equation}
Here,
\begin{equation*}
    V_h^{\mathrm{HH}}
    =
    \left\{
    v\in H^1(\Omega_0):
    v|_{\Omega_e}\in \mathcal{L}_1(\Omega_e),
    \quad e=1,\ldots,N_e
    \right\},
\end{equation*}
where $\mathcal{L}_1(\Omega_e)$ denotes the local first-order Lagrange finite-element space on $\Omega_e$. For simplex elements, this corresponds to $\mathbb{P}_1(\Omega_e)$; for quadrilateral or hexahedral elements, it corresponds to the usual first-order tensor-product Lagrange space. No essential boundary condition is imposed on $\tilde{q}$, so \cref{eq:HHFilter} corresponds to the natural homogeneous Neumann condition: $\nabla\tilde{q}\cdot\bn = 0$ on $\partial\Omega_0$.

\subsubsection{Density field}\label{sss:densityField}

The density field defines the local occupancy of structural material within the design domain. With the convention introduced above, the element-wise density values $\{\rho_e\}_{e=1}^{N_e}$ define the piecewise-constant field
\begin{equation*}
    \rho(\bX)
    =
    \sum_{e=1}^{N_e}
    \rho_e\,\chi_e(\bX),
\end{equation*}
where $\chi_e$ is the characteristic function of element $\Omega_e$. The design space for the raw density field is $Q_h$ defined in \cref{eq:rawDesignSpace}. We additionally impose the element-wise bounds
\begin{equation*}
    \rho_{\min}\leq \rho_e \leq 1,
    \qquad e=1,\ldots,N_e .
\end{equation*}
In the implementation, a small positive lower bound $\rho_{\min}\ll 1$ is used for the raw density variable to avoid singular stiffness matrices.

The density field is regularized using the Helmholtz-filter operator defined in \cref{eq:HHFilter}. Specifically,
\begin{equation*}
    \tilde{\rho}
    =
    \mathcal{H}(\rho;r_\rho),
\end{equation*}
where $r_\rho$ is the density filter radius.

The filtered density field is then passed through a Heaviside projection to obtain the physical density field $\bar{\rho}$:
\begin{equation}\label{eq:rhoProjection}
    \bar{\rho}
    = \mathcal{P}_\rho(\tilde{\rho}) := 
    \frac{
    \tanh(\beta_\rho\eta_\rho)
    +
    \tanh\!\left(\beta_\rho(\tilde{\rho}-\eta_\rho)\right)
    }
    {
    \tanh(\beta_\rho\eta_\rho)
    +
    \tanh\!\left(\beta_\rho(1-\eta_\rho)\right)
    } .
\end{equation}
Here, $\eta_\rho$ is the projection threshold, and $\beta_\rho$ controls the sharpness of the transition. During optimization, $\beta_\rho$ is progressively increased to drive $\bar{\rho}$ toward a nearly binary distribution. The physical density field $\bar{\rho}$ is the density variable that enters the free-energy interpolation introduced in \cref{ss:energyInterpolation}.

\subsubsection{Magnetic particle volume fraction field}\label{sss:phiField}

The magnetic particle volume fraction field controls the local concentration of hard-magnetic particles within the design domain. As in the case of the density field,
\begin{equation*}
    \phi(\bX)
    =
    \sum_{e=1}^{N_e}
    \phi_e\,\chi_e(\bX) \qquad 
    \in Q_h,
\end{equation*}
where $\{\phi_e\}_{e=1}^{N_e}$ are the element-wise values of the raw magnetic particle volume fraction field. The local element-wise bounds are imposed as
\begin{equation*}
    0\leq \phi_e \leq \phi_{\max},
    \qquad e=1,\ldots,N_e,
\end{equation*}
Here, $\phi_{\max}$ is the imposed local upper bound on particle volume fraction, set to $\phi_{\max}=0.30$ in the optimization examples. This is below the $\phi=0.4$ singularity of the \textbf{Hill} relation discussed in \cref{ss:Gmodels}; the physical condition $\phi<1$ alone would not ensure that all the candidate $G(\phi)$ relations remain finite.

The magnetic particle volume fraction field is regularized using the same Helmholtz-filter operator defined in \cref{eq:HHFilter}. Specifically,
\begin{equation*}
    \phi_{\mathrm{phys}}
    =
    \mathcal{H}(\phi;r_\phi),
\end{equation*}
where $r_\phi$ is the filter radius for the magnetic particle volume fraction field.

Unlike the density field, the magnetic particle volume fraction field is not passed through a Heaviside projection and is not SIMP-penalized. This choice reflects the interpretation of $\phi_{\mathrm{phys}}$ as \blue{a continuous particle volume fraction field} rather than a void--solid structural density. The filtered field $\phi_{\mathrm{phys}}$ enters the free-energy interpolation introduced in \cref{ss:energyInterpolation}: it modifies the effective shear modulus through the selected structure--property relation and scales the magnetic interaction according to the local concentration of magnetic particles.

\subsubsection{Remanent magnetization direction field}\label{sss:thetaField}

The remanent magnetization direction field controls the local orientation of the embedded remanent magnetic flux density. As before,
\begin{equation*}
    \theta_{B^r}(\bX)
    =
    \sum_{e=1}^{N_e}
    \theta_{B^r,e}\,\chi_e(\bX) \qquad
    \in Q_h,
\end{equation*}
where $\{\theta_{B^r,e}\}_{e=1}^{N_e}$ are the element-wise values of the raw remanent magnetization direction field. \blue{The two-dimensional examples use the local element-wise bounds}
\begin{equation*}
    -\pi\leq \theta_{B^r,e}\leq \pi,
    \qquad e=1,\ldots,N_e\,.
\end{equation*}

Similar to the previous design fields, the remanent magnetization direction field is regularized using the Helmholtz-filter operator to obtain a filtered field $\theta_{B^r,\mathrm{phys}}$:
\begin{equation}\label{eq:thetaFiltered}
    \theta_{B^r,\mathrm{phys}}
    =
    \mathcal{H}(\theta_{B^r};r_\theta),
\end{equation}
where $r_\theta$ is the filter radius.

\begin{remark}
The Helmholtz filter in \cref{eq:thetaFiltered} is applied directly to the scalar angular field $\theta_{B^r}$, with angles represented in the interval $[-\pi,\pi]$. \blue{This convention is used for the filtered direction fields in the two-dimensional examples.} If optimized directions contain neighboring values near both ends of this interval, a periodic representation based on filtering $\cos\theta_{B^r}$ and $\sin\theta_{B^r}$ would be more appropriate.
\end{remark}

The filtered direction field is then used to define the local remanent magnetic flux density:
\begin{equation}\label{eq:BrThetaOptim}
    \bB^r\!\left(\theta_{B^r,\mathrm{phys}}\right)
    =
    B^r\, \blue{\left[
        \cos\theta_{B^r,\mathrm{phys}}\;, \;
        \sin\theta_{B^r,\mathrm{phys}}
    \right]^T}.
\end{equation}
For the three-dimensional example in \cref{ss:design3d}, the same scalar angle parameterizes a direction in the prescribed $x$--$y$ plane by appending a zero third component. \blue{In that example, $\theta_{B^r}$ is retained as an unfiltered cell-wise field.}
Similar to $\phi$, $\theta_{B^r}$ is not subjected to Heaviside projection or SIMP penalization, allowing continuous variation of the remanent direction. The corresponding physical direction field enters the free energy through $\bB^r$.

\subsection{Energy interpolation and design variable coupling}\label{ss:energyInterpolation}

The filtered and projected design fields from \cref{ss:designVariableParameterization} enter the nonlinear finite-element formulation through the design-dependent total free-energy density. This is the step where the physical coupling among structural density, magnetic particle concentration, and remanent magnetization direction is imposed. For a load case $\ell$ with applied magnetic flux density $\bB^{a,(\ell)}$, the design-dependent free-energy density is written as
\begin{equation}\label{eq:WhatSplit}
    \widehat{W}^{(\ell)}
    (
        \bF;
        \bar{\rho},
        \phi_{\mathrm{phys}},
        \theta_{B^r,\mathrm{phys}},
        \bB^{a,(\ell)}
    )
    =
    \widehat{W}_{\mathrm{elas}}
    (
        \bF;
        \bar{\rho},
        \phi_{\mathrm{phys}}
    )
    +
    \widehat{W}_{\mathrm{magn}}^{(\ell)}
    (
        \bF;
        \bar{\rho},
        \phi_{\mathrm{phys}},
        \theta_{B^r,\mathrm{phys}},
        \bB^{a,(\ell)}
    ).
\end{equation}
The two terms are defined below to make explicit how each design field enters the elastic and magnetic parts of the energy.

\subsubsection{Density-penalized moduli and effective material response}\label{sss:densityPenalizedModuli}

The density field is incorporated through a SIMP-type interpolation of the reference elastic moduli. Define
\begin{equation*}
    s_\rho(\bar{\rho})
    =
    \epsilon_\rho
    +
    (1-\epsilon_\rho)\bar{\rho}^{p_\rho},
\end{equation*}
where $\epsilon_\rho>0$ is a small stiffness floor and $p_\rho$ is the density penalization exponent. The density-penalized matrix shear and bulk moduli are
\begin{equation*}
    \widehat{G}_0(\bar{\rho})
    =
    s_\rho(\bar{\rho})G_0,
    \qquad
    \widehat{K}_0(\bar{\rho})
    =
    s_\rho(\bar{\rho})K_0 .
\end{equation*}

The selected effective shear-modulus relation is then evaluated using the density-penalized matrix shear modulus. The design-dependent effective shear modulus is written in the general form
\begin{equation}\label{eq:effectiveGDesignGeneral}
    \widehat{G}
    \left(
        \bar{\rho},
        \phi_{\mathrm{phys}}
    \right)
    =
    G_{\mathrm{sel}}
    \left(
        \phi_{\mathrm{phys}};
        \widehat{G}_0(\bar{\rho})
    \right),
\end{equation}
where $G_{\mathrm{sel}}$ denotes the selected effective shear-modulus relation. Based on the model-selection study in \cref{s:modelSelection}, the optimization examples use the Mooney relation. Therefore,
\begin{equation*}
    \widehat{G}
    \left(
        \bar{\rho},
        \phi_{\mathrm{phys}}
    \right)
    =
    \widehat{G}_0(\bar{\rho})
    \exp\left(
        \frac{k_E\phi_{\mathrm{phys}}}
        {1-1.35\phi_{\mathrm{phys}}}
    \right).
\end{equation*}
Consistent with the constitutive formulation in \cref{ss:strainEnergyModels}, the effective bulk modulus is not modified by $\phi_{\mathrm{phys}}$ and is taken as
\begin{equation*}
    \widehat{K}(\bar{\rho})
    =
    \widehat{K}_0(\bar{\rho}).
\end{equation*}

Using the selected \textbf{NH2} strain-energy density, the design-dependent elastic free-energy density is
\begin{equation}\label{eq:WhatElasNH2}
    \widehat{W}_{\mathrm{elas}}
    \left(
        \bF;
        \bar{\rho},
        \phi_{\mathrm{phys}}
    \right)
    =
    \frac{
        \widehat{G}
        \left(
            \bar{\rho},
            \phi_{\mathrm{phys}}
        \right)
    }{2}
    \left(
        I_1-3-2\ln J
    \right)
    +
    \frac{
        \widehat{K}(\bar{\rho})
    }{2}
    (J-1)^2 .
\end{equation}

\subsubsection{Magnetic energy coupling}\label{sss:magneticEnergyCoupling}

For load case $\ell$, the design-dependent magnetic free-energy density is
\begin{equation}\label{eq:WhatMagn}
    \widehat{W}_{\mathrm{magn}}^{(\ell)}
    \left(
        \bF;
        \bar{\rho},
        \phi_{\mathrm{phys}},
        \theta_{B^r,\mathrm{phys}},
        \bB^{a,(\ell)}
    \right)
    =
    -
    \frac{1}{\mu_0}
    \bar{\rho}\,
    \phi_{\mathrm{phys}}\,
\blue{    \bM(\bF)
    \bB^r
    \left(
        \theta_{B^r,\mathrm{phys}}
    \right)}
    \cdot
    \bB^{a,(\ell)} .
\end{equation}
Here, $\bB^r(\theta_{B^r,\mathrm{phys}})$ is defined in \cref{eq:BrThetaOptim}. The factor $\bar{\rho}$ attenuates the magnetic energy in low-density regions, while $\phi_{\mathrm{phys}}$ scales the magnetic interaction according to the local magnetic particle concentration. Thus, magnetic actuation is active only where structural material and magnetic particles are present, while $\theta_{B^r,\mathrm{phys}}$ controls the local remanent direction.

The relations \Cref{eq:WhatElasNH2,eq:WhatMagn}, together with \cref{eq:WhatSplit}, complete the definition of the design-dependent total free-energy density.

\subsection{Optimization formulation}\label{ss:optimizationFormulation}

This subsection defines the finite-dimensional optimization problem in terms of the raw design fields introduced in \cref{ss:designVariableParameterization}. The physical fields that enter the equilibrium equations are
\begin{equation*}
    \bar{\rho}
    =
    \mathcal{P}_\rho\!\left(\mathcal{H}(\rho;r_\rho)\right),
    \qquad
    \phi_{\mathrm{phys}}
    =
    \mathcal{H}(\phi;r_\phi),
    \qquad
    \theta_{B^r,\mathrm{phys}}
    =
    \mathcal{H}(\theta_{B^r};r_\theta),
\end{equation*}
where $\mathcal{H}(\cdot;r)$ denotes the Helmholtz filter operator with radius $r$ in \cref{eq:helmholtzFilter} and $\mathcal{P}_\rho$ denotes the Heaviside projection in \cref{eq:rhoProjection}. These physical fields determine the design-dependent free-energy density through \cref{eq:WhatSplit,eq:WhatElasNH2,eq:WhatMagn}.

For each load case $\ell=1,\ldots,N_\ell$, the applied magnetic flux density $\bB^{a,(\ell)}$, body force $\bb^{(\ell)}$, prescribed traction $\bar{\bt}^{(\ell)}$, and prescribed displacement $\bar{\bu}^{(\ell)}$ are specified. The displacement field $\bu^{(\ell)}$ satisfies the design-dependent residual equation
\begin{equation}\label{eq:weakResidualHat}
    \widehat{\mathcal{R}}^{(\ell)}
    \left(
        \bu^{(\ell)},\delta\bu;
        \bar{\rho},
        \phi_{\mathrm{phys}},
        \theta_{B^r,\mathrm{phys}},
        \bB^{a,(\ell)},
        \bb^{(\ell)},
        \bar{\bt}^{(\ell)},
        \bar{\bu}^{(\ell)}
    \right)
    =
    0,
    \qquad
    \forall \delta\bu\in V_v .
\end{equation}
Here, $\widehat{\mathcal{R}}^{(\ell)}$ is obtained from the residual functional in \cref{eq:weakResidual} by replacing $W$ with the design-dependent free-energy density $\widehat{W}^{(\ell)}$ in \cref{eq:WhatSplit}. Equivalently, the first Piola--Kirchhoff stress in load case $\ell$ is
\begin{equation*}
    \widehat{\bP}^{(\ell)}
    =
    \frac{\partial \widehat{W}^{(\ell)}}{\partial \bF}.
\end{equation*}
After finite-element discretization, \cref{eq:weakResidualHat} gives the nonlinear algebraic residual equation
\begin{equation}\label{eq:discreteResidualHat}
    \widehat{\br}^{(\ell)}
    \left(
        \bu^{(\ell)};
        \bar{\rho},
        \phi_{\mathrm{phys}},
        \theta_{B^r,\mathrm{phys}},
        \bB^{a,(\ell)},
        \bb^{(\ell)},
        \bar{\bt}^{(\ell)},
        \bar{\bu}^{(\ell)}
    \right)
    =
    \mathbf{0},
    \qquad
    \ell=1,\ldots,N_\ell .
\end{equation}

The optimization problem is written in terms of the raw design variables $\rho$, $\phi$, and $\theta_{B^r}$:
\begin{equation}\label{eq:optimizationProblem}
\begin{aligned}
    &\min_{\rho,\,\phi,\,\theta_{B^r}}
    &&
    f
    \left(
        \bu^{(1)},\ldots,\bu^{(N_\ell)}
    \right)
    \\
    &\text{subject to}
    &&
    \widehat{\br}^{(\ell)}
    \left(
        \bu^{(\ell)};
        \bar{\rho},
        \phi_{\mathrm{phys}},
        \theta_{B^r,\mathrm{phys}},
        \bB^{a,(\ell)},
        \bb^{(\ell)},
        \bar{\bt}^{(\ell)},
        \bar{\bu}^{(\ell)}
    \right)
    =
    \mathbf{0},
    \qquad
    \ell=1,\ldots,N_\ell,
    \\
    &&
    &
    \frac{1}{|\Omega_0|}
    \sum_{e=1}^{N_e}
    \bar{\rho}_e v_e
    \leq
    v_{\max},
    \\
    &&
    &
    \frac{1}{|\Omega_0|}
    \sum_{e=1}^{N_e}
    \phi_{\mathrm{phys},e} v_e
    \leq
    \Phi_{\max},
    \\
    &&
    &
    \left.
    \begin{aligned}
        \rho_{\min} &\leq \rho_e \leq 1, \\
        0 &\leq \phi_e \leq \phi_{\max}, \\
        -\pi &\leq \theta_{B^r,e} \leq \pi
    \end{aligned}
    \right\}
    \qquad
    e=1,\ldots,N_e .
\end{aligned}
\end{equation}
Here, $v_e=|\Omega_e|$ denotes the measure (area or volume) of element $\Omega_e$, and $|\Omega_0|=\sum_{e=1}^{N_e}v_e$. The quantities $\bar{\rho}_e$ and $\phi_{\mathrm{phys},e}$ denote element-wise averages of the corresponding physical fields. The parameter $v_{\max}$ specifies the allowable structural material volume fraction, $\Phi_{\max}$ bounds the \blue{domain-averaged magnetic particle volume fraction}, $\rho_{\min}$ is the lower bound on the raw density variable used to avoid singular stiffness matrices, and $\phi_{\max}$ specifies the \blue{local upper bound on magnetic particle volume fraction}.

The objective function $f$ is problem-dependent. In the examples considered in \cref{s:applications}, it is assembled from one or more load-case responses. A typical form is
\begin{equation*}
    f
    \left(
        \bu^{(1)},\ldots,\bu^{(N_\ell)}
    \right)
    =
    \sum_{\ell=1}^{N_\ell}
    w_\ell
    f^{(\ell)}
    \left(
        \bu^{(\ell)}
    \right),
\end{equation*}
where $w_\ell$ is a prescribed load-case weight. Other choices, such as max-type objectives or single-load-case objectives, can be used without changing the design-variable parameterization or the equilibrium formulation.

Although \cref{eq:optimizationProblem} is written with all three design variables active, the same formulation applies when only a subset of the fields is optimized. Inactive fields are prescribed and held fixed, and MMA updates only the remaining active design variables. \blue{The formulation supports optimization of individual design fields, optimization of particle volume fraction and remanent direction within a fixed geometry, and joint optimization of all three fields.}

\subsection{Adjoint-based gradient computation}\label{ss:sensitivity}

The optimization problem in \cref{eq:optimizationProblem} is solved using the Method of Moving Asymptotes (MMA), a gradient-based optimization algorithm that requires derivatives of the objective and constraint functions with respect to the raw design variables. The objective derivatives are computed using an adjoint method, following the general framework described in \cite{jia2024fenitop,zhao2022topology}, while the derivatives of the state-independent volume constraints are computed directly.

Let $f$ denote the objective function introduced in \cref{eq:optimizationProblem}. For each load case $\ell$, let $\mathbf{u}^{(\ell)}$ denote the discrete displacement vector and let
\begin{equation*}
    \widehat{\br}^{(\ell)}
    \left(
        \mathbf{u}^{(\ell)};
        \bar{\rho},
        \phi_{\mathrm{phys}},
        \theta_{B^r,\mathrm{phys}},
        \bB^{a,(\ell)},
        \bb^{(\ell)},
        \bar{\bt}^{(\ell)},
        \bar{\bu}^{(\ell)}
    \right)
    =
    \mathbf{0}
\end{equation*}
denote the discrete residual equation from \cref{eq:discreteResidualHat}. The tangent stiffness matrix for load case $\ell$ is
\begin{equation*}
    \widehat{\mathbf{K}}_{\mathrm{T}}^{(\ell)}
    =
    \frac{\partial \widehat{\br}^{(\ell)}}
    {\partial \mathbf{u}^{(\ell)}} .
\end{equation*}
The corresponding adjoint vector $\boldsymbol{\lambda}^{(\ell)}$ is obtained by solving
\begin{equation*}
    \left(
    \widehat{\mathbf{K}}_{\mathrm{T}}^{(\ell)}
    \right)^T
    \boldsymbol{\lambda}^{(\ell)}
    =
    -
    \frac{\partial f}
    {\partial \mathbf{u}^{(\ell)}} .
\end{equation*}
For the objective function, this adjoint system is solved once per load case. The resulting adjoint vector is then used to evaluate objective derivatives with respect to all active design variables.

\blue{Let $\eta_i$ denote a physical design degree of freedom, where}
\begin{equation*}
    \blue{\eta_i}
    \in
    \left\{
    \blue{\bar{\rho}_i},\,
    \blue{\phi_{\mathrm{phys},i}},\,
    \blue{\theta_{B^r,\mathrm{phys},i}}
    \right\}.
\end{equation*}
The total derivative of $f$ with respect to $\blue{\eta_i}$ is
\begin{equation}\label{eq:adjointGradientPhysical}
    \frac{df}{d\blue{\eta_i}}
    =
    \left.
    \frac{\partial f}{\partial \blue{\eta_i}}
    \right|_{\mathbf{u}^{(1)},\ldots,\mathbf{u}^{(N_\ell)}}
    +
    \sum_{\ell=1}^{N_\ell}
    \left(
    \boldsymbol{\lambda}^{(\ell)}
    \right)^T
    \frac{\partial \widehat{\br}^{(\ell)}}
    {\partial \blue{\eta_i}}.
\end{equation}
The first term accounts for any explicit dependence of the objective on the design field, while the second term accounts for the dependence through the equilibrium equations. If the objective depends on only one load case, the summation in \cref{eq:adjointGradientPhysical} is restricted to that load case. For state-independent constraints, such as the volume constraints in \cref{eq:optimizationProblem}, the adjoint contribution is zero and their derivatives are computed directly.

The adjoint calculation gives derivatives with respect to the physical fields that enter the finite-element formulation. Since MMA updates the raw design variables, these derivatives are propagated backward through the filtering and projection operations using the chain rule.

For the density field, the raw, filtered, and projected variables are related by
\begin{equation*}
    \tilde{\rho}
    =
    \mathcal{H}(\rho;r_\rho),
    \qquad
    \bar{\rho}
    =
    \mathcal{P}_\rho(\tilde{\rho}) .
\end{equation*}
In vector form, let $\boldsymbol{\rho}$, $\tilde{\boldsymbol{\rho}}$, and $\bar{\boldsymbol{\rho}}$ denote the corresponding discrete fields. The chain rule gives
\begin{equation}\label{eq:chainRhoVector}
    \frac{df}{d\boldsymbol{\rho}}
    =
    \left(
    \frac{\partial \tilde{\boldsymbol{\rho}}}
    {\partial \boldsymbol{\rho}}
    \right)^T
    \left(
    \frac{\partial \bar{\boldsymbol{\rho}}}
    {\partial \tilde{\boldsymbol{\rho}}}
    \right)^T
    \frac{df}{d\bar{\boldsymbol{\rho}}}.
\end{equation}
The projection map $\mathcal{P}_\rho$ is applied pointwise to the filtered density field (see \cref{eq:rhoProjection}). Therefore, its discrete derivative is diagonal:
\begin{equation*}
    \frac{\partial \bar{\rho}_i}{\partial \tilde{\rho}_j}
    =
    \mathcal{P}_\rho'(\tilde{\rho}_i)\delta_{ij},
\end{equation*}
where $\delta_{ij}$ is the Kronecker delta. Thus, the projection part of the chain rule reduces to an entry-wise multiplication.

The Helmholtz filter is a linear PDE-based map. In discrete form, the filtered density is obtained from
\begin{equation}\label{eq:discreteHelmholtzFilterRho}
    \mathbf{A}_{\rho}
    \tilde{\boldsymbol{\rho}}
    =
    \mathbf{B}_{\rho}
    \boldsymbol{\rho},
\end{equation}
where $\mathbf{A}_{\rho}$ is the assembled finite-element matrix associated with the Helmholtz-filter bilinear form, and $\mathbf{B}_{\rho}$ is the assembled matrix whose product with the raw element-wise density vector $\boldsymbol{\rho}$ gives the discretized right-hand side. Therefore, the discrete filter map is
\begin{equation*}
    \tilde{\boldsymbol{\rho}}
    =
    \mathbf{H}_{\rho}
    \boldsymbol{\rho},
    \qquad
    \mathbf{H}_{\rho}
    =
    \mathbf{A}_{\rho}^{-1}\mathbf{B}_{\rho}.
\end{equation*}
Substituting this relation into \cref{eq:chainRhoVector} gives
\begin{equation*}
    \frac{df}{d\boldsymbol{\rho}}
    =
    \mathbf{H}_{\rho}^{T}
    \mathbf{D}_{\rho}
    \frac{df}{d\bar{\boldsymbol{\rho}}},
    \qquad
    \mathbf{D}_{\rho}
    =
    \mathrm{diag}
    \left(
    \mathcal{P}_\rho'(\tilde{\rho}_1),
    \ldots,
    \mathcal{P}_\rho'(\tilde{\rho}_{n_\rho})
    \right).
\end{equation*}
Equivalently, this can be evaluated without explicitly forming $\mathbf{H}_{\rho}$. First, compute
\begin{equation*}
    \mathbf{w}_{\rho}
    =
    \mathbf{D}_{\rho}
    \frac{df}{d\bar{\boldsymbol{\rho}}},
\end{equation*}
then solve the transpose Helmholtz-filter system
\begin{equation*}
    \mathbf{A}_{\rho}^{T}\mathbf{z}_{\rho}
    =
    \mathbf{w}_{\rho},
\end{equation*}
and finally compute
\begin{equation}\label{eq:rhoBackwardFilter}
    \frac{df}{d\boldsymbol{\rho}}
    =
    \mathbf{B}_{\rho}^{T}
    \mathbf{z}_{\rho}.
\end{equation}
\blue{These operations apply the chain rule through the density filter and projection.}

For the magnetic particle volume fraction and remanent magnetization direction, the procedure is the same except that no Heaviside projection is applied. Thus, for
\begin{equation*}
    \phi_{\mathrm{phys}}=\mathcal{H}(\phi;r_\phi),
    \qquad
    \theta_{B^r,\mathrm{phys}}
    =
    \mathcal{H}(\theta_{B^r};r_\theta),
\end{equation*}
the gradients are propagated only through \blue{the transposes of the corresponding discrete filter operators}. In other words, the same steps in \cref{eq:discreteHelmholtzFilterRho,eq:rhoBackwardFilter} are used with $\rho$ replaced by $\phi$ or $\theta_{B^r}$ and with the projection derivative omitted.

The resulting gradients with respect to the raw design variables $\rho_e$, $\phi_e$, and $\theta_{B^r,e}$ are supplied to MMA for the next design update. When a design field is inactive, its gradient is not used and the corresponding field is held fixed during optimization.

The complete optimization procedure is summarized in the flowchart in \cref{fig:OptLoop}. The next section presents numerical examples demonstrating how the formulation handles different active design fields, objectives, and loading scenarios.

\definecolor{designFill}{RGB}{232,242,252}
\definecolor{analysisFill}{RGB}{255,242,224}
\definecolor{updateFill}{RGB}{241,235,252}
\definecolor{decisionFill}{RGB}{255,249,213}
\definecolor{finalFill}{RGB}{232,246,232}
\definecolor{yesColor}{RGB}{85,125,45}
\definecolor{noColor}{RGB}{175,55,45}

\begin{figure}[H]
    \centering
    \resizebox{\linewidth}{!}{%
\begin{tikzpicture}[
    font=\sffamily\small,
    node distance=8mm and 10mm,
    process/.style={
        rounded corners=2.5pt,
        draw=black,
        line width=0.7pt,
        minimum height=9mm,
        text width=29mm,
        align=center,
        inner sep=4pt
    },
    wide/.style={
        process,
        text width=35mm
    },
    design/.style={
        process,
        fill=designFill
    },
    designwide/.style={
        wide,
        fill=designFill
    },
    analysiswide/.style={
        wide,
        fill=analysisFill
    },
    update/.style={
        process,
        fill=updateFill
    },
    updatewide/.style={
        wide,
        fill=updateFill
    },
    decision/.style={
        diamond,
        aspect=2.1,
        draw=black,
        line width=0.7pt,
        fill=decisionFill,
        text width=22mm,
        align=center,
        inner sep=1.5pt
    },
    final/.style={
        rounded corners=2.5pt,
        draw=black,
        line width=0.8pt,
        fill=finalFill,
        minimum height=9mm,
        text width=25mm,
        align=center,
        inner sep=4pt
    },
    arrow/.style={
        -{Latex[length=2.2mm,width=1.7mm]},
        line width=0.75pt,
        draw=black!80
    },
    yesarrow/.style={
        -{Latex[length=2.2mm,width=1.7mm]},
        line width=0.9pt,
        draw=yesColor
    },
    noarrow/.style={
        -{Latex[length=2.2mm,width=1.7mm]},
        line width=0.9pt,
        draw=noColor,
        rounded corners=5pt
    },
    lab/.style={
        font=\sffamily\footnotesize,
        fill=white,
        inner sep=1pt
    },
    yeslab/.style={
        lab,
        text=yesColor
    },
    nolab/.style={
        lab,
        text=noColor
    }
]

\node[design] (init)
{Initialize raw\\design variables\\[-1pt]$\rho,\,\phi,\,\theta_{B^r}$};

\node[design,right=of init] (filter)
{Filter and projection\\[-1pt]$\mathcal{H},\,\mathcal{P}_{\rho}$};

\node[designwide,right=of filter] (phys)
{Physical design fields\\[-1pt]$\bar{\rho},\,\phi_{\mathrm{phys}},\,\theta_{B^r,\mathrm{phys}}$};

\node[analysiswide,right=of phys] (solve)
{Nonlinear finite-element solve\\[-1pt]$\widehat{\mathbf{r}}^{(\ell)}=\mathbf{0}$};

\node[analysiswide,below=15mm of solve] (eval)
{Objective and\\constraint evaluation\\[-1pt]$f,\,g_i$};

\node[updatewide,left=of eval] (adj)
{Adjoint sensitivities\\and chain rule};

\node[update,left=of adj] (mma)
{MMA update\\of raw variables};

\node[decision,left=of mma] (conv)
{Converged?};

\node[final,below=13mm of conv] (final)
{Final design};

\draw[arrow] (init) -- (filter);
\draw[arrow] (filter) -- (phys);
\draw[arrow] (phys) -- (solve);
\draw[arrow] (solve) -- (eval);
\draw[arrow] (eval) -- (adj);
\draw[arrow] (adj) -- (mma);
\draw[arrow] (mma) -- (conv);

\draw[yesarrow] (conv) -- node[yeslab,right] {yes} (final);

\draw[noarrow]
    (conv.north) -- ++(0,1.00) -|
    node[nolab,pos=0.25,above] {no}
    (filter.south);

\end{tikzpicture}
}
    \caption{Flowchart of the joint material--structural optimization procedure. Starting from the raw design variables, filtering and projection are applied to obtain the physical design fields. The nonlinear finite-element problem is then solved, objective and constraint values are evaluated, adjoint sensitivities are computed, and MMA updates the raw design variables. The loop continues until the convergence criterion is satisfied.}
    \label{fig:OptLoop}
\end{figure}

\blue{\subsection{Fixed design regions}\label{ss:fixedDesignRegions}

The parameterizations and gradient formulas above describe active fields whose values are free throughout $\Omega_0$. A field may also be held fixed in part of the domain. For the field under consideration, let $\Omega_{\mathrm{fixed}}\subset\Omega_0$ denote this region. In the discretization, $\Omega_{\mathrm{fixed}}$ is represented by a union of elements whose indices form $\mathcal E_{\mathrm{fixed}}$:
\begin{equation*}
    \Omega_{\mathrm{fixed}}
    =
    \bigcup_{e\in\mathcal E_{\mathrm{fixed}}}\Omega_e.
\end{equation*}
For a raw field $q\in\{\rho,\phi,\theta_{B^r}\}$, the element values $q_e$ in $\mathcal E_{\mathrm{fixed}}$ remain at their specified values and are excluded from the optimization vector. Filtering is still performed over the full domain $\Omega_0$, using both the free and fixed raw values.

The physical values on the fixed elements must also remain fixed. Let $\mathcal I_{\mathrm{fixed}}$ contain the indices of the physical degrees of freedom belonging to the elements in $\mathcal E_{\mathrm{fixed}}$. For continuous physical fields, these are the nodes of those elements, including nodes shared with neighboring elements. For an unfiltered element-wise field, they are the physical element values.

Let $\mathcal T$ denote filtering, projection, or their composition for the field under consideration. Then $\mathcal T(q)$ is the field obtained before the fixed physical values are imposed. For example, $\mathcal T(\rho)=\mathcal P_\rho\!\left(\mathcal H(\rho;r_\rho)\right)$ is the projected density. For the unfiltered direction field in the three-dimensional example in \cref{ss:design3d}, $\mathcal T(\theta_{B^r})=\theta_{B^r}$. The physical field used in the finite-element problem is then defined by:
\begin{equation*}
    q_{\mathrm{phys},i}
    =
    \begin{cases}
        q_i^{\mathrm{fixed}},
        &i\in\mathcal I_{\mathrm{fixed}},\\
        [\mathcal T(q)]_i,
        &i\notin\mathcal I_{\mathrm{fixed}}.
    \end{cases}
\end{equation*}
Here, $q_i^{\mathrm{fixed}}$ is the specified fixed physical value, and $q_{\mathrm{phys}}$ denotes $\bar\rho$, $\phi_{\mathrm{phys}}$, or $\theta_{B^r,\mathrm{phys}}$ for the field being considered. The fixed region and its values may differ among the three fields. If no region is held fixed, $\mathcal E_{\mathrm{fixed}}$ and $\mathcal I_{\mathrm{fixed}}$ are empty.

To differentiate this assignment, note that fixed physical values do not vary with the free raw values. Therefore, for a free element value,
\begin{equation*}
    \frac{df}{dq_e}
    =
    \sum_{i\notin\mathcal I_{\mathrm{fixed}}}
    \frac{df}{dq_{\mathrm{phys},i}}
    \frac{\partial[\mathcal T(q)]_i}{\partial q_e},
    \qquad e\notin\mathcal E_{\mathrm{fixed}}.
\end{equation*}
For density, the diagonal matrix $\mathbf D_\rho$ in \cref{ss:sensitivity} is replaced by
\begin{equation*}
    (\mathbf D_\rho)_{ii}
    =
    \begin{cases}
        0,&i\in\mathcal I_{\mathrm{fixed}},\\
        \mathcal P_\rho'(\tilde\rho_i),
        &i\notin\mathcal I_{\mathrm{fixed}}.
    \end{cases}
\end{equation*}
The remaining operations in \cref{eq:chainRhoVector,eq:discreteHelmholtzFilterRho,eq:rhoBackwardFilter} are unchanged. For particle volume fraction and remanent direction, no projection derivative is required. Gradient entries corresponding to fixed physical degrees of freedom are set to zero before applying the transpose of the corresponding discrete Helmholtz filter operator. No filter operation is required for an unfiltered direction field.

Only derivatives with respect to the free raw element values are supplied to MMA. The same treatment applies to objective and constraint gradients.
}

\section{Design of hMSM structures using joint material--structural optimization}
\label{s:applications}

This section demonstrates the joint material--structural optimization framework developed in \cref{s:topOptimFormulation}. All examples use the constitutive model selected in \cref{s:modelSelection}, namely the \textbf{NH2} strain-energy density together with the \textbf{Mooney} effective shear-modulus relation. \blue{The examples progress from optimizing particle volume fraction and remanent direction in fixed geometries to optimizing all three design fields, as summarized in \cref{tab:optimizationExamplesSummary}.} The wheel and translational examples use prescribed geometries, while the restorative beam and three-dimensional morphing surface also vary structural density. These cases show how the \blue{selected $G(\phi)$ relation} and the separate design fields enter the objectives considered here.

\begin{table}[h]
\centering
\caption{Summary of the optimization examples considered in this section.}
\label{tab:optimizationExamplesSummary}
\vspace{-5pt}
\small
\setlength{\tabcolsep}{6pt}
\renewcommand{\arraystretch}{1.15}
\begin{tabularx}{\linewidth}{@{}l c X@{}}
\toprule
\textbf{Example} & \textbf{Optimized fields} & \textbf{Main demonstration} \\
\midrule
Rotational actuator
&
$\phi,\ \theta_{B^r}$
&
Magnetic design in a fixed wheel geometry: redistribute magnetic material and orient remanent magnetization to increase \blue{tangential displacement near the outer rim}.
\\
\midrule
Translational actuator
&
$\phi,\ \theta_{B^r}$
&
Magnetic design in a fixed scissor-like geometry: produce \blue{a prescribed horizontal displacement of the output plate} while suppressing undesired vertical displacement.
\\
\midrule
Restorative beam
&
$\rho,\ \phi,\ \theta_{B^r}$
&
Full joint material--structural design: optimize structural density, magnetic particle distribution, and remanent magnetization direction for bidirectional restoration.
\\
\midrule
3D morphing surface
&
$\rho,\ \phi,\ \theta_{B^r}$
&
Three-dimensional joint design: track a prescribed surface shape with the remanent direction constrained to the $x$--$y$ plane.
\\
\bottomrule
\end{tabularx}
\end{table}

\subsection{Key computational details}

\Cref{tab:optimizationParameters} summarizes the key parameters used in the four optimization examples. The implementation is based on the FEniTop library\footnote{FEniTop: \url{https://github.com/missionlab/fenitop}} \cite{jia2024fenitop} and the FEniCSx finite-element library (version 0.10.0) \cite{baratta2023dolfinx}. 

For each optimization iteration and load case, the nonlinear equilibrium equation is solved for the displacement field using the DOLFINx Newton solver. The Jacobian is obtained by automatic differentiation of the nonlinear residual. An incremental convergence criterion is used, with the absolute and relative tolerances both set to $10^{-4}$. To improve robustness for the finite-deformation response, the applied magnetic field and mechanical traction are increased linearly from zero to their target values over $N_{\mathrm{step}}$ equal load increments. The converged displacement from each increment is used as the initial guess for the following increment. At the beginning of each load case, the displacement field and applied loads are reset to zero so that the load cases are solved independently.

Further implementation details are available in the \lstinline{CEADpx/MatTO} software repository\footnote{\lstinline{CEADpx/MatTO}: \blue{\url{https://github.com/CEADpx/matto/tree/v0.3.0}}} \cite{galloway2026matto}.

\begin{table}[h]
\centering
\caption{Optimization parameters that differ across the four design examples. \blue{Common settings are the \textbf{NH2} strain-energy density, Mooney effective shear-modulus relation, $G_0=100~\mathrm{kPa}$, $\phi_{\max}=0.30$, $\rho_{\min}=0.05$, and $p_\rho=3$. The first three examples use $\epsilon_\rho=10^{-6}$, an optimization tolerance of $10^{-5}$, and at most $100$ optimization iterations.} \blue{For the 3D morphing surface, results are reported at iteration $500$ with $\epsilon_\rho=10^{-2}$ and a design-change tolerance of $10^{-3}$. Its density-projection parameters are $\beta_{\max}=8$ and $\beta_{\mathrm{int}}=30$; the restorative beam uses $\beta_{\max}=4$ and $\beta_{\mathrm{int}}=25$.} In the rotational and translational actuator examples, $\rho\equiv1$ is fixed and $v_{\max}=1.00$ therefore denotes a fully solid structure rather than an active volume constraint. Here, $\mathcal{D}_{\mathrm{act}}$ is the set of active design fields, $B^r$ is the fixed magnitude of the remanent magnetic flux density, $\bB^a$ is the applied magnetic flux-density vector, $\mathbf{t}$ is the applied traction vector, $N_{\mathrm{LC}}$ is the number of load cases, $N_{\mathrm{step}}$ is the number of load steps per load case, $v_{\max}$ is the \blue{upper bound on the structural material volume fraction}, $\Phi_{\max}$ is the \blue{upper bound on the domain-averaged magnetic particle volume fraction}, $r$ is the filter radius, and $m_{\mathrm{MMA}}$ is the MMA move limit. The entries for $B^r$ and $\bB^a$ are reported in $\mathrm{mT}$, $\mathbf{t}$ in $\mathrm{mN/mm^2}$, and $r$ in $\mathrm{mm}$. Entries on corresponding lines in the $\bB^a$ and $\mathbf{t}$ columns define paired load cases. For the 3D example, the entries in the $r$ column give $r_\rho$, $r_\phi$, and $r_\theta$, respectively; a dash indicates that the direction field is unfiltered.}
\label{tab:optimizationParameters}
\vspace{-5pt}
\scriptsize
\setlength{\tabcolsep}{2.2pt}
\renewcommand{\arraystretch}{1.2}
\resizebox{\textwidth}{!}{%
\begin{tabular}{@{}l c c c c c c c c c c l@{}}
\toprule
\textbf{Example}
&
$\mathcal{D}_{\mathrm{act}}$
&
$B^r$
&
$\bB^a$
&
$\mathbf{t}$
&
$N_{\mathrm{LC}}$
&
$N_{\mathrm{step}}$
&
$v_{\max}$
&
$\Phi_{\max}$
&
$r$
&
$m_{\mathrm{MMA}}$
&
$f$
\\
\midrule
\makecell[l]{Rotational\\actuator}
&
$\{\phi,\theta_{B^r}\}$
&
$100$
&
$\blue{[0,100]^T}$
&
$\blue{[0,0]^T}$
&
$1$
&
$100$
&
$1.00$
&
$0.30$
&
$1.00$
&
$0.01$
&
\makecell[l]{\blue{tangential displacement}\\\blue{near the outer rim}}
\\
\midrule
\makecell[l]{Translational\\actuator}
&
$\{\phi,\theta_{B^r}\}$
&
$100$
&
$\blue{[125,0]^T}$
&
$\blue{[0,0]^T}$
&
$1$
&
$40$
&
$1.00$
&
$0.15$
&
$0.20$
&
$0.01$
&
\makecell[l]{displacement\\tracking}
\\
\midrule
\makecell[l]{Restorative\\beam}
&
$\{\rho,\phi,\theta_{B^r}\}$
&
$200$
&
\makecell[c]{$\blue{[0,25]^T}$\\$\blue{[0,-25]^T}$}
&
\makecell[c]{$\blue{[0,-0.50]^T}$\\$\blue{[0,0.50]^T}$}
&
$2$
&
$50$
&
$0.50$
&
$0.10$
&
$1.00$
&
$\blue{0.05}$
&
compliance
\\
\midrule
\makecell[l]{3D morphing\\surface}
&
$\{\rho,\phi,\theta_{B^r}\}$
&
$200$
&
\blue{$[0,0,200]^T$}
&
$\blue{[0,0,-0.50]^T}$
&
$1$
&
$10$
&
$0.60$
&
$0.10$
&
\makecell[c]{\blue{$1.00$}\\\blue{$1.00$}\\\blue{--}}
&
$0.05$
&
\makecell[l]{surface-shape\\tracking}
\\
\bottomrule
\end{tabular}%
}
\end{table}

\subsection{Rotational actuator}\label{ss:rotationalActuator}

\begin{figure}[H]
    \centering
    \includegraphics[width=0.85\linewidth]{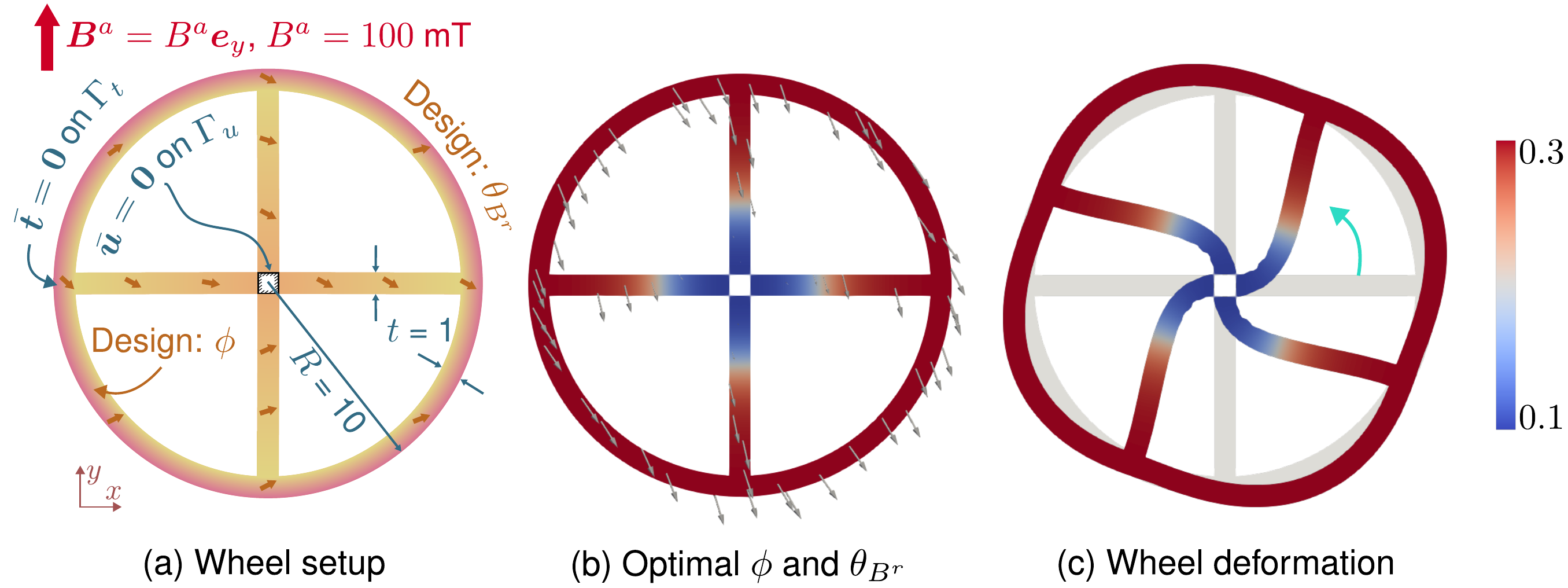}
    \caption{Optimization of the rotational actuator. (a) Fixed wheel geometry with prescribed structural density $\rho\equiv1$, remanent magnetic flux-density magnitude $B^r=100~\mathrm{mT}$, and uniform upward applied magnetic field $\bB^a=\blue{[0,100]^T}~\mathrm{mT}$. The magnetic particle volume fraction $\phi$ and remanent direction $\theta_{B^r}$ are optimized. (b) Optimized magnetic particle volume fraction and remanent magnetization direction. The arrows indicate the optimized local direction $\theta_{B^r}$. (c) Deformed configuration under the applied magnetic field, with the undeformed configuration shown in the background.}
    \label{fig:wheelOptimization}
\end{figure}

The first example optimizes the magnetic material distribution and remanent magnetization direction within the fixed wheel geometry shown in \cref{fig:wheelOptimization}(a). The structural density is prescribed as $\rho\equiv1$, while the magnetic particle volume fraction $\phi$ and remanent direction $\theta_{B^r}$ are treated as design fields. Their raw fields are initialized uniformly at $\phi=0.30$ and $\theta_{B^r}=0$; the initial \blue{particle volume fraction} equals \blue{the prescribed bound on the domain-averaged particle volume fraction}. The objective is to maximize the counterclockwise tangential displacement near the outer rim under a uniform upward applied magnetic field, $\bB^a=\blue{[0,100]^T}~\mathrm{mT}$.

To quantify the rotational response, the tangential displacement of material points located in a narrow annular band near the outer rim of the wheel is measured. This avoids relying on a small set of boundary nodes and gives a smoother objective for optimization. For a material point $\bX=\blue{[X,Y]^T}$, the counterclockwise unit tangent vector about the wheel center is
\begin{equation*}
    \btau(\bX)
    =
    \frac{1}{\sqrt{X^2+Y^2}}\, \blue{\left[-Y, X \right]^T} .
\end{equation*}
The rotational objective is then defined as
\begin{equation}\label{eq:wheelObjective}
    f_{\mathrm{rot}}(\bu)
    =
    -
    \int_{\Omega_0}
    w_{\mathrm{rim}}(\bX)\,
    \bu(\bX)\cdot\btau(\bX)\,\dd V .
\end{equation}
Here, $w_{\mathrm{rim}}$ is a smooth radial weighting function given by
$w_{\mathrm{rim}}=\exp[-((r-r_0)/\sigma)^2]$, where
$r=\sqrt{X^2+Y^2}$, $r_0=0.95R$, and $\sigma=0.75~\mathrm{mm}$.
This weighting localizes the objective to a narrow annular band near the outer rim.
The quantity $\bu\cdot\btau$ is the counterclockwise tangential displacement. Therefore, the negative sign in \cref{eq:wheelObjective} makes the minimization problem favor larger counterclockwise rotation. Other computational parameters are listed in \cref{tab:optimizationParameters}.

The optimized fields are shown in \cref{fig:wheelOptimization}(b). Despite the fixed structural geometry, the optimization produces a nonuniform magnetic particle distribution and a spatially varying remanent magnetization direction. Because $\phi$ influences both the magnetic coupling and the effective shear modulus through the Mooney relation, its distribution controls the local balance between magnetic actuation and mechanical stiffness. Simultaneously, $\theta_{B^r}$ controls the local orientation of the remanent magnetic flux density. The optimizer therefore coordinates the two fields to promote the desired counterclockwise motion rather than producing a uniformly magnetized wheel.

The resulting deformation is shown in \cref{fig:wheelOptimization}(c). Near the wheel axis, one arm bends under magnetic actuation, and its compressed edge develops a localized inward deflection. This deformation is consistent with local buckling, but a stability analysis was not performed to establish a buckling threshold. For comparison, the intuitive wheel design introduced in \cref{s:WheelGripperBench} is used as the reference design. This design has a uniform magnetic particle volume fraction $\phi=\phi_{\max}=0.30$ and a uniform horizontal remanent magnetization direction, $\theta_{B^r}=0~\mathrm{rad}$. Although the optimization is driven by the weighted tangential-displacement objective in \cref{eq:wheelObjective}, the final designs are compared using the total displacement magnitude, $\|\bu(\bX_m)\|$, at the material point $\bX_m=\blue{[0,10]^T}~\mathrm{mm}$ located at the top of the undeformed wheel. When evaluated under the same applied magnetic loading, the intuitive design produces a displacement magnitude of $6.35~\mathrm{mm}$, whereas the optimized design achieves $10.14~\mathrm{mm}$. This represents a $59.7\%$ increase in the measured displacement magnitude and demonstrates that optimizing $\phi$ and $\theta_{B^r}$ can substantially enhance rotational actuation even when the structural geometry is fixed.

\subsection{Translational actuator}\label{ss:translationalActuator}

\begin{figure}[H]
\centering
\includegraphics[width=0.95\linewidth]{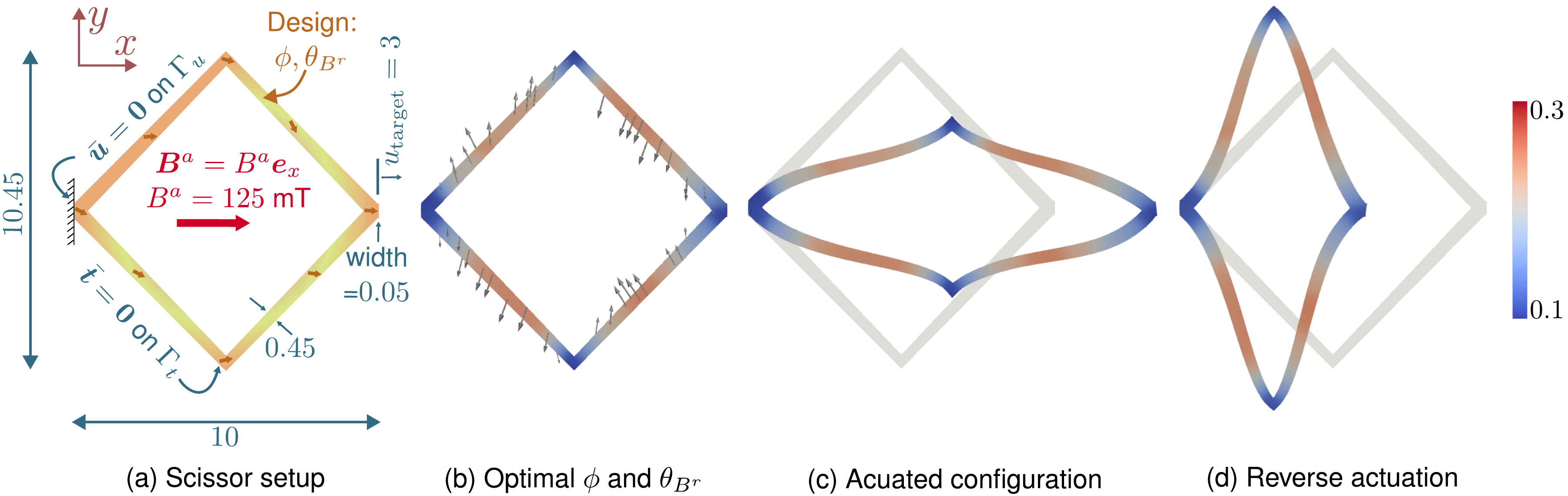}
\caption{Optimization of the translational actuator. (a) Fixed scissor-like geometry with prescribed structural density $\rho\equiv1$, remanent magnetic flux-density magnitude $B^r=100~\mathrm{mT}$, and uniform rightward applied magnetic field $\bB^a=\blue{[125,0]^T}~\mathrm{mT}$. The magnetic particle volume fraction $\phi$ and remanent direction $\theta_{B^r}$ are optimized to produce the target output displacement $\bu_{\mathrm{target}}=\blue{[3,0]^T}~\mathrm{mm}$. (b) Optimized magnetic particle volume fraction and remanent magnetization direction. The arrows indicate the optimized local direction $\theta_{B^r}$. (c) Deformed configuration under the rightward applied magnetic field. (d) Deformed configuration under the reversed applied magnetic field $\bB^a=\blue{[-125,0]^T}~\mathrm{mT}$.}
\label{fig:Scissor}
\end{figure}

The second example optimizes the magnetic material distribution and remanent magnetization direction within the fixed scissor-like geometry shown in \cref{fig:Scissor}(a). As in the rotational actuator, the structural density is prescribed as $\rho\equiv1$, while $\phi$ and $\theta_{B^r}$ are treated as design fields. The raw fields start from $\phi=0.15$ and $\theta_{B^r}=0$ uniformly, except that $\phi=0$ is imposed in the nonmagnetic output plate. \blue{The uniform value equals the prescribed bound on the domain-averaged particle volume fraction}, while fixing the output plate at zero leaves the initial domain average below that bound. Under the uniform rightward applied magnetic field $\bB^a=\blue{[125,0]^T}~\mathrm{mT}$, the objective is to produce a prescribed rightward displacement of the output plate while suppressing undesired vertical motion. The corresponding optimization parameters are listed in the translational-actuator row of \cref{tab:optimizationParameters}.

A displacement-tracking objective localized near the output point is used to optimize $\phi$ and $\theta_{B^r}$ so that the output plate undergoes the prescribed horizontal displacement while suppressing vertical motion:
\begin{equation}\label{eq:translationObjective}
    f_{\mathrm{tr}}(\bu)
    =
    \int_{\Omega_0}
    w_{\mathrm{out}}(\bX)
    \left\|
        \bu(\bX)-\bu_{\mathrm{target}}
    \right\|^2
    \dd V ,
\end{equation}
where $w_{\mathrm{out}}$ is a smooth Gaussian weighting function centered at the output location:
\begin{equation*}
    w_{\mathrm{out}}(\bX)
    =
    \exp\!\left(
    -\frac{\norm{\bX-\bX_{\mathrm{out}}}^2}{\sigma_{\mathrm{out}}^2}
    \right),
    \qquad
    \bX_{\mathrm{out}}=\blue{[9.99,7]^T}~\mathrm{mm},
    \qquad
    \sigma_{\mathrm{out}}=0.35~\mathrm{mm}.
\end{equation*}
In the implementation, the target displacement is
\begin{equation*}
    \bu_{\mathrm{target}}
    = \blue{\left[3~\mathrm{mm}, 0\right]^T}.
\end{equation*}
Thus, minimizing \cref{eq:translationObjective} drives the displacement field toward the desired motion while penalizing deviations from the prescribed displacement vector.

Several non-intuitive features emerge from the optimized design. As shown in \cref{fig:Scissor}(b), magnetic material is concentrated near the central portions of the scissor members, with comparatively little material placed in the compliant hinge regions. The output plate is prescribed as a nonmagnetic region, $\phi=0$, and is therefore excluded from the magnetic material optimization. The optimizer also produces two dominant remanent magnetization orientations, approximately aligned with the upper-left and lower-left diagonal directions. Together, the optimized particle distribution and remanent directions generate a coordinated deformation response that promotes the prescribed horizontal output motion while reducing parasitic vertical displacement.

\blue{The deformation shown in \cref{fig:Scissor}(c) is evaluated} at the material point $\bX_m$ located at the vertical midpoint of the outer edge of the output plate. Under the rightward applied magnetic field $\bB^a=\blue{[125,0]^T}~\mathrm{mT}$, the optimized design produces $\bu(\bX_m)=\blue{[2.62,0.039]^T}~\mathrm{mm}$, confirming that the output motion is predominantly horizontal. \blue{For the evaluation shown} in \cref{fig:Scissor}(d), the optimized distributions of $\phi$ and $\theta_{B^r}$ are held fixed while the applied field is reversed to $\bB^a=\blue{[-125,0]^T}~\mathrm{mT}$. The resulting displacement is $\bu(\bX_m)=\blue{[-4.00,-0.075]^T}~\mathrm{mm}$, indicating motion in the opposite horizontal direction. The forward and reversed displacement magnitudes are not expected to be equal because the response is geometrically nonlinear and \blue{only the load case with the rightward applied field} is included in the optimization. Nevertheless, the reversed response demonstrates directional reversibility within the model and shows that the proposed framework \blue{can optimize particle distributions and remanent directions} for prescribed and controllable modes of motion.

\subsection{Restorative beam}\label{ss:restorativeBeam}

The third example applies the formulation to a restorative beam by varying structural density $\rho$, magnetic particle volume fraction $\phi$, and remanent magnetization direction $\theta_{B^r}$ together. In contrast to the rotational and translational actuators, the structural geometry is no longer prescribed. The goal is to obtain a beam that resists mechanical loading while using magnetic actuation to restore the structure toward its undeformed configuration. The raw fields are initialized uniformly at $\rho=0.50$, $\phi=0.10$, and $\theta_{B^r}=0$; the first two values match the domain-average volume bounds. Alternative starts are examined in Section~6.6 and \labelcref{s:appBeamSensitivity}.

\begin{figure}[H]
\centering
\includegraphics[width=0.85\textwidth]{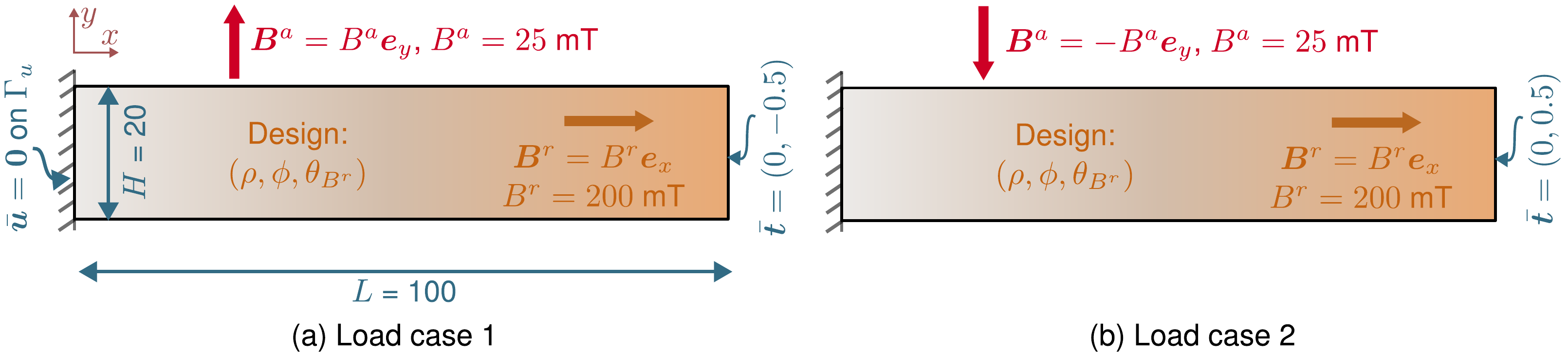}
\caption{Multi-load-case restorative beam setup with $B^r=200~\mathrm{mT}$. (a) Downward traction $\bar{\bt}=\blue{[0,-0.50]^T}~\mathrm{kPa}$ paired with an upward applied magnetic field $\bB^a=\blue{[0,25]^T}~\mathrm{mT}$. (b) Reversed loading with upward traction $\bar{\bt}=\blue{[0,0.50]^T}~\mathrm{kPa}$ and downward applied magnetic field $\bB^a=\blue{[0,-25]^T}~\mathrm{mT}$. The design variables are the structural density $\rho$, magnetic particle volume fraction $\phi$, and remanent magnetization direction $\theta_{B^r}$. Lengths are in millimeters.}
\label{fig:beamOptSetup}
\end{figure}

The problem setup is presented in \cref{fig:beamOptSetup}. The key feature of this example is the use of two opposing load cases. In the first load case, a downward traction is paired with an upward applied magnetic field; in the second, both directions are reversed. This multi-load-case formulation prevents the design from being specialized to a single bending direction and instead requires the optimized structural and magnetic fields to provide restorative behavior under loading in either direction. The optimization parameters for this case are listed in the restorative-beam row of \cref{tab:optimizationParameters}.

The objective is the weighted compliance over the load cases,
\begin{equation}\label{eq:beamCompliance}
    f_{\mathrm{comp}}
    =
    \sum_{\ell=1}^{N_\ell}
    w_\ell
    \int_{\Gamma_t^{(\ell)}}
    \bu^{(\ell)}
    \cdot
    \bar{\bt}^{(\ell)}
    \,\dd S ,
\end{equation}
where $\bu^{(\ell)}$ and $\bar{\bt}^{(\ell)}$ are the displacement field and applied traction for load case $\ell$, respectively, and $w_\ell$ is the corresponding load-case weight. Equal weights, $w_1=w_2=1$, are assigned to the two opposing load cases. Minimizing \cref{eq:beamCompliance} promotes a structure that remains stiff under the applied tractions while \blue{using the optimized particle distribution and remanent directions to counteract} the mechanical loading.

\begin{figure}[H]
\centering
\includegraphics[width=0.9\textwidth]{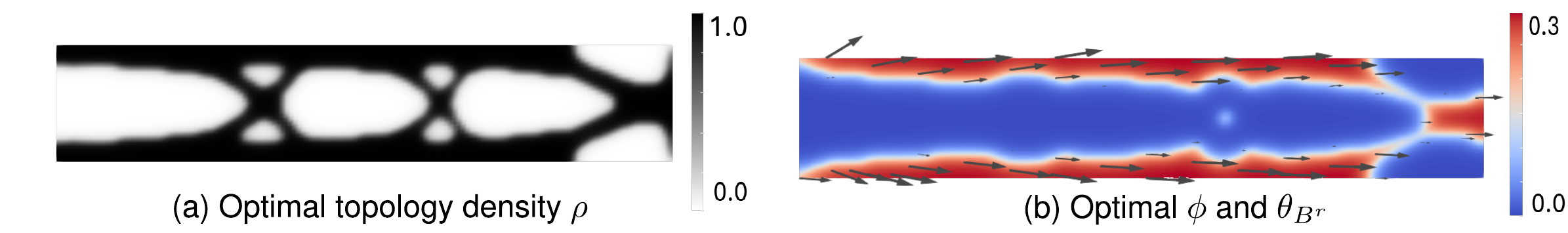}
\caption{Optimized restorative beam. (a) Optimized structural density $\rho$. (b) Optimized magnetic particle volume fraction $\phi$ and remanent magnetization direction $\theta_{B^r}$. Arrows indicate the optimized local direction $\theta_{B^r}$.}
\label{fig:beamOptimization}
\end{figure}

\cref{fig:beamOptimization} shows the optimized design. Because the two load cases are equal and opposite and are assigned the same weight, the objective does not favor either bending direction. The resulting structural and magnetic distributions are therefore nearly symmetric about the beam centerline. The optimized structural density forms continuous load-carrying paths along the upper and lower portions of the beam, while material is removed from regions that contribute less effectively to the restorative response. \blue{In the magnetic energy, the product $\bar{\rho}\phi_{\mathrm{phys}}$ vanishes as the projected structural density approaches zero.} Consequently, magnetic material is confined to structurally occupied regions rather than being placed within the void.

Within the solid regions, magnetic material is concentrated along portions of the upper and lower load paths. This placement serves two coupled purposes: increasing $\phi$ raises the local effective shear modulus through the Mooney relation, thereby stiffening key load-carrying regions, while also strengthening the local magnetic coupling where actuation can generate a restoring response. The spatially varying remanent directions further coordinate this response by allowing different regions to counteract the two opposing mechanical loads under their corresponding applied-field directions.

\begin{figure}[H]
\centering
\includegraphics[width=0.98\textwidth]{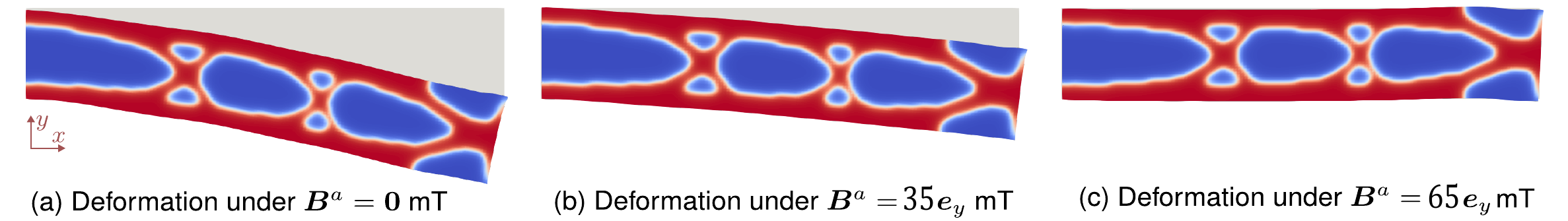}
\caption{Evaluation of the optimized beam under the fixed downward traction $\bar{\bt}=\blue{[0,-1.00]^T}~\mathrm{kPa}$, which is twice the magnitude used during optimization. Selected configurations are shown as the upward applied magnetic field is increased from $0$ to $75~\mathrm{mT}$. The magnetically induced response progressively counteracts the increased mechanical loading and restores the beam toward its undeformed configuration.}
\label{fig:beamLoadSteps}
\end{figure}

To evaluate the response of the optimized design beyond the loading conditions used during optimization, the beam is subjected to a downward traction $\bar{\bt}=\blue{[0,-1.00]^T}~\mathrm{kPa}$, which is twice the magnitude of the corresponding optimization traction. The applied magnetic field remains upward and is increased from $0$ to $75~\mathrm{mT}$ in increments of $5~\mathrm{mT}$. Selected configurations from this field sweep are shown in \cref{fig:beamLoadSteps}. With no applied magnetic field, the beam bends downward under the mechanical traction. As the applied magnetic field magnitude increases, the magnetically induced response progressively counteracts the mechanical load. At approximately $|\bB^a|=65~\mathrm{mT}$, the beam is restored close to its undeformed configuration despite the increased mechanical loading.

\blue{The optimized distributions of structural density, particle volume fraction, and remanent direction enable the beam to counteract mechanical loading in either direction;} \blue{the evaluation at twice the optimization traction further demonstrates restoration close to the undeformed configuration at an applied field of approximately $65~\mathrm{mT}$ (\cref{fig:beamLoadSteps}).}

\subsection{Design of a three-dimensional morphing surface}\label{ss:design3d}

\blue{In the final example, a morphing structure is designed to attain a prescribed shape under an applied magnetic field and mechanical load. The design domain is the slab $\Omega_0 = [-12, 12]\times[-12,12]\times[0,3]$ (length in mm), with four side faces clamped. The upper $0.5~\mathrm{mm}$ layer is a solid, nonmagnetic skin with $\rho=1$ and $\phi=0$ in both the raw and physical design fields. The physical values are enforced after filtering and projection. All three design fields are excluded from optimization in the skin, with the remanent direction held at its initial value. The structural density, particle volume fraction, and remanent direction are optimized below the skin.}

\blue{Mirror symmetry about $X=0$ and $Y=0$ is imposed on the design and displacement fields. The computational domain is the quadrant $\Omega_q=[0,12]\times[0,12]\times[0,3]~\mathrm{mm}^3$, discretized using $24\times24\times6$ hexahedral elements. The outer faces $X=12~\mathrm{mm}$ and $Y=12~\mathrm{mm}$ are clamped. On the symmetry planes, $u_x=0$ at $X=0$ and $u_y=0$ at $Y=0$, with the remaining displacement components unconstrained. The full slab is obtained by reflecting $\Omega_q$ across $X=0$ and $Y=0$. The scalar design fields, displacement components, and remanent directions are reflected across the same planes.}

The displacement field is fully three-dimensional. The remanent magnetization is constrained to the $x$--$y$ plane,
\begin{equation*}
    \bB^r
    =
    B^r [\cos\theta_{B^r}, \sin\theta_{B^r}, 0]^T, 
    \qquad B^r=200~\mathrm{mT},
\end{equation*}
while the uniform applied field is \blue{$\bB^a=[0,0,200]^T~\mathrm{mT}$}. Thus, this example demonstrates three-dimensional deformation and spatial optimization of all three design fields, while retaining the single-angle parameterization by prescribing the plane in which the remanent direction lies. \blue{The raw density and particle volume fraction fields below the skin are initialized uniformly at $\rho=0.50$ and $\phi=0.10$, with the prescribed skin values imposed as described above.}

The desired vertical displacement of the top surface is prescribed as the shallow dome
\begin{equation*}
    w^\star(X,Y)
    =
    h_\star
    \left[1-\left(X/12\right)^2\right]^2
    \left[1-\left(Y/12\right)^2\right]^2,
    \qquad
    h_\star=1~\mathrm{mm},
\end{equation*}
where $X$ and $Y$ are measured in millimeters from the center of the slab. A single load case applies a downward traction of $0.50~\mathrm{kPa}$ to the top surface under the applied magnetic field defined above. The objective is the normalized mean-square difference between the vertical surface displacement and the target shape \blue{over the quadrant's top surface},
\begin{equation*}
\blue{    f_{\mathrm{dome}}
    =
    \frac{1}{A_q h_\star^2}
    \int_{\Gamma_{\mathrm{top},q}}
    \left(u_z-w^\star\right)^2\,\dd S,
    \qquad
    A_q=12^2~\mathrm{mm}^2.}
\end{equation*}
\blue{The magnetic energy in \cref{eq:WhatMagn} couples the remanent direction to the gradient of vertical displacement with respect to $X$ and $Y$. The initial remanent direction follows the gradient of the target surface, and the corresponding angular field is optimized element-wise without filtering, with bounds $[-3\pi/2,\pi/2]$.}

\blue{The material-volume constraints are evaluated over the computational quadrant and include the prescribed skin:}
\begin{equation*}
\blue{    \frac{1}{|\Omega_q|}\int_{\Omega_q}\bar{\rho}\,\dd V\leq0.60,
    \qquad
    \frac{1}{|\Omega_q|}\int_{\Omega_q}\phi_{\mathrm{phys}}\,\dd V\leq0.10.}
\end{equation*}
\blue{By symmetry, evaluating these averages over the full slab gives the same values.}

\begin{figure}[H]
\centering
\includegraphics[width=0.99\textwidth]{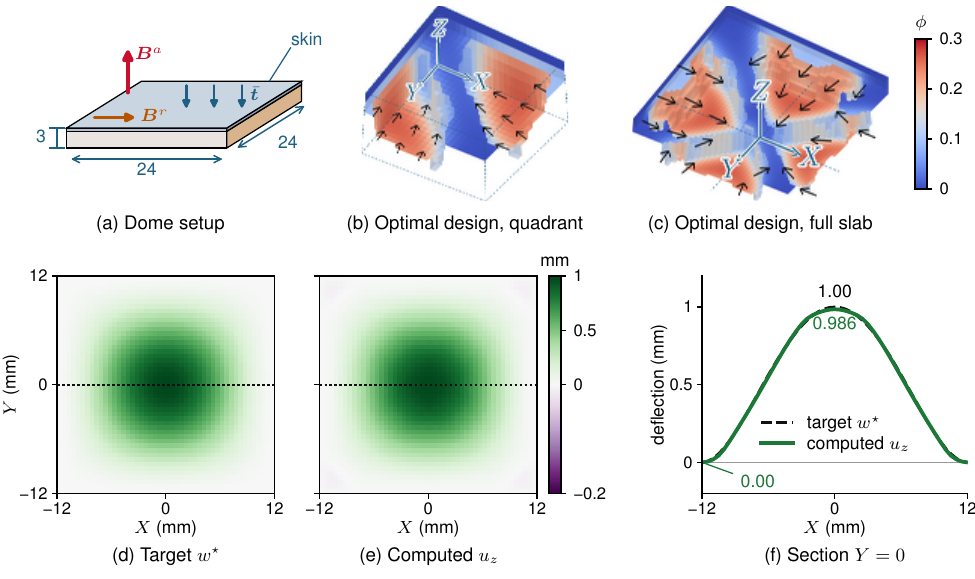}
\caption{\blue{Three-dimensional hMSM morphing surface at iteration $500$. (a) A $24\times24\times3~\mathrm{mm}$ slab with a $0.5~\mathrm{mm}$ solid, nonmagnetic skin, clamped on its side faces and subjected to a downward traction of $0.50~\mathrm{kPa}$ and an applied magnetic field of $200~\mathrm{mT}$ along $z$. The remanent flux density has magnitude $200~\mathrm{mT}$ and lies in the $x$--$y$ plane. (b, c) Optimized design viewed from below, including the skin: computed quadrant (b) and full slab obtained by reflection (c). The dashed box outlines the quadrant; the cut faces in (b) and dashed lines in (c) mark the symmetry planes. Cells are shown where physical density exceeds $0.5$; color denotes particle volume fraction, which is zero in the skin. Arrows indicate remanent directions on bottom faces where particle volume fraction is at least $0.1$. (d, e) Target and computed top-surface displacements on the same scale in millimeters. (f) Corresponding profiles along $Y=0$, marked by dotted lines in (d, e).}}
\label{fig:domeOptimization}
\end{figure}

{\blue{At iteration $500$, the computed central displacement is $0.986~\mathrm{mm}$ for a target of $1.00~\mathrm{mm}$, and the root-mean-square tracking error over the top surface is $0.00472~\mathrm{mm}$. Both volume constraints are satisfied. \Cref{fig:domeOptimization} shows the optimized design and compares the target and computed displacements over the surface and along $Y=0$.}}

\subsection{Sensitivity of the restorative-beam design}\label{ss:beamSensitivity}

\blue{The restorative beam is used to examine sensitivity to the effective shear-modulus relation, the initial design fields, and the mesh, and to compare the designs obtained with the $\bF$- and $\bR$-based magnetic energies.} This problem optimizes all three fields, $\rho$, $\phi$, and $\theta_{B^r}$. \blue{For the shear-modulus, initialization, and mesh studies,} each final design is transferred to a common $200\times40$ mesh and its two-load-case compliance is reevaluated without further optimization. The study settings and complete results are given in \labelcref{s:appBeamSensitivity}.

Designs obtained using the Mooney, Guth, and Hill $G(\phi)$ relations yield similar compliance values when evaluated under each fixed relation. For each fixed evaluation relation, the relative range $(f_{\mathrm{comp}}^{\max}-f_{\mathrm{comp}}^{\min})/f_{\mathrm{comp}}^{\min}$ across the three designs is at most $3.8\%$. Thus, for this beam problem, the attained compliance is robust with respect to the three tested $G(\phi)$ relations.

\blue{To assess sensitivity to initialization, the beam was optimized from four initial designs on the same $150\times30$ mesh using the Mooney relation: two uniform initializations differing only in remanent direction, a uniform initialization with smaller $\rho$ and $\phi$, and an initialization with spatially varying fields. The detailed initial fields are given in \labelcref{s:appBeamSensitivity}. The first two initializations give compliances of $51.52$ and $49.83$ when evaluated on the $200\times40$ mesh; the initialization with smaller $\rho$ and $\phi$ and the spatially varying initialization give $53.69$ and $45.48$, respectively. Thus, the two initializations differing only in remanent direction give similar compliance values, while the wider range across all four cases demonstrates greater sensitivity when the other initial fields also vary.}

To examine mesh sensitivity, we optimized the same Mooney beam from the baseline uniform initialization on $100\times20$, $150\times30$, and $200\times40$ meshes, holding the loads, constraints, and optimizer settings fixed. We then evaluated all three final designs on the same $200\times40$ mesh. Their compliances are $94.61$, $91.37$, and $90.73$, respectively. The difference falls from $4.3\%$ between the coarsest and finest designs to $0.7\%$ between the $150\times30$ and $200\times40$ designs, relative to the finest value. This suggests that the reported compliance becomes less sensitive to refinement over the two finer meshes; the full settings and results are given in \labelcref{s:appBeamSensitivity}.

\begin{figure}[htbp]
\centering
\includegraphics[width=0.9\textwidth]{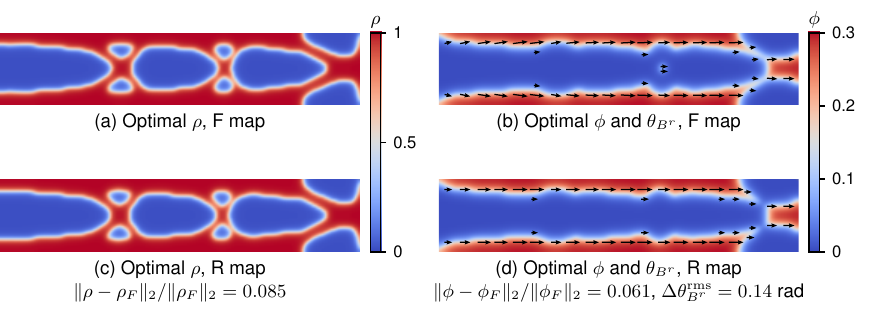}
\caption{\blue{Restorative-beam designs after $100$ optimization iterations using the $\bF$-based energy (a, b) and the $\bR$-based energy (c, d). Panels (a, c) show physical structural density; panels (b, d) show physical particle volume fraction and remanent direction. Arrows indicate $\theta_{B^r}$ where $\bar\rho\phi_{\mathrm{phys}}>0.03$; their length increases with $\bar\rho\phi_{\mathrm{phys}}$. The difference measures shown below (c, d) are defined in \labelcref{s:appMagneticMaps}.}}
\label{fig:beamMapDesigns}
\end{figure}

\blue{To compare the magnetic energies, we repeat the beam optimization using $\bR$ in place of $\bF$ on the same $150\times30$ mesh, with the initial fields, filters, density projection, material parameters, two load cases, volume bounds, and MMA settings unchanged. \Cref{fig:beamMapDesigns} shows the two designs after $100$ optimization iterations. They are similar, with small local differences in structural density, particle volume fraction, and remanent direction. The relative discrete $L^2$ differences are $0.085$ for physical density and $0.061$ for particle volume fraction, and the weighted root-mean-square angular difference is $0.14~\mathrm{rad}$ ($8.3^\circ$). Both designs satisfy the prescribed volume bounds. The comparison measures are defined in \labelcref{s:appMagneticMaps}. The similar designs obtained for this beam do not imply that the two energies give the same optimized design for other problems.}

\section{Conclusion}\label{s:conclusion}

This work developed a model-informed framework for the analysis and joint material--structural optimization of hard-magnetic soft materials. The first part of the work placed classical rigid-inclusion relations, a Hill self-consistent relation, and constrained-kinematics models into a common effective shear-modulus form $G(\phi)$. The numerical studies showed that, for the actuation problems considered, the choice of strain-energy density function has a relatively small effect on the predicted deformation, whereas the effective shear-modulus relation can have a significant effect. This sensitivity is strongest when magnetic material overlaps with regions undergoing substantial deformation, as observed in the uniformly magnetized beam and wheel examples, and is much weaker when magnetic material is separated from the dominant deformation region, as observed in the tip-magnetized beam and gripper examples. These results highlight the importance of selecting an appropriate structure--property relation when particle concentration influences both stiffness and magnetic actuation.

Experimental stress--strain data were then used to select a representative $G(\phi)$ relation. For the dataset and metrics considered, the \textbf{Mooney} relation gave the lowest errors, with \textbf{Guth} and \textbf{Hill} very close behind. The \textbf{Mooney} relation was used with the \textbf{NH2} strain-energy density in the optimization studies.

Building on this selected constitutive model, a joint material--structural optimization formulation was developed in which structural density, magnetic particle volume fraction, and remanent magnetization direction are coupled through the nonlinear finite-strain magneto-mechanical energy. The rotational, translational, restorative, and three-dimensional morphing examples showed that the framework can optimize different active subsets of the design fields, handle different objective functions and loading scenarios, and produce non-intuitive hMSM designs with prescribed deformation responses. \blue{For the restorative beam, the $\bF$- and $\bR$-based optimizations give similar designs, with small local differences in physical density, particle volume fraction, and remanent direction. The design differences may be larger for other problems.} \blue{For fabrication, these fields would be converted into a geometry, particle volume fractions, and magnetization directions that can be realized with the selected manufacturing process.}

Although the examples focused on hMSMs, the formulation is more broadly applicable to filled and field-responsive elastomers in which local material composition affects both stiffness and actuation. The same computational structure can be extended to dielectric elastomers, magnetorheological elastomers, and shape-memory-alloy-based elastomers by replacing the magnetic energy and effective-modulus relation with the corresponding electro-mechanical, magneto-mechanical, or thermomechanical material model. In this sense, the open-source \texttt{CEADpx/MatTO} implementation provides \blue{a starting point for a more general nonlinear joint optimization framework for coupled material composition, structural density, and actuation fields \citep{galloway2026framework}.}

Future work will focus on further validation, model extension, and computational acceleration. The present formulation uses an isotropic, concentration-dependent stiffness relation and a prescribed applied magnetic field; extending it to account for remanence-induced mechanical anisotropy and self- and demagnetizing fields would require an enriched free energy and a coupled magnetostatic state problem satisfying Maxwell's equations. Additional experimental datasets are needed to test the \blue{selected effective shear-modulus relations} under different filler types, volume fractions, loading modes, and actuation conditions. Extensions to viscoelastic and dissipative material models will be important for capturing rate dependence, hysteresis, and cyclic response in practical soft-material systems. Since nonlinear finite-strain optimization with multiple design fields and load cases can be computationally expensive, another important direction is the development of neural-operator and other surrogate models to accelerate repeated forward and adjoint solves. Recent work on neural operators, residual-based error correction, and neural-operator acceleration of inverse and optimization problems provides useful tools for this direction \cite{cao2022residual,jha2024residual,jha2025theory}. Together, these directions will strengthen the use of experimentally informed constitutive modeling and joint material--structural optimization as a systematic route for designing field-responsive soft structures with tailored actuation behavior.

\section*{Acknowledgments}
PKJ acknowledges support from the National Science Foundation through the Engineering Research Initiation (ERI) program under award No.~2502279, and from the South Dakota Board of Regents Competitive Research Grant (SDBOR CRG) program. These awards supported IG's work on this project at different stages. Any opinions, findings, and conclusions or recommendations expressed in this material are those of the authors and do not necessarily reflect the views of the funding agencies.

\section*{\blue{Use of AI tools}}
\blue{ChatGPT and Claude were used to review and edit the manuscript and assist with postprocessing of simulation results. Claude also assisted with software implementation and documentation. The authors reviewed and verified the resulting text, code, and results and take responsibility for the content of this work.}


\appendix

\section{Rigid-inclusion limit of Hill's self-consistent shear-modulus relation}
\label{s:appHill}

This section derives the rigid-inclusion limit of Hill's self-consistent shear-modulus relation used in \cref{ss:Gmodels}. Let $K_0$ and $G_0$ denote the bulk and shear moduli of the matrix, and let $K_1$ and $G_1$ denote the \blue{bulk and shear moduli of the inclusions}. The particle volume fraction is denoted by $\phi$. The effective bulk and shear moduli of the composite are denoted by $K(\phi)$ and $G(\phi)$, respectively.

For an isotropic dispersion of spherical inclusions, Hill's self-consistent equations \cite{hill1965self} can be written in the present notation as
\begin{equation*}
    \frac{1}{K(\phi)+\frac{4}{3}G(\phi)}
    =
    \frac{\phi}{K_1+\frac{4}{3}G(\phi)}
    +
    \frac{1-\phi}{K_0+\frac{4}{3}G(\phi)}
\end{equation*}
and
\begin{equation*}
    \frac{\phi}{G(\phi)-G_0}
    +
    \frac{1-\phi}{G(\phi)-G_1}
    =
    \frac{\beta(\phi)}{G(\phi)}.
\end{equation*}
The parameter $\beta(\phi)$ is defined through
\begin{equation*}
    \alpha(\phi)
    =
    3 - 5\beta(\phi),
    \qquad
    \alpha(\phi)
    =
    \frac{K(\phi)}
    {K(\phi)+\frac{4}{3}G(\phi)} .
\end{equation*}
Equivalently, combining the above two equations gives
\begin{equation}\label{eq:HillBetaAppendix}
    \beta(\phi) =  \frac{3 - \alpha(\phi)}{5} =
    \frac{6\left(K(\phi)+2G(\phi)\right)}
    {5\left(3K(\phi)+4G(\phi)\right)}.
\end{equation}

The bulk-modulus equation can be rearranged as
\begin{equation}\label{eq:HillBulkSolvedAppendix}
    K(\phi)
    =
    \frac{
    K_0-4\phi A_K(\phi)G(\phi)
    }
    {
    1+3\phi A_K(\phi)
    },
    \qquad
    A_K(\phi)
    =
    \frac{K_0-K_1}{3K_1+4G(\phi)}.
\end{equation}
Similarly, the shear-modulus equation can be written as
\begin{equation}\label{eq:HillShearSolvedAppendix}
    G(\phi)
    =
    \frac{G_0}{1+\phi A_G(\phi)},
    \qquad
    A_G(\phi)
    =
    \frac{G_0-G_1}
    {(1-\beta(\phi))G(\phi)+\beta(\phi)G_1}.
\end{equation}
\blue{\Cref{eq:HillBulkSolvedAppendix,eq:HillShearSolvedAppendix}} remain implicit because $A_K(\phi)$ and $A_G(\phi)$ depend on the effective moduli.

Next, consider the rigid-inclusion limit,
\begin{equation*}
    K_1 \rightarrow \infty,
    \qquad
    G_1 \rightarrow \infty.
\end{equation*}
In this limit,
\begin{equation*}
    A_K(\phi) \rightarrow -\frac{1}{3},
    \qquad
    A_G(\phi) \rightarrow -\frac{1}{\beta(\phi)}.
\end{equation*}
Substitution into \cref{eq:HillBulkSolvedAppendix,eq:HillShearSolvedAppendix} gives
\begin{equation}\label{eq:HillRigidBulkAppendix}
    K(\phi)
    =
    \frac{K_0+\frac{4}{3}\phi G(\phi)}{1-\phi},
\end{equation}
and
\begin{equation}\label{eq:HillRigidShearAppendix}
    G(\phi)
    =
    \frac{G_0}{1-\phi/\beta(\phi)}.
\end{equation}

Finally, take the nearly incompressible matrix limit, $K_0\rightarrow\infty$. From \cref{eq:HillRigidBulkAppendix}, this implies $K(\phi)\rightarrow\infty$ for $0\leq \phi<1$. Substituting this limit into \cref{eq:HillBetaAppendix} gives
\begin{equation*}
    \beta(\phi) \rightarrow \frac{2}{5}.
\end{equation*}
Therefore, \cref{eq:HillRigidShearAppendix} reduces to
\begin{equation*}
    G(\phi)
    =
    \frac{G_0}{1-\frac{5}{2}\phi}.
\end{equation*}
This is the \textbf{Hill} effective shear-modulus relation used in \cref{eq:GmodelsStrategy1}. The expression is singular at $\phi=0.4$, so it is used only for $\phi<0.4$. In the optimization studies, the maximum magnetic particle volume fraction is $\phi_{\max}=0.30$, which remains below this singular limit.

\section{Mesh convergence studies}\label{s:appConvergence}

Mesh convergence studies were performed to verify that the numerical predictions presented throughout this work were not significantly influenced by spatial discretization. For each benchmark problem, progressively refined meshes were analyzed, and the resulting displacement responses were compared against a reference solution obtained using the finest mesh considered. The mesh resolutions adopted in the main text were selected to provide a balance between computational cost and solution accuracy.

\subsection{Beam benchmark}

The cantilever beam geometry introduced in \cref{ss:beamSetup} was analyzed using characteristic mesh sizes $h_c=0.15$, $0.10$, $0.08$, and $0.06~\mathrm{mm}$, where $h_c$ denotes the target element size prescribed during mesh generation. The domain was discretized using unstructured first-order triangular finite elements.

Convergence was assessed using the average displacement magnitude measured along the right edge of the beam. Two representative cases were considered: the uniformly magnetized beam subjected to the largest applied field, $B^a=400~\mathrm{mT}$, and the tip-magnetized beam shown in \cref{fig:beamSetup}(b). The resulting displacement predictions and relative differences with respect to the reference mesh are summarized in \cref{tab:meshConvBeam}.

\begin{table}[!tbp]
\centering
\caption{Beam benchmark convergence study.}
\label{tab:meshConvBeam}
\vspace{-5pt}
\small
\setlength{\tabcolsep}{6pt}
\renewcommand{\arraystretch}{1.15}
\begin{tabular}{@{}c r c c c c@{}}
\toprule
&
&
\multicolumn{2}{c}{\textbf{Uniform magnetization}}
&
\multicolumn{2}{c}{\textbf{Tip-magnetized}}
\\
\cmidrule(lr){3-4}
\cmidrule(l){5-6}
$h_c$ (mm)
&
Elements
&
$|\bu|_{\mathrm{avg}}$
&
Diff. (\%)
&
$|\bu|_{\mathrm{avg}}$
&
Diff. (\%)
\\
\midrule
0.15 & 2,334  & 7.76 & 15.4      & 2.91 & 29.8      \\
0.10 & 4,934  & 9.15 & 0.2       & 4.32 & 4.4       \\
0.08 & 7,562  & 9.23 & 0.6       & 4.36 & 5.3       \\
0.06 & 13,494 & 9.17 & Reference & 4.14 & Reference \\
\bottomrule
\end{tabular}
\end{table}

For the uniformly magnetized beam, the selected mesh differed from the reference solution by only $0.6\%$. For the tip-magnetized beam, the corresponding difference was $5.3\%$. In contrast, the coarsest mesh produced errors of $15.4\%$ and $29.8\%$, respectively. Although the displacement predictions do not converge monotonically with mesh refinement, the results indicate that the deformation response becomes largely mesh independent for mesh sizes of $h_c \le 0.10~\mathrm{mm}$.

Based on these results, a characteristic mesh size of $h_c=0.08~\mathrm{mm}$ was adopted for all subsequent beam evaluation studies. The selected mesh contained approximately $3{,}780$ nodes and $7{,}562$ elements, while the reference mesh contained approximately $6{,}746$ nodes and $13{,}494$ elements. This discretization provided a suitable balance between computational cost and solution accuracy while maintaining a consistent mesh across all constitutive and structure--property model comparisons.

\subsection{Wheel and gripper benchmarks}

The wheel and gripper geometries shown in \cref{fig:wheelGripperSetup} were analyzed using progressively refined meshes generated with Gmsh. The domains were discretized using unstructured first-order triangular finite elements, and the finest mesh for each benchmark was used as the reference solution.

Convergence was assessed using the average displacement magnitude measured along the evaluation boundary associated with each benchmark. For the wheel benchmark, characteristic mesh sizes of $h_c=0.10$, $0.08$, $0.06$, and $0.045~\mathrm{mm}$ were considered. For the gripper benchmark, characteristic mesh sizes of $h_c=0.10$, $0.08$, and $0.06~\mathrm{mm}$ were considered. The resulting displacement predictions and relative differences with respect to the reference meshes are summarized in \cref{tab:meshConvWheelGripper}.

\begin{table}[!tbp]
\centering
\caption{Wheel and gripper benchmark convergence study.}
\label{tab:meshConvWheelGripper}
\vspace{-5pt}
\small
\setlength{\tabcolsep}{5pt}
\renewcommand{\arraystretch}{1.15}
\begin{tabular}{@{}c r c c @{\hspace{1.0cm}} c r c c@{}}
\toprule
\multicolumn{4}{c}{\textbf{Wheel}}
&
\multicolumn{4}{c}{\textbf{Gripper}}
\\
\cmidrule(lr){1-4}
\cmidrule(l){5-8}
$h_c$ (mm)
&
Elements
&
$|\bu|_{\mathrm{avg}}$
&
Diff. (\%)
&
$h_c$ (mm)
&
Elements
&
$|\bu|_{\mathrm{avg}}$
&
Diff. (\%)
\\
\midrule
0.10  & 24,717  & 6.73 & 15.4      & 0.10 & 68,514  & 17.87 & 6.0       \\
0.08  & 36,973  & 7.16 & 10.0      & 0.08 & 107,268 & 18.48 & 2.8       \\
0.06  & 65,583  & 8.02 & 0.8       & 0.06 & 190,252 & 19.00 & Reference \\
0.045 & 113,483 & 7.96 & Reference & --   & --      & --    & --        \\
\bottomrule
\end{tabular}
\end{table}

For the wheel benchmark, the selected mesh differed from the reference solution by $0.8\%$, whereas the coarsest mesh produced an error of $15.4\%$. For the gripper benchmark, the selected mesh differed from the reference solution by $2.8\%$, while the coarsest mesh produced an error of $6.0\%$. These results indicate that both benchmark problems exhibit a largely mesh-independent response for the selected discretizations.

Based on this study, characteristic mesh sizes of $h_c=0.06~\mathrm{mm}$ and $h_c=0.08~\mathrm{mm}$ were adopted for the wheel and gripper benchmarks, respectively. The selected wheel mesh contained approximately $32{,}460$ nodes and $65{,}583$ elements, while the selected gripper mesh contained approximately $53{,}627$ nodes and $107{,}268$ elements. These discretizations provided a suitable balance between computational cost and solution accuracy while maintaining consistent meshes across all constitutive and structure--property model comparisons.

\section{Details of constitutive model selection}\label{s:appModelSelection}

The data-processing and simulation details used in \cref{s:modelSelection} are provided here.

\subsection{Experimental data extraction and processing}\label{ss:gonzalezData}

\begin{figure}[h]
    \centering
    \includegraphics[width=\linewidth]{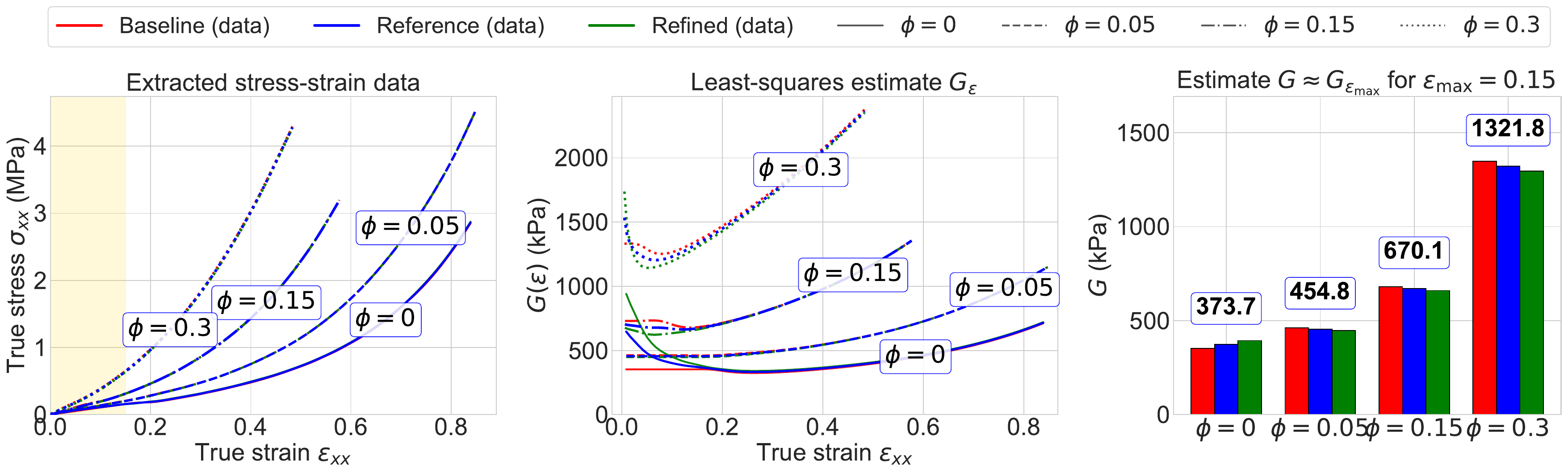}
    \caption{Experimental data used for constitutive model selection. Left: extracted true stress--strain curves at four particle volume fractions, with the shaded region indicating the calibration window $\varepsilon_{xx}\in[0,\,0.15]$. Center: least-squares shear-modulus estimate as a function of the upper strain limit. Right: shear-modulus estimates at $\varepsilon_{\max}=0.15$; the labels correspond to the reference dataset.}
    \label{fig:gonzalezDataExtract}
\end{figure}

In \cite[Fig.~1]{garcia2021influence}, true stress--strain curves are reported for dog-bone specimens with a mixing ratio of $5{:}1$ and four particle volume fractions $\phi \in \{0,\,0.05,\,0.15,\,0.30\}$. The figure was digitized by calibrating the figure axes and then tracing the curves. Because digitization is sensitive to axis scaling and to missing or noisy points at very small strain, two independent digitization passes were constructed and combined rather than relying on a single extraction. These two digitization passes are referred to as the baseline and refined datasets. The extracted data were cleaned and regularized to obtain true stress--strain curves over a common strain range and strain interval. The two datasets were then averaged to obtain the reference dataset. The extracted curves and resulting modulus estimates are presented in \cref{fig:gonzalezDataExtract}.

The shear modulus is estimated using $G=E/3$, corresponding to an incompressible material, where Young's modulus is obtained using the relation below:
\begin{equation*}
    E = \frac{\dd \sigma}{\dd \varepsilon}.
\end{equation*}
For a fixed $\phi \in \{0,\,0.05,\,0.15,\,0.30\}$, the modulus calculation is restricted to the small-strain regime by setting $\varepsilon_{\max}=0.15$. A direct derivative of the stress--strain curve near zero strain can be sensitive to noise and to the choice of the first nonzero data point. Therefore, the modulus is estimated by fitting a line through the origin over the interval $(0,\bar{\varepsilon}]$.

For a fixed upper strain limit $\bar{\varepsilon}$, let $\{(\varepsilon_i,\sigma_i)\}$ denote all data points with $\varepsilon_i\in(0,\bar{\varepsilon}]$. The line $\sigma(\varepsilon)=E^{\mathrm{data}}_{\bar{\varepsilon}}\varepsilon$ is obtained by minimizing
\begin{equation*}
    \sum_{i,\,\varepsilon_i \in (0,\bar{\varepsilon}]} \left( \sigma_i - E^{\mathrm{data}}_{\bar{\varepsilon}} \varepsilon_i \right)^2.
\end{equation*}
The normal equation gives
\begin{equation*}
    E^{\mathrm{data}}_{\bar{\varepsilon}}
    =
    \frac{\sum_{i,\,\varepsilon_i \in (0,\bar{\varepsilon}]} \varepsilon_i \sigma_i}
         {\sum_{i,\,\varepsilon_i \in (0,\bar{\varepsilon}]} \varepsilon_i^2},
    \qquad
    G^{\mathrm{data}}_{\bar{\varepsilon}}
    =
    \frac{E^{\mathrm{data}}_{\bar{\varepsilon}}}{3}
    =
    \frac{1}{3}
    \frac{\sum_{i,\,\varepsilon_i \in (0,\bar{\varepsilon}]} \varepsilon_i \sigma_i}
         {\sum_{i,\,\varepsilon_i \in (0,\bar{\varepsilon}]} \varepsilon_i^2}.
\end{equation*}
Repeating this calculation for each volume fraction gives $E^{\mathrm{data}}_{\bar{\varepsilon}}(\phi)$ and $G^{\mathrm{data}}_{\bar{\varepsilon}}(\phi)$. The shear moduli $G^{\mathrm{data}}_{\varepsilon_{\max}}(\phi)$ with $\varepsilon_{\max}=0.15$ are shown in \cref{fig:gonzalezDataExtract}. For $\phi=0$, the shear modulus of the host matrix is set to
\begin{equation*}
    G_0 = G^{\mathrm{data}}_{\varepsilon_{\max}}(0) = 373.7\,\mathrm{kPa}.
\end{equation*}
This value is used as the reference matrix shear modulus in Metric~1, while Metric~2 considers this value together with two nearby values of $G_0$ to assess whether the selected model remains robust to modest variations in the matrix shear modulus.

\subsection{Metric~1 details}\label{ss:metric1Details}

Metric~1 quantifies how well each candidate $G(\phi)$ relation reproduces the extracted shear modulus $G^{\mathrm{data}}_{\varepsilon_{\max}}(\phi)$. For each model, the mean and standard deviation of \cref{eq:metric1rE} are computed over $\phi \in \{0,\,0.05,\,0.15,\,0.30\}$. The results are shown in \cref{tab:modelSummary} under the column heading ``Metric~1''. The reference dataset is used to report the error values and to set $G_0$. The baseline and refined datasets are included in \cref{fig:metric1Gmodels} to show the small variation due to digitization.

\subsection{Uniaxial simulation setup for Metric~2}\label{ss:uniaxialSetup}

\begin{figure}[h]
    \centering
    \includegraphics[width=\linewidth]{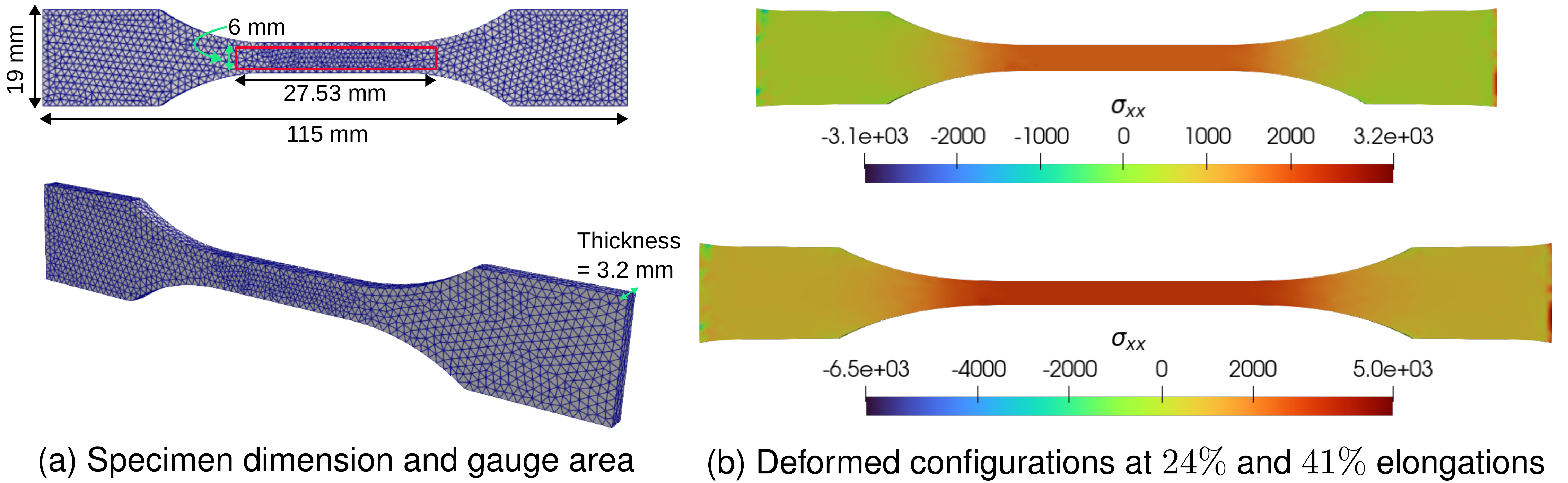}
    \caption{Uniaxial dog-bone validation setup. \textbf{Left}: geometry based on the ISO~I / ASTM Type~I profile used in \citet{garcia2021influence,garcia2019magneto,barba2020temperature}; boundary conditions; and finite-element mesh (11\,976 elements, 3\,384 nodes) with gauge-region refinement. \textbf{Right}: Cauchy stress $\sigma_{xx}$ at two load steps, corresponding to percentage elongations of $24\%$ and $41\%$. The results correspond to the Mooney model, \textbf{NH2} energy, uniform particle volume fraction $\phi = 0.3$, and matrix shear modulus $G_0 = 373.7$\,kPa.}
    \label{fig:uniaxialSetup}
\end{figure}

To test whether the same models reproduce the full uniaxial response, the dog-bone experiments of \citet{garcia2021influence,garcia2019magneto,barba2020temperature} are simulated with the finite-element model described in \cref{s:setup}. \cref{fig:uniaxialSetup} summarizes the problem: an ISO~I / ASTM Type~I tensile specimen (115\,mm $\times$ 19\,mm $\times$ 3.2\,mm overall; gauge section 27.53\,mm $\times$ 6\,mm $\times$ 3.2\,mm) is meshed with refinement in the gauge ($h_c = 1.0$\,mm) and coarser discretization in the grips ($h_c = 2.5$\,mm). The discretization contains 11\,976 elements and 3\,384 nodes. The left end is fixed, and the right end is subjected to a displacement-controlled stretch in the loading ($x$) direction with zero applied magnetic field ($\bB^a=\bzero$). Gauge true strain and average Cauchy stress $\sigma_{xx}$ in the gauge section are recorded during the test. The $\sigma_{xx}$ field in \cref{fig:uniaxialSetup} shows that the stress is approximately uniform in the gauge region.

Based on the sensitivity studies in \cref{s:modelDependence}, the strain-energy density is fixed to the neo-Hookean model \textbf{NH2} in \cref{eq:NH2}. The bulk modulus is taken as $K(\phi)=K_0$, consistent with the constitutive formulation in \cref{ss:strainEnergyModels}. All seven $G(\phi)$ models in \cref{tab:modelSummary}, including \textbf{Default}, are evaluated at $\phi \in \{0,\,0.05,\,0.15,\,0.30\}$. Simulations are run for \blue{three values of the matrix shear modulus},
\begin{equation*}
    G_0 \in \{373.7,\,400,\,420\}\,\mathrm{kPa},
\end{equation*}
which allows the model comparison to assess sensitivity to modest variations in the matrix shear modulus. The target engineering strain on the overall specimen length is $200\%$, applied in 58 equal displacement increments, with $\Delta u_x = 4$\,mm per step. No magnetic loading is applied, so the test isolates the mechanical $G(\phi)$ model embedded in \textbf{NH2}.

\subsection{Metric~2 averaging definitions}\label{ss:computeRMSE}

For each $\phi$, $G_0$, and $G(\phi)$ model, the simulated curve is compared with the reference data using
\begin{equation*}
    \mathrm{rRMSE}(\phi,G_0,\mathrm{model})
    =
    \sqrt{
    \frac{1}{N}
    \sum_{i=1}^{N}
    \left(
    \frac{
    \sigma^{\mathrm{model}}(\varepsilon_i^{\mathrm{data}})
    -
    \sigma^{\mathrm{data}}_i
    }
    {\sigma^{\mathrm{data}}_i}
    \right)^2
    },
\end{equation*}
where $\sigma^{\mathrm{model}}(\varepsilon_i^{\mathrm{data}})$ denotes the simulated stress interpolated at the experimental strain value $\varepsilon_i^{\mathrm{data}}$. Points with $|\sigma^{\mathrm{data}}| \le 10^{-3}\,\mathrm{MPa}$ are excluded to avoid division by values near zero.

Averaging over \blue{the three values of $G_0$} at fixed $\phi$ gives
\begin{equation*}
    \bar{\mathrm{rRMSE}}(\phi, \mathrm{model})
    =
    \frac{1}{3}
    \sum_{i=1}^{3}
    \mathrm{rRMSE}(\phi, G_{0,i}, \mathrm{model}),
\end{equation*}
with standard deviation $\bar{\mathrm{rRMSE}}_\sigma(\phi, \mathrm{model})$ computed from the same set $\{\mathrm{rRMSE}(\phi, G_{0,i}, \mathrm{model})\}_{i=1}^{3}$. This pair summarizes model error at a given $\phi$ while accounting for uncertainty in $G_0$.

The double mean over $\phi$ and $G_0$ is
\begin{equation*}
    \hat{\bar{\mathrm{rRMSE}}}(\mathrm{model})
    = \frac{1}{4}\sum_{j=1}^{4} \bar{\mathrm{rRMSE}}(\phi_j, \mathrm{model}),
\end{equation*}
with standard deviation $\hat{\bar{\mathrm{rRMSE}}}_\sigma(\mathrm{model})$ taken over the same four values $\{\bar{\mathrm{rRMSE}}(\phi_j, \mathrm{model})\}_{j=1}^{4}$.

The Metric~2 columns in \cref{tab:modelSummary} report $\hat{\bar{\mathrm{rRMSE}}}(\mathrm{model})$ and $\hat{\bar{\mathrm{rRMSE}}}_\sigma(\mathrm{model})$ for each $G(\phi)$ model. \blue{At $\phi=0$, all seven relations give $G(0)=G_0$. With the same \textbf{NH2} energy and $G_0$, they produce the same stress--strain response and the same rRMSE.}

\begin{figure}[h]
    \centering
    \includegraphics[width=0.7\linewidth]{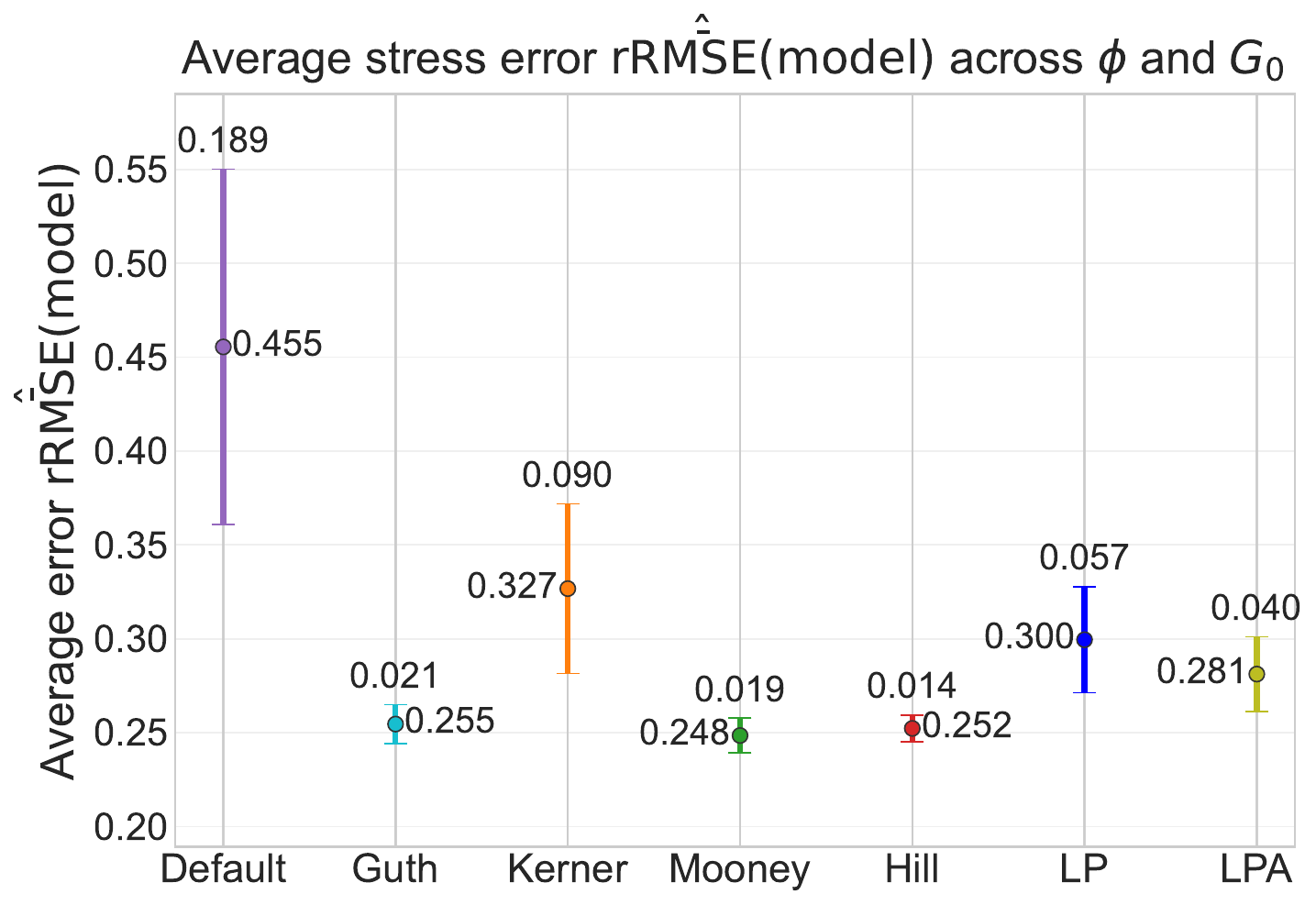}
    \caption{Summary of the rRMSE for seven $G(\phi)$ models. The plot reports the mean rRMSE over $\phi\in\{0,\,0.05,\,0.15,\,0.30\}$ and $G_0\in\{373.7,\,400,\,420\}\,\mathrm{kPa}$. Labels denote the mean values, and vertical bars denote the standard deviation over $\phi$.}
    \label{fig:metric2Summary}
\end{figure}

\cref{fig:metric2Summary} summarizes the full stress--strain comparison after averaging over the tested volume fractions and \blue{values of the matrix shear modulus}. The \textbf{Default} model gives the largest mean rRMSE, confirming that a $\phi$-dependent modulus relation is required. \blue{\textbf{Mooney} gives the smallest mean rRMSE ($0.248$), followed by \textbf{Hill} ($0.252$) and \textbf{Guth} ($0.255$).}

\begin{figure}[H]
    \centering
    \includegraphics[width=0.8\linewidth]{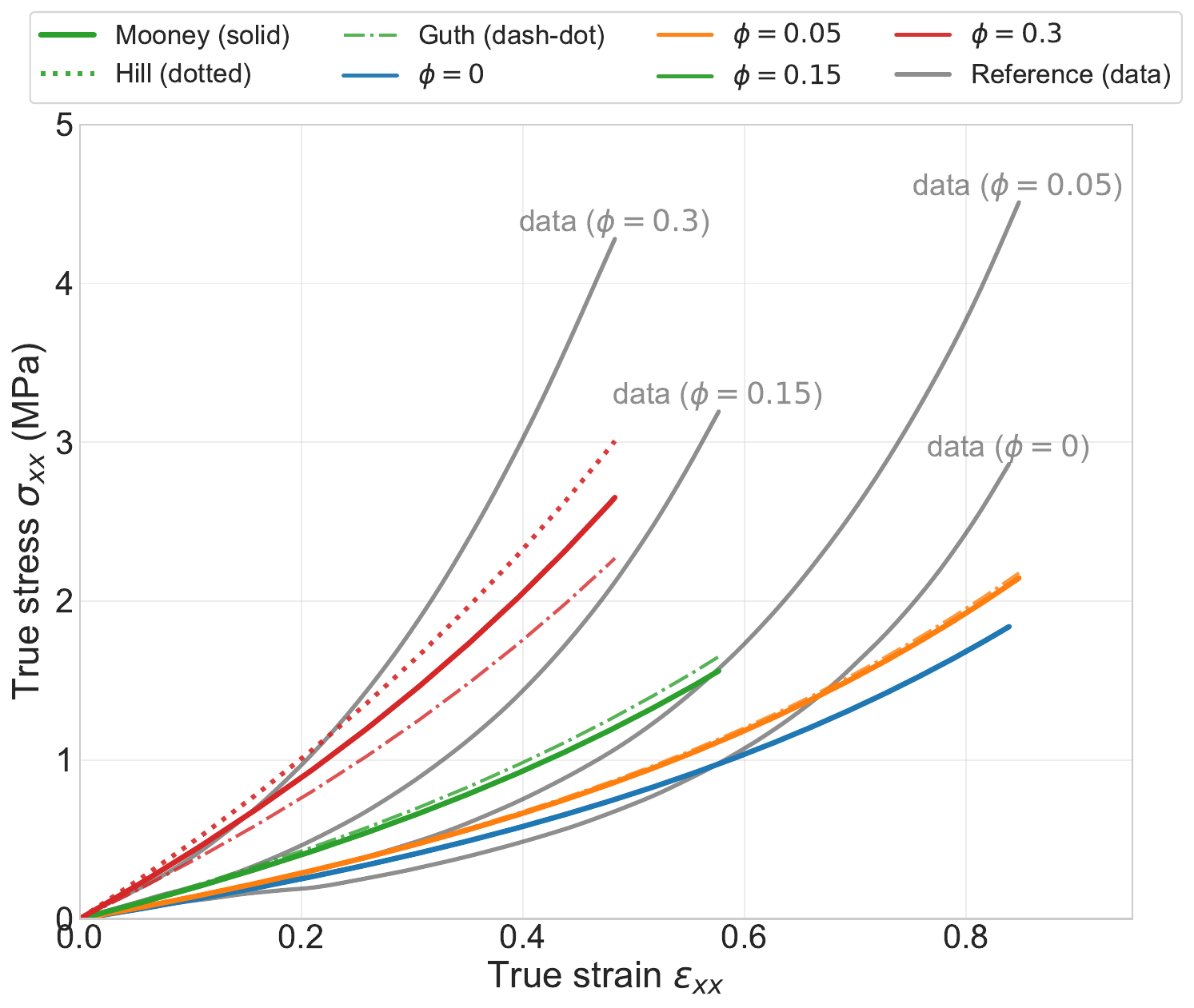}
    \caption{Reference stress--strain data and \textbf{NH2} simulations at $G_0=373.7\,\mathrm{kPa}$ for Mooney, Hill, and Guth. Color encodes the particle volume fraction $\phi$, and line style encodes the $G(\phi)$ model.}
    \label{fig:metric2StressStrain}
\end{figure}

\cref{fig:metric2StressStrain} shows representative stress--strain curves for the \blue{three relations with the smallest mean rRMSE}, namely \textbf{Mooney}, \textbf{Hill}, and \textbf{Guth}, using $G_0=373.7\,\mathrm{kPa}$. The three models are close at small and moderate volume fractions, but their differences become more visible at $\phi=0.30$. \textbf{Hill} gives a slightly stiffer response at the largest volume fraction, while \textbf{Mooney} provides the best overall balance across the full set of volume fractions and $G_0$ values.

\section{Restorative-beam optimization sensitivity}\label{s:appBeamSensitivity}

{
\color{black}
The restorative-beam problem in \cref{ss:restorativeBeam} was solved with the same geometry, two load cases, volume bounds, and density projection as in the main example. The comparison uses the \textbf{NH2} strain-energy function and $G_0=100~\mathrm{kPa}$. Each saved final physical field is interpolated onto a $200\times40$ quadrilateral mesh and evaluated with the indicated $G(\phi)$ relation, without further optimization. The tables report the resulting two-load-case compliance on this common mesh, so designs from different optimization meshes can be compared using the same discretization.

\blue{The $G(\phi)$ study uses an MMA move limit of $0.05$ and $100$ optimization iterations. The mesh study uses a move limit of $0.005$ and $100$ iterations, while the initialization study uses the same move limit and $200$ iterations. Within each study, these settings are held fixed across all cases.}

\subsection{Effective shear-modulus relation}
The beam was optimized separately with the Mooney, Guth, and Hill relations on a $150\times30$ mesh, starting each run from the same uniform raw fields $(\rho,\phi,\theta_{B^r})=(0.50,0.10,0)$. The resulting physical fields were held fixed and each design was evaluated under all three relations. \blue{\Cref{tab:beamGcross} reports the compliances evaluated on the $200\times40$ mesh; rows identify the relation used in optimization and columns the relation used for evaluation.}

\begin{table}[!tbp]
\centering
\caption{Two-load-case compliance of restorative-beam designs obtained using three $G(\phi)$ relations. Each design is evaluated using each relation on the common $200\times40$ mesh.}
\label{tab:beamGcross}
\vspace{-5pt}
\small
\setlength{\tabcolsep}{7pt}
\renewcommand{\arraystretch}{1.15}
\begin{tabular}{@{}l c c c@{}}
\toprule
Optimization relation & \multicolumn{3}{c}{Evaluation relation} \\
\cmidrule(l){2-4}
& Mooney & Guth & Hill \\
\midrule
Mooney & 49.25 & 50.76 & 47.47 \\
Guth   & 48.28 & 49.26 & 46.80 \\
Hill   & 49.37 & 51.15 & 47.42 \\
\bottomrule
\end{tabular}
\end{table}

\blue{For each column of} \cref{tab:beamGcross}, the relative compliance range is computed as $(f_{\mathrm{comp}}^{\max}-f_{\mathrm{comp}}^{\min})/f_{\mathrm{comp}}^{\min}$ over the three designs. The largest value is $3.8\%$, obtained when the designs are evaluated using the Guth relation.

\subsection{Initialization}
Four feasible \blue{initializations of the raw design fields} were tested on the $150\times30$ mesh using the Mooney relation. The first two hold $\rho=0.50$ and $\phi=0.10$ uniformly, at their domain-average volume bounds, and differ only in the constant initial remanent direction. \blue{The third initialization uses uniform values $\rho=0.35$ and $\phi=0.05$.} For the spatially varying start, with $L=100~\mathrm{mm}$, $H=20~\mathrm{mm}$, and $\blue{[x,y]^T}$ measured from the lower left corner, the raw fields are
\begin{equation*}
\begin{aligned}
\rho_0(x,y)&=0.50+0.15\cos(2\pi x/L)\cos(\pi y/H),\\
\phi_0(x,y)&=0.10+0.05\sin(2\pi x/L)\cos(\pi y/H),\\
\theta_{B^r,0}(x,y)&=\frac{\pi}{2}\sin(2\pi x/L)\cos(\pi y/H).
\end{aligned}
\end{equation*}
The four outcomes are compared in \cref{tab:beamInit}. In particular, the spatially varying start yields the lowest compliance of this set; similar results from the \blue{two uniform initializations differing only in remanent direction} should not be interpreted as independence from initialization more generally.

\begin{table}[!tbp]
\centering
\caption{Initialization study. Compliance is evaluated on the common $200\times40$ mesh using the Mooney relation. The spatially varying fields are defined in the text.}
\label{tab:beamInit}
\vspace{-5pt}
\small
\setlength{\tabcolsep}{7pt}
\renewcommand{\arraystretch}{1.15}
\begin{tabular}{@{}l c c c c@{}}
\toprule
Initial fields & $\rho_0$ & $\phi_0$ & $\theta_{B^r,0}$ & Compliance \\
\midrule
Uniform, horizontal & 0.50 & 0.10 & $0$ & 51.52 \\
Uniform, vertical & 0.50 & 0.10 & $\pi/2$ & 49.83 \\
\blue{Uniform, lower $\rho_0$ and $\phi_0$} & 0.35 & 0.05 & $0$ & 53.69 \\
Spatially varying & see text & see text & see text & 45.48 \\
\bottomrule
\end{tabular}
\end{table}

\subsection{Optimization mesh}
\blue{The beam was optimized using the Mooney relation from the uniform initial fields specified in \cref{ss:restorativeBeam} on three structured quadrilateral meshes. \Cref{tab:beamMesh} reports the compliances of the three final designs evaluated on the same $200\times40$ mesh, allowing comparison using a common discretization. The compliances of the designs optimized on the $150\times30$ and $200\times40$ meshes differ by approximately $0.7\%$.}

\begin{table}[H]
\centering
\caption{Optimization mesh study. All saved designs are evaluated on the common $200\times40$ mesh.}
\label{tab:beamMesh}
\vspace{-5pt}
\small
\setlength{\tabcolsep}{9pt}
\renewcommand{\arraystretch}{1.15}
\begin{tabular}{@{}c c c@{}}
\toprule
Optimization mesh & Element size (mm) & Compliance \\
\midrule
$100\times20$ & 1.00 & 94.61 \\
$150\times30$ & 0.67 & 91.37 \\
$200\times40$ & 0.50 & 90.73 \\
\bottomrule
\end{tabular}
\end{table}
}

\blue{\section{Magnetic-map comparison for the restorative beam}\label{s:appMagneticMaps}

\subsection{Polar rotation and differentiation}
For $\det\bF>0$, the right polar decomposition gives $\bR=\bF\bU^{-1}$ with $\bU=(\bF^T\bF)^{1/2}$ positive definite. This definition applies in two and three dimensions. The two-dimensional beam uses the closed form
\begin{equation}
\bR(\bF)=\frac{1}{\sqrt{a^2+b^2}}
\begin{bmatrix}a&-b\\b&a\end{bmatrix},
\qquad a=F_{11}+F_{22},\qquad b=F_{21}-F_{12}.
\end{equation}
The selected magnetic energy is differentiated by automatic differentiation to obtain the stress, tangent, and design sensitivities. In the rotation-based model this includes the derivative of $\bR(\bF)$.

\subsection{Measures of design difference}
The measures use the final physical nodal fields of both designs on the common $150\times30$ mesh. Subscripts $F$ and $R$ denote the designs optimized with the corresponding magnetic energies. For $q=\bar{\rho}$ or $\phi_{\mathrm{phys}}$, the relative discrete $L^2$ difference is
\begin{equation}
 d_q=\frac{\left[\sum_i(q_{R,i}-q_{F,i})^2\right]^{1/2}}
 {\left[\sum_iq_{F,i}^2\right]^{1/2}}.
\end{equation}
The angular difference accounts for the $2\pi$ periodicity of the remanent direction:
\begin{equation}
\delta\theta_i=\operatorname{atan2}\!\left(\sin(\theta_{R,i}-\theta_{F,i}),\cos(\theta_{R,i}-\theta_{F,i})\right),
\qquad
w_i=\frac{(\bar{\rho}\phi_{\mathrm{phys}})_{R,i}+(\bar{\rho}\phi_{\mathrm{phys}})_{F,i}}{2},
\end{equation}
\begin{equation}
\Delta\theta_{\mathrm{rms}}=
\left(\frac{\sum_i w_i\delta\theta_i^2}{\sum_i w_i}\right)^{1/2}.
\end{equation}
These nodal measures give $d_\rho=0.08473$, $d_\phi=0.06126$, and $\Delta\theta_{\mathrm{rms}}=0.14414~\mathrm{rad}$ ($8.26^\circ$). The norms and angular averages use nodal sums.}

\end{document}